\documentclass[prc,reprint,nofootinbib,preprintnumbers,floatfix,unsortedaddress,superscriptaddress,longbibliography,times]{revtex4-2}
\ifx\pdfsuppresswarningpagegroup\undefined\else\pdfsuppresswarningpagegroup=1\fi\relax
\PassOptionsToPackage{unicode, pdfprintscaling=None, colorlinks}{hyperref}

\usepackage{orcidlink}
\usepackage{multirow}
\usepackage{siunitx}
\usepackage{colortbl}
\usepackage{dcolumn}

\usepackage{tikz}
\usetikzlibrary{decorations.markings, arrows, decorations.pathmorphing}

\usepackage{graphicx}
\usepackage{latexsym}
\usepackage{amsmath,amssymb}
\usepackage{amsfonts}
\usepackage{bm}
\usepackage{bbm}
\usepackage{microtype}
\usepackage{xspace}
\usepackage{placeins}

\usepackage{hyperref}
\usepackage[capitalise]{cleveref}

\usepackage{mathtools}

\usepackage[export]{adjustbox}

\newlength{\FigureWidth}
\usepackage{xcolor}

\usepackage{makecell}
\usepackage{tabularx}
\newcolumntype{L}[1]{>{\raggedright\arraybackslash}p{#1}}
\newcolumntype{C}[1]{>{\centering\arraybackslash}p{#1}}
\newcolumntype{R}[1]{>{\raggedleft\arraybackslash}p{#1}}
\newcommand\TopRule{\Xhline{0.08em}}

\newcommand\MidRule{\Xhline{0.03em}}
\newcommand\BotRule{\Xhline{0.08em}}

\newcommand{\br}{{\mathbf{r}}}
\newcommand{\bbr}{{\bar{\mathbf{r}}}}
\newcommand{\bd}{{\pmb{\delta}}}
\newcommand{\bk}{{\mathbf{k}}}
\newcommand{\bq}{{\mathbf{q}}}
\newcommand{\bbq}{{\bar{\mathbf{q}}}}
\newcommand{\bp}{{\mathbf{p}}}
\newcommand{\bn}{{\mathbf{n}}}

\newcommand{\sgn}{{\gamma}}
\newcommand{\bwt}{{\omega}}
\newcommand{\Mel}{{\cal M}}

\newcommand{\He}{{^3\text{He}}}
\newcommand{\nopi}{\ensuremath{\pi\hskip-0.40em /}}
\newcommand{\eftnopi}{EFT$_{\nopi}$\xspace}
\newcommand{\ua}{{\uparrow}}
\newcommand{\da}{{\downarrow}}

\renewcommand\vec\mathbf

\newcommand\Nud[1]{{N^\dagger_{#1}}}
\newcommand\tNud[1]{{\tilde{N}^\dagger_{#1}}}
\newcommand\NuT[1]{{N^T_{#1}}}
\newcommand\Nu[1]{{N^{\phantom{\dagger}}_{#1}}}

\newcommand\ptd[1]{{p^\dagger_{#1}}}
\newcommand\pt[1]{{p^{\phantom{\dagger}}_{#1}}}
\newcommand\ntd[1]{{n^\dagger_{#1}}}
\newcommand\nt[1]{{n^{\phantom{\dagger}}_{#1}}}

\newcommand\ProjS[1]{\bar{P}^{#1}}
\newcommand\ProjT[1]{P^{#1}}

\newcommand{\mN}{{m_N}}
\newcommand{\Mp}{{M_{\rm phys}}}

\newcommand{\Lp}{{L_{\rm phys}}}
\newcommand{\Lpsq}{{L^2_{\rm phys}}}
\newcommand{\Czerosing}{\ensuremath{C_0^{(^1 \! S_0)}}\xspace}
\newcommand{\Czerotrip}{\ensuremath{C_0^{(^3 \! S_1)}}\xspace}
\newcommand{\Ctb}{{C_{\rm 3B}}}

\usepackage{relsize}

\usepackage[acronyms, nohypertypes={acronym}, nopostdot, style=super, nonumberlist, toc]{glossaries}
\setacronymstyle{long-sc-short}
\glsenableentrycount
\makeglossaries  

\newcommand\ac[1]{\gls{#1}}

\newacronym{PNA}{pna}{particle-number-algorithm}
\newacronym{SFA}{sfa}{spin-flip-algorithm}

\newacronym{WF}{wf}{Wilson-Fisher}
\newacronym{AF}{af}{asymptotically free}

\newacronym{RG}{rg}{renormalization group}

\newacronym{QIS}{qis}{Quantum Information Science}
\newacronym{PPT}{ppt}{positive-semidefinite partial transpose}
\newacronym{NPT}{npt}{negative partial transpose}

\newacronym[longplural={conformal field theories}]{CFT}{cft}{conformal field theory}
\newacronym[longplural={lattice field theories}]{LFT}{lft}{lattice field theory}
\newacronym[longplural={effective field theories}]{EFT}{eft}{effective field theory}
\newacronym[longplural={quantum field theories}]{QFT}{qft}{quantum field theory}

\newacronym[]{DMRG}{dmrg}{Density Matrix Renormalization Group}
\newacronym[]{TFIM}{tfim}{Transverse Field Ising Model}

\newacronym[]{LOCC}{locc}{Local Operations and Classical Communicaton}
\newacronym[]{OBC}{obc}{open boundary conditions}

\newacronym{MPS}{mps}{matrix product states}

\newacronym{JLP}{jlp}{Jordan-Lee-Preskill}

\newacronym{BBN}{bbn}{big bang nucleosynthesis}

\newacronym{LEC}{lec}{low-energy constant}

\newacronym{QCD}{qcd}{quantum chromodynamics}
\newacronym{MC}{mc}{Monte Carlo}

\newacronym{IR}{ir}{infrared}
\newacronym{UV}{uv}{ultraviolet}

\newacronym{QED}{qed}{quantum electrodynamics}
\newacronym{SNR}{snr}{signal-to-noise ratio}

\newacronym{NLSM}{nlsm}{nonlinear sigma model}

\newacronym{CL}{cl}{Complex Langevin}

\newacronym{CSA}{csa}{Cartan subalgebra}

\newacronym{SSB}{ssb}{spontaneous symmetry breaking}

\newacronym{AFQMC}{afqmc}{auxiliary field quantum Monte Carlo}
\newacronym{iHMC}{ihmc}{imaginary-mass Hybrid Monte Carlo}

\newacronym{MCMC}{mcmc}{Markov Chain Monte Carlo}

\newacronym{QI}{qi}{quantum information}

\begin{document}


\title{Worldline Monte Carlo method for few body nuclear physics}

\author{Shailesh Chandrasekharan\orcidlink{0000-0002-3711-4998}}
\email{sch27@duke.edu}
\affiliation{ Department of Physics, Box 90305, Duke University, Durham, North Carolina 27708, USA}
\author{Son T. Nguyen\orcidlink{0000-0002-6104-7035}}%
 \email{snguyen@wlu.edu}
\affiliation{ Department of Physics, Box 90305, Duke University, Durham, North Carolina 27708, USA}
\affiliation{
Department of Physics and Engineering, Washington and Lee University, Lexington, Virginia 24450, USA
}
\author{Thomas R.~Richardson\orcidlink{0000-0001-6314-7518}}
\email{richardt@uni-mainz.de}
\affiliation{ Department of Physics, Box 90305, Duke University, Durham, North Carolina 27708, USA}
\affiliation{Institut f\"ur Kernphysik and PRISMA$^+$ Cluster of Excellence, Johannes Gutenberg-Universit\"at, 55128 Mainz, Germany}%

\date{\today}

\begin{abstract}
In this work we introduce a worldline based fermion Monte Carlo algorithm for studying few body quantum mechanics of self-interacting fermions in the Hamiltonian lattice formulation. Our motivation to construct the method comes from our interest in studying renormalization of chiral nuclear effective field theory with lattice regularization. In particular we wish to apply our method to compute the lattice spacing dependence of local lattice interactions as we take the continuum limit of the lattice theory. Our algorithm can compute matrix elements of the operator $\exp(-\beta H)$ where $H$ is the lattice Hamiltonian and $\beta$ is a free real parameter. These elements help us compute deep bound states that are well separated from scattering states even at values of $\beta$ which are not very large. Computing these bound state energies accurately can help us study renormalization of the lattice theory. In addition to developing the algorithm, in this work we also introduce a finite volume renormalization scheme for the lattice Hamiltonian of the leading pionless effective field theory and show how it would work in the one and two body sectors.
\end{abstract}

\maketitle

\section{Introduction}
\label{sec1}

Fermion Monte Carlo methods have a long history \cite{PhysRevB.16.3081} and are known to be notoriously difficult to design due to the fermion sign problem \cite{PhysRevLett.94.170201}. Over the years, several approaches have been proposed to either circumvent or alleviate the challenges, and new ideas continue to emerge even today. In this work, we refine the worldline approach \cite{Wiese:1992np,PhysRevD.99.074511} and show its potential for solving a certain class of problems in few-body nuclear physics. While these problems can perhaps also be studied by other already well-known methods, we believe that our method will be a useful addition to the literature and may be refined in the future to solve new problems that may be inaccessible to the existing methods as we discuss below.

One of the most popular methods that is widely applicable is the so-called constrained path Monte Carlo method \cite{Zhang:1996us}. While this method suffers from systematic errors, recent progress suggests that these errors can be controlled in many cases and even eliminated in some parameter regimes. Reviews of the method with several new applications can be found in the recent literature \cite{zhang2013,RevModPhys.87.1067,Curry:2023sxh}. Unfortunately, constrained path methods are not easily extendable to all kinds of quantum many-body problems. For example, when fermions interact with gauge fields, constrained path algorithms can become tricky and, as far as we know, have not yet been explored. In addition to traditional gauge theories, when chiral symmetry is realized nonlinearly, pions interact with nuclons as gauge fields \cite{Chandrasekharan:2003wy}. We believe our worldline algorithms can be extended to such theories more easily.

There are of course well known exact fermion Monte Carlo methods that do not rely on the constrained path approximation. These methods are based on integrating out the fermions entirely since all problems can be formulated as though fermions are freely moving in the background of some bosonic fields like scalar, gauge, or auxiliary fields \cite{PhysRevD.24.2278,PhysRevB.34.7911,PhysRevB.36.8632,Duane1987216,Sorella1989}. Reviews of these exact fermion Monte Carlo methods from the perspective of various communities can also be found in the recent literature \cite{Kennedy:2004ae,Assaad2008,Lee:2008fa,Drut_2013}. Unfortunately, the exact fermion Monte methods face their own difficulties when sign problems remain unsolved. To solve the sign problems, some type of pairing of fermion degrees of freedom must be manifest in the formulation. New types of pairings  continue to be discovered, like the recent example where the physics of a Dirac fermion was viewed as the paired physics of two Majorana fermions \cite{PhysRevB.89.111101,PhysRevB.91.241117,PhysRevLett.117.267002,PhysRevLett.116.250601}. However, many interesting problems do not have any natural pairing mechanisms and hence suffer from sign problems and beyond the scope of these methods. Recent research has focused on new ideas for solving the sign problems like the complex Langevin method \cite{Berger:2019odf} or  contour deformation methods \cite{Alexandru:2020wrj}. Our method offers another possible approach for few body physics since we do not try to solve the sign problem, but can still extract useful numbers.

In addition to sign problems, the exact fermion Monte Carlo methods are also known to scale poorly with system size especially as one approaches critical points or continuum limits. One of the bottlenecks here is the non-local nature of the fermion determinant in the Boltzmann weight that is necessary to solve sign problems. This quantity not only fluctuates a lot near criticality, but also introduces numerical instabilities \cite{ALF:2020tyi}. Still, impressive large scale calculations using parallel supercomputers is the norm  in fields like lattice QCD today, when sign problems can be solved. Our approach is qualitatively different since we do not have large fermion determinants and hence are free of such numerical instabilities.
The reason for not having to deal with fermion determinants is that we view fermions as hard core bosons and have to only sample their worldlines. We include the fermion permutation sign into observables. Thus, while in principle our method can suffer from sign problems, we will shown in this work that in the few body sector this sign problem is rather mild. We give a heuristic argument for this below.

It is well known that bosonic Monte Carlo methods allows one to study large system sizes accurately as long as no sign problems are present \cite{RevModPhys.67.279}. This is also the reason why constrained path methods are efficient, since they too  convert fermionic problems into bosonic problems, with the caveat that the physics of the fermions must be captured well within the constrained path framework. The method of treating fermions as hard core bosons was was introduced long ago as a possible fermion Monte Carlo method \cite{Wiese:1992np}. Since then, our understanding of bosonic algorithms in the worldline formulation has matured considerably with the discovery of worm and loop algorithms \cite{PhysRevLett.87.160601,PhysRevE.66.046701,Adams:2003cc}. Such algorithms an be easily extended to hard core bosons. We can also include gauge fields rather easily \cite{Frank:2019jzv}. Thus, the worldline fermion Monte Carlo methods combines ideas from constrained path methods (nodes when two identical fermions come together are exactly implemented) with ideas of auxiliary field methods (free fermion worldlines can be summed into fermion determinants). Furthermore, it is straightforward to couple fermions with scalar and gauge fields if necessary. On the other hand, tackling the fermion sign problem head on is necessary. We note that the worldline approach we are exploring has some connections to the pinhole algorithm that has been recently proposed \cite{PhysRevLett.119.222505,PhysRevLett.125.192502}.

While our current work does not solve the fermion sign problem, sometimes they can be solved in the worldline approach. The first complete solution to the fermion sign problem in the worldline formulation was discovered using the idea of the meron-cluster algorithm \cite{PhysRevLett.83.3116}. Here space-time was split into loops such that the fermion determinant of each loop turned out to be either $0$ (meron cluster) or $2$ (normal cluster). The fermion bag algorithm extended this idea by splitting space-time into more complex bags and the fermion determinant in each bag turned out to be either zero or positive \cite{PhysRevD.82.025007}. When the fermion bag algorithms were extended to Hamiltonian problems, a simplification was observed at high temperatures \cite{PhysRevD.101.074501}. Equivalently, a dilute system of fermions would have a mild sign problem up to sufficiently low temperatures so as to be able to access the low energy physics. Fermions only permute with other fermions within some neighborhood \cite{Lee:2002sy}. This suggests that the study of few body physics on a large lattice would be a natural place to explore new applications for the fermionic wordline methods \cite{PhysRevD.99.074511}. In this work we explore this possibility further.

One place where our proposed worldline fermion Monte Carlo method could be useful is in understanding renormalization of nuclear effective field theories \cite{Kaplan:1998tg,Kaplan:1996xu,Bedaque:1998kg,Bedaque:1998km,Bedaque:1999ve,Beane:2003da} (see Refs.~\cite{Epelbaum:2008ga, Hammer:2019poc, Epelbaum:2019kcf} for recent reviews). For example, one could understand the physics of few nucleons in a finite physical volume by discretizing space on a lattice and taking the lattice spacing to zero. These calculations naturally require studies on large lattices with a fixed number of particles which means we are exploring ultra-dilute systems. In the presence of contact interactions like in chiral nuclear effective field theories, the continuum limit of the lattice theory is usually ill defined without a renormalization procedure.
Addressing the challenges one encounters in the various procedures of renormalization is an interesting field of research \cite{Epelbaum:2016ffd,Epelbaum:2018zli,Epelbaum:2020maf}. Recent work even suggests that  continuum limits may not even exist in some cases since the theory may not have a well defined limit as the cutoff is removed \cite{Gasparyan:2022isg}. Renormalization of contact interactions in non-relativistic field theory are also interesting from other perspectives \cite{Korber:2019cuq}.

From a phenomenological perspective, lattice regularization of chiral nuclear effective field theory in the Lagrangian approach on a space-time lattice is very well developed and has been used for \textit{ab initio} calculations of a wide variety of nuclei using a simple lattice model \cite{Lee:2004qd, Lee:2004si, Borasoy:2005yc, Borasoy:2006qn, Borasoy:2007vi, Borasoy:2007vk,Borasoy:2007vy, Lee:2005is, Lee:2005it, Lee:2008xsa, Lee:2020meg, Lee:2021wey, Epelbaum:2009pd, Epelbaum:2010xt, Epelbaum:2011md, Epelbaum:2009rkz, Lahde:2015ona, Klein:2015vna, Klein:2018iqa, Klein:2018lqz, Alarcon:2017zcv, Li:2018ymw, Lu:2018bat, Lu:2019nbg, Lu:2021tab, Elhatisari:2022qfr}. The results are astoundingly consistent with experimental observations, especially in light nuclei. However, most of these lattice calculations are done at one or a few lattice spacings, without a continuum extrapolation. Without such an extrapolation, one can wonder if we have discovered an excellent model for nuclear physics on the lattice or can we understand the results within the framework of continuum effective field theory. The latter necessarily requires us to understand the lattice spacing dependence of the lattice results. 

One recent example that drives home the above point is the disagreement in the theoretical calculations of the ${}^4$He monopole transition form factor between the more traditional \textit{ab initio} nuclear calculations \cite{Bacca:2012xv,PhysRevLett.130.152502} and the lattice calculations \cite{Meissner:2023cvo}; the former disagrees with the experimental data by nearly one hundred percent while the latter reproduces the data almost exactly. While the continuum calculations employ sophisticated potentials from chiral nuclear effective field theory, the lattice calculations are performed with a simple Wigner SU(4) invariant contact interaction at a single lattice spacing. It would be interesting to understand if this excellent agreement of the lattice calculation with experiments continues even in the continuum limit. Perhaps it can then teach us the limitations of the current continuum calculations.

In this work we take a step in this direction by formulating the leading order pionless effective field theory (\eftnopi) in the worldline using a lattice discretization of the continuum Hamiltonian. Our current goal is only to introduce the lattice Hamiltonian and the corresponding Monte Carlo method. We plan to carry out the full renormalization program of pionless EFT in later publications. One feature of our method is that it does not rely on any special solution to the sign problem. It only relies on being able to compute several matrix elements of the operator $\exp(-\beta H)$ accurately at some reasonable value of $\beta$. As we explain in this paper, we can extract the deeply bound well separated energy eigenvalues of the lattice Hamiltonian which we need to compute the renormalization of the couplings. More generally, our method could have applications in other fields of physics like quantum chemistry or condensed matter physics that involve repulsive interactions among fermions and those that include scalar and gauge field interactions.

Our work is organized as follows. In \cref{sec2} we introduce our lattice model and explain how we can tune the lattice parameters to reproduce a hypothetical continuum nuclear physics model. In \cref{sec3} we carry out the renormalization program exactly in the one body and two body sectors since the calculations can be done without resorting Monte Carlo calculations. With three or more nucleons, the required calculations become difficult and we need a method to compute the lowest energy eigenstate on the lattice. In \cref{sec4} we introduce the transfer matrix and discuss how we can extract the low lying spectrum using it. We then introduce our Monte Carlo method and algorithm in \cref{sec5} and devote \cref{sec6} to test our algorithm in various situations up to four particles. We choose a $2^3$ lattice to show that our method can reproduce the exactly computable matrix elements and discuss how these can help in computing the low lying spectrum. In \cref{sec7} we argue that our method easily extends to larger lattices in the  three body and four body sectors. There we also explain some of the challenges we have to overcome when we explore smaller lattice spacings. Finally in \cref{sec8} we present our conclusions.

\section{The Lattice Model}
\label{sec2}

Our lattice model is constructed in the Fock space formulation, using nucleon annihilation and creation operators $\Nu{\br,f}$ and $\Nud{\br,f}$ where $\br$ is the lattice site on a periodic cubic spatial lattice with $L$ sites in each direction and $f = 1,2,3,4$ are the four flavors that label neutrons ($n$) and protons ($p$) with spin-half components $\ua$ and $\da$. In flavor suppressed notation we can write
\begin{align}
\Nud{\br} = 
\begin{pmatrix}
\ntd{\br,\ua} & \ntd{\br,\da} & \ptd{\br,\ua} & \ptd{\br,\da}
\end{pmatrix},\quad
\Nu{\br} = 
\begin{pmatrix}
\nt{\br,\ua} \\ \nt{\br,\da} \\ \pt{\br,\ua} \\ \pt{\br,\da}
\end{pmatrix}.
\end{align}
Our lattice Hamiltonian is a sum of two terms
\begin{align}
H\ &=\ H_0+H_{\rm int},
\label{eq:Hfull}
\end{align}
where $H_0$ is the free lattice Hamiltonian that describes hopping of nucleons on the lattice and is given by 
\begin{align}
H_0 \ & =\  \ \varepsilon \ \sum_{\br,\hat{\alpha}} 
\Big(2 \Nud{\br} \Nu{\br} - \Nud{\br}\Nu{\br+\hat{\alpha}} - 
\Nud{\br+\hat{\alpha}}\Nu{\br}\Big),
\label{eq:Hfree}
\end{align}
with $\hat{\alpha}$ representing the three lattice unit vectors in the positive direction. The interactions among the nucleons are encoded in the single site interaction term 
\begin{align}
H_{\rm int}\ & =\ \varepsilon \Czerosing \ \sum_{\br,a} \ 
 \Big(\NuT{\br}(\ProjS{a})\Nu{\br}\Big)^\dagger \Big(\NuT{\br}(\ProjS{a})\Nu{\br}\Big)^{\phantom{dagger}}
\nonumber \\
& + \ \varepsilon \Czerotrip \ \sum_{\br,a} \ 
 \Big(\NuT{\br}(\ProjT{a})\Nu{\br}\Big)^\dagger \Big(\NuT{\br}(\ProjT{a})\Nu{\br}\Big)^{\phantom{dagger}}
\nonumber \\
& + \ \varepsilon C_{3B} \ \sum_{\br}\ \Big\{
\Big(\ptd{\br,\ua}\pt{\br,\ua}+
\ptd{\br,\da}\pt{\br,\da}\Big)
\ntd{\br,\da}\ntd{\br,\ua}\nt{\br,\ua}\nt{\br,\da}
\nonumber \\
& \qquad\ \ \ + \Big(\ntd{\br,\ua}\nt{\br,\ua}+
\ntd{\br,\da}\nt{\br,\da}\Big)
\ptd{\br,\da}\ptd{\br,\ua}\pt{\br,\ua}\pt{\br,\da}\ \Big\}.
\label{eq:Hint}
\end{align}
These are the three interaction couplings at leading order in \eftnopi, written is a flavor suppressed notation.

The operators $\ProjS{a}, \ProjT{a}, a=x,y,z$ are projections on the spin-singlet-isospin-triplet sector and the spin-triplet-isospin-singlet sectors and given by
\begin{equation}
\ProjS{a} = \frac{1}{2\sqrt{2}} \tau^2 \tau^a \sigma^2,\qquad
\ProjT{a} = \frac{1}{2\sqrt{2}} \sigma^2 \sigma^a \tau^2,
\label{eq:projectors}
\end{equation}
where $\tau^a$ are the Pauli matrices on the isospin space and $\sigma^a$ are the same matrices but act on the spin space. 

\begin{table}[!htb]
\centering
\renewcommand{\arraystretch}{1.4}
\setlength{\tabcolsep}{4pt}
\begin{tabular}{l|c}
\TopRule
$\Lp$ & $3.4$ fm \\
 $\Mp$ & $1634$ MeV \\
 $E^{(N)}_1$ & $40.56$ MeV \\
 $E^{(nn)}_0$ & $-17.8$ MeV \\
$E^{(pn)}_0$ & $-25.4$ MeV \\
$E^{(pnn)}_0$ & $-65.6$ MeV \\
\BotRule 
\end{tabular}
\caption{Physical parameters we use to fix our four lattice parameters in a hypothetical world with $m_N=1634$ MeV and $m_\pi= 806$ MeV. We take these values from \cite{Eliyahu:2019nkz}, which were obtained from Lattice QCD calculations performed in Ref.~\cite{NPLQCD:2012mex}. \label{tab:physparams}}
\end{table}

The parameter $\varepsilon$ absorbs the energy scale so that the parameters $\Czerosing$, $\Czerotrip$ and $\Ctb$ are dimensionless constants. Let us now explain how we fix these four parameters to connect the lattice physics with continuum physics. We first assume that our model on a cubical lattice with $L$ sites in each direction describes a continuum physical system in a periodic cubical box of physical length $\Lp$. This implies that the lattice spacing $a$ is given by $a = \Lp/L$ and the continuum limit is reached when $L$ becomes large. Next, in order to fix the four free parameters $\varepsilon$, $\Czerosing$, $\Czerotrip$ and $\Ctb$ in \cref{eq:Hfull}, we find four energy levels of physical states in the continuum periodic cubical box of physical length $\Lp$ and match them with the corresponding energies obtained from the lattice Hamiltonian $H$ for each given lattice size $L$. The four free energy levels we choose are the lowest non-zero energy of a single free nucleon of mass $\Mp$ in a periodic cubical box of length $\Lp$, which is given by 
\begin{align}
E_1^{(N)} = \frac{4\pi^2 \hbar^2}{2\mN \Lpsq},
\label{eq:1pec}
\end{align}
the lowest energy state in the di-nuetron channel with $S_z=0$ ($n^\dagger_\uparrow n^\dagger_\downarrow$) which we label as $E^{(nn)}_0$, the lowest energy state in the deuteron channel with $S_z=0$ ($n^\dagger_\uparrow n^\dagger_\downarrow$) which we label as $E^{(nn)}_0$, the triplet deuteron ground state with $S_z=1$ ($p^\dagger_\uparrow n^\dagger_\uparrow$) which we label as $E^{(pn)}_0$, and the spin half triton ground state with $S_z=1/2$ (which is a superpposition of $p^\dagger_\uparrow n^\dagger_\uparrow n^\dagger_\downarrow$ and $p^\dagger_\downarrow n^\dagger_\uparrow n^\dagger_\uparrow$) which we label as $E^{(pnn)}_0$. While these energies cannot be determined experimentally, they can be determined using lattice QCD. For example in Ref.~\cite{NPLQCD:2012mex}, lattice QCD calculations were performed in a periodic box of $\Lp = 3.4$ fm with a very heavy pion mass ($m_\pi = 806$ MeV), so that pionless EFT should be a good approximation of the theory.\footnote{A similar approach was adopted in Ref.~\cite{PhysRevLett.114.052501} to use lattice QCD data to predict the binding energies of other nuclei, and Refs.~\cite{Detmold:2021oro, Sun:2022frr, Detmold:2023lwn, Bazak:2022mjh, Eliyahu:2019nkz} have performed similar calculations in order to use the EFT to extrapolate finite volume results to infinite volume. }
The nucleon mass was calculated to be $m_N=1634$ MeV. Substituting this into \cref{eq:1pec} we obtain $E^{(N)}_1=40.56$ MeV. The ground state energy in the di-nuetron channel was found to be $E^{(nn)}_0=-17.8$ MeV. The deuteron ground state had an energy of $E^{(pd)}_0=-25.4$ MeV, and the spin half triton ground turned out to have an energy of $E^{(pnn)}_0=-65.6$ MeV. Note that the negative sign implies binding, which can be an artifact of the finite volumes and the large pion mass used in the simulations. These physical parameters are summarized in the \cref{tab:physparams}, and we will use them to fix our four lattice parameters in this work.

\section{Renormalization}
\label{sec3}

In our work we take a very simple definition of renormalization based on its application in quantum mechanics \cite{Lepage:1997cs}. We define it as the existence of the continuum limit of a lattice theory when the lattice parameters are tuned as a function of the lattice spacing $a$ while keeping a few low energy levels fixed. In this section we study the renormalization of our lattice theory in the one and two particle sectors where we can perform calculations without the need for a Monte Carlo method. 

Focusing on the one particle sector we can fix  $\varepsilon$ by computing the lowest non-zero energy of a single free nucleon in a periodic cubical box. In the continuum this was computed in \cref{eq:1pec}. On the lattice it is given by 
\begin{align}
E_{1,{\rm lat}}^{(N)} = 2\ \varepsilon(a)\ \big(1-\cos(2\pi/L)\big).
\label{eq:1pel}
\end{align}
By matching $E_{1,{\rm latt}}^{(N)} = E_1^{(N)}$ we determine 
\begin{align}
\varepsilon(a) \ =\ \frac{1}{\big(1-\cos(2\pi a/\Lp)\big)}\ \frac{\pi^2\hbar^2}{\mN \Lpsq}. 
\label{eq:varepsilon}
\end{align}
Notice that $\varepsilon(a)$ has dimensions of energy and sets the energy scale as a function of the lattice spacing.

We can similarly determine the renormalization of two particle couplings $\Czerosing(a)$, $\Czerotrip(a)$ if we can compute the lowest energy levels on a finite lattice in the dineutron and deuteron channels and match those energies to the physical values given in \cref{tab:physparams}. Both these calculations require us to consider the generic lattice Hamiltonian of the form $H = H_0 + H_{\rm int}$ where, considering the the di-neutron channel, we can redefine $H_0$ and $H_{\rm int}$ for this calculation as
\begin{align}
&H_0 = \varepsilon(a)\ 
\sum_{\br,\alpha,\sigma} 2 n^\dagger_{\br,\sigma}  n_{\br,\sigma} - (
n^\dagger_{\br,\sigma}  n_{\br+\hat{\alpha},\sigma} + n^\dagger_{\br+\hat{\alpha},\sigma} n_{\br,\sigma}) \\
&H_{\rm int} = \ \varepsilon(a) \ C(a) \ \sum_\br n^\dagger_{\br,\uparrow}n^\dagger_{\br,\downarrow} n_{\br,\downarrow}n_{\br,\uparrow}.
\end{align}
Note we have already computed $\varepsilon(a)$ in \cref{eq:varepsilon}  and the coupling $C(a)=\Czerosing(a)$. For the deuteron calculation we simply replace $n_{\br,\downarrow}$ with $p_{\br,\uparrow}$ and set $C(a)=\Czerotrip(a)$. Hence the calculation will be the same for both the channels except when the matching to the bound state energy is performed at the end using \cref{tab:physparams}.

We can reduce the two body problem into a one body problem, by defining the complete set of two particle states as
\begin{align}
|\bq, \bd \rangle \ =\ \frac{1}{\sqrt{L^3}}\sum_{\br}  n^\dagger_{\br,\uparrow}\ 
n^\dagger_{\br+\bd,\downarrow}\ e^{i (2\pi/L) \bq\cdot \br}\ |0\rangle,
\end{align}
where we have labeled the states with the center of mass momentum $\bq$ and the relative coordinate $\bd$ both of which are integer vectors. These states satisfy the orthonormality relation $\langle \bq,\bd|\bq',\bd'\rangle = \delta_{\bq,\bq'},\delta_{\bd,\bd'}$. It is then easy to verify that 
\begin{align}
\sum_\br \ n^\dagger_{\br,\sigma}  n_{\br,\sigma} |\bq, \bd \rangle & \ =\ 2|\bq, \bd \rangle \\
\sum_\br\ n^\dagger_{\br,\da}  n_{\br+\hat{\alpha},\da} |\bq, \bd \rangle & \ =\ |\bq,\bd-\hat\alpha\rangle 
\\ 
\sum_\br\ n^\dagger_{\br+\hat\alpha,\da}  n_{\br,\da} |\bq, \bd \rangle & \ =\ |\bq,\bd+\hat\alpha\rangle
\end{align}
\begin{align}
\sum_\br \ n^\dagger_{\br,\ua}  n_{\br+\hat\alpha,\ua} |\bq, \bd \rangle & \ =\ e^{i(2\pi/L)q_\alpha} |\bq,\bd+\hat\alpha\rangle \\
\sum_\br \ n^\dagger_{\br+\hat\alpha,\ua}  n_{\br,\ua} |\bq, \bd \rangle & \ =\ e^{-i(2\pi/L)q_\alpha} |\bq,\bd-\hat\alpha\rangle
\end{align}
Using these results we can show that
\begin{align}
& H |\bq,\bd\rangle \ =\   \varepsilon(a)\ C(a)\ \delta_{\bd,0}\ |\bq,\bd\rangle \nonumber \\
& + \varepsilon(a) \sum_\alpha \Big( 4 |\bq,\bd\rangle - z_\alpha |\bq,\bd+\hat{\alpha}\rangle - z^*_\alpha|\bq,\bd-\hat{\alpha}\rangle\Big),
\label{eq:2ph}
\end{align}
where $z_\alpha(\bq) = (1+e^{i(2\pi/L)q_\alpha})$. As expected the Hamiltonian is block diagonal for every value of $\bq$. 

\begin{figure}
\includegraphics[width=0.48\textwidth]{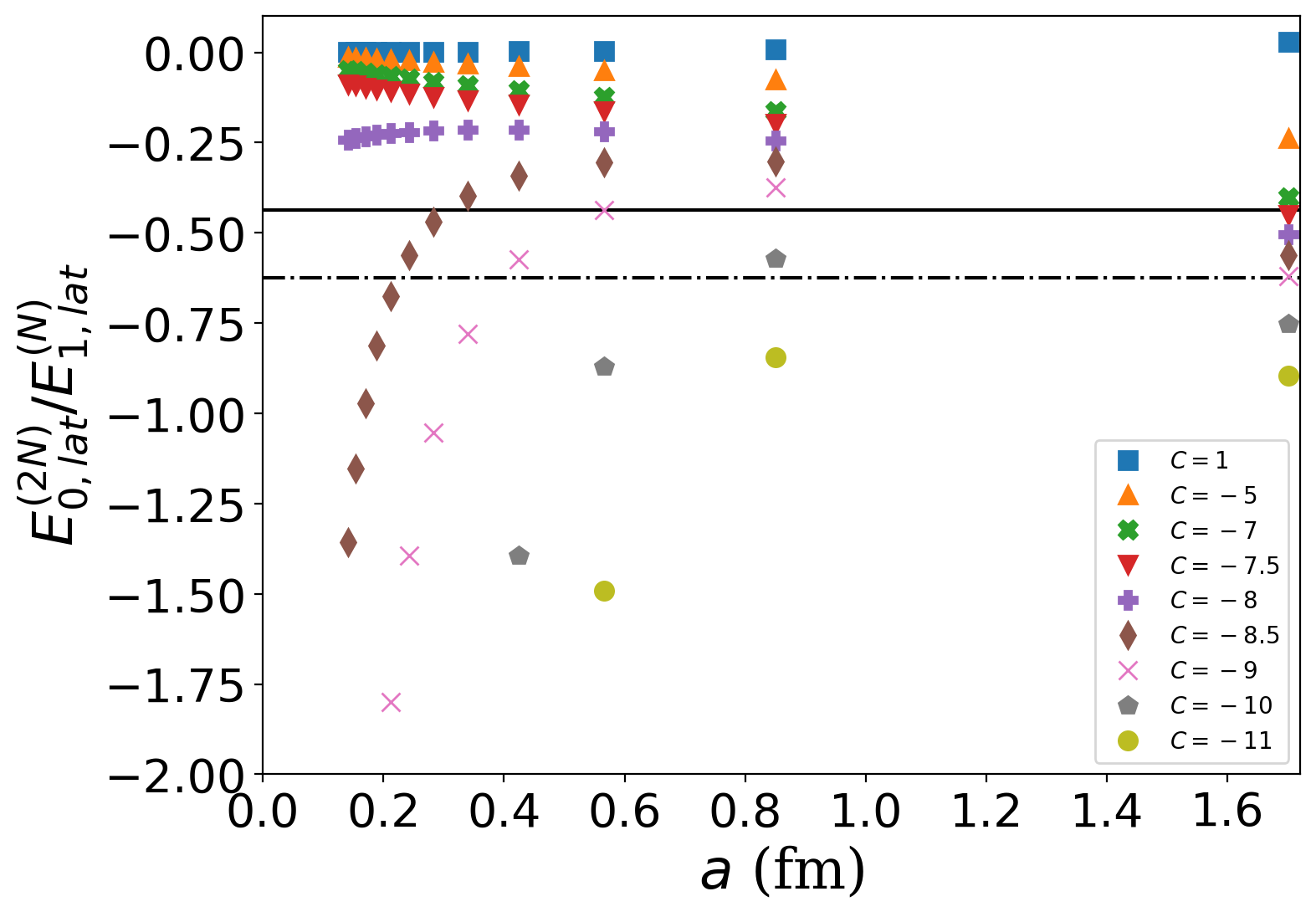}
\caption{The plot of $E_{0,{\rm lat}}^{(2N)}/E_{1,lat}^{(N)}$ as a function of the lattice spacing $a$ for different values of $C$. The solid and dot-dashed horizontal lines are physical values from \cref{tab:physparams} for the dineutron and deuteron channels respectively. For each fixed $a$, the value of $C(a)$ is the one that matches the horizontal line. 
\label{fig:2pspec}}
\end{figure}

In order to compute $C(a)$ we have to equate the ground state energy obtained from \cref{eq:2ph} when $\bq=0$ to the fixed physical value $E$ in a cubical box with side $\Lp$ as we vary the lattice spacing $a = \Lp/L$. Since the ground state will appear in the sector with $\bq=0$ in \cref{eq:2ph}, we can focus on that sector. We have used exact diagonalization to compute the ground state energy, which we refer to as $E_{0,{\rm lat}}^{(2N)}$, for $L\leq 24$ and various values of $C$. Our results are shown in \cref{fig:2pspec}. 

In a theory with a contact interaction, we can compute the spectrum in the two body channel using a different method which allows us to access much larger lattice sizes. To understand how, we simplify the notation by defining $|\bd\rangle \equiv |\bq=0,\bd\rangle$. In this notation we can write
\begin{align}
H_0 |\bd\rangle \ & = \
2 \varepsilon(a) \sum_\alpha
(2|\bd\rangle -  |\bd+\hat{\alpha}\rangle - 2|\bd-\hat{\alpha}\rangle),
\label{eq:2ph0-q0} \\
H_{\rm int} |\bd\rangle \ & =\  \varepsilon(a) C(a)\ \delta_{\bd,0}\ |\bd\rangle.
\label{eq:2phi-q0}
\end{align}
Consider the matrix $G(E,a) = (E-H)^{-1}$ with $H$ given by \cref{eq:2ph}. For a fixed value of $a$, the poles of $G(E,a)$ as a function of $E$ are at the eigenvalues of $H$. For the contact interaction in our problem we can derive the identity
\begin{align}
(E-&H)^{-1} \ =\ (E-H_0)^{-1} \ + \ 2\ \varepsilon(a)\  {\cal G}(E,a) \nonumber \\
& \times\  (E-H_0)^{-1}|\bd=0\rangle\langle\bd=0|(E-H_0)^{-1},
\end{align}
where we define ${\cal G}(E,a) = ((2/C(a)) - I_{\rm lat}(E,a))^{-1}$ and 
\begin{align}
I_{\rm lat}(a,E)\ =\ 2\varepsilon(a) \langle \bd=0|(E-H_0)^{-1}|\bd=0\rangle.
\end{align}
Note that the pole at the lowest physical bound state is now transformed to the pole in ${\cal G}(E,a)$, which implies that
\begin{align}
C(a) = 2/I_{\rm lat}(E,a).
\label{eq:cren}
\end{align}
In the expression above, $E$ is the physical energy from \cref{tab:physparams} that is being matched. For the dineutron channel we choose $E=-17.8$ MeV while for the deuteron channel we choose $E=-25.4$ MeV to obtain $\Czerosing(a)$ and $\Czerotrip(a)$ respectively.

\begin{table}[!t]
\renewcommand{\arraystretch}{1.4}
\setlength{\tabcolsep}{6pt}
\begin{tabular}{r|c|c|c|c}
\TopRule
 $L$ & $a$ (fm) & $\varepsilon$ (MeV) & $\Czerosing$ & $\Czerotrip$ \\
\MidRule
 2 & 1.70 & 10.14 & $ -7.373$ & $-9.044 $ \\
 4 & 0.85 & 20.28 & $-9.374$ & $-10.221$  \\
 6 & 0.5667 & 40.56 & $-9.006$ & $-9.502$\\
 8 & 0.4250 & 69.23 & $-8.742$ &$-9.085$  \\
 16 & 0.2125 & 266.4 & $-8.318$ & $-8.468$ \\
 24 & 0.1417 & 595.11 & $-8.178$ & $-8.273$\\
 48 & 0.0708 & 2370.26 & $-8.043$ & $-8.088$ \\ 
 96 & 0.0354 & 9470.87  & $-7.978$ & $-8.000$\\
 192 & 0.0177 & 37873.36 & $-7.945$ & $-7.956$\\
 384 & 0.0089 & 151483.28& $-7.929$ & $-7.935$ \\
 768 & 0.0045 &605922.99 & $-7.921$ & $-7.924$\\
\BotRule 
\end{tabular}
\caption{The values of $\varepsilon$, $\Czerosing$ and $\Czerosing$ as a function of $a$ for the hypothetical physical system with physical parameters given in \cref{tab:physparams}. \label{tab:lat2pparams}}
\end{table}

\begin{figure*}
\includegraphics[width=0.47\textwidth]{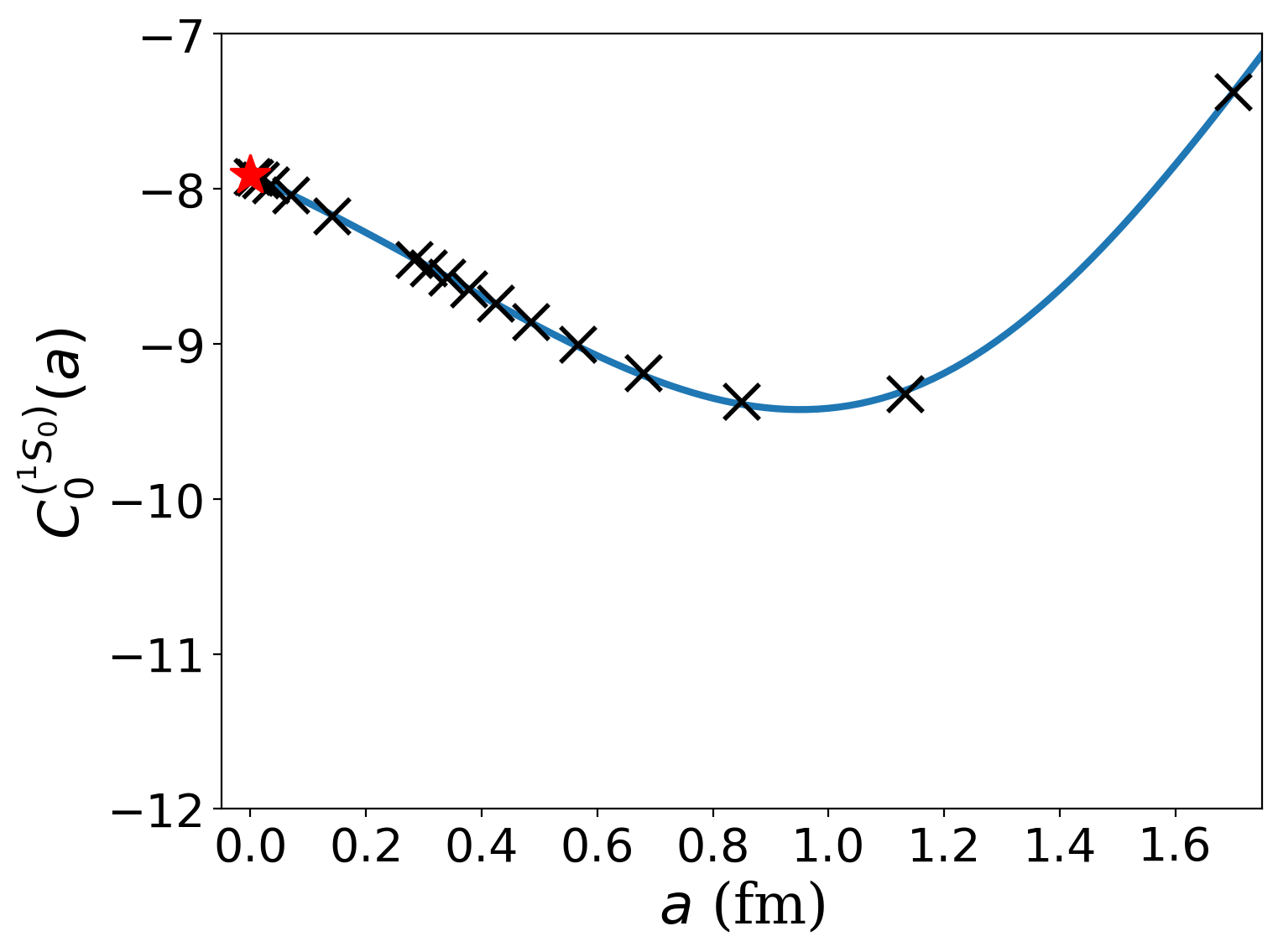}
\includegraphics[width=0.47\textwidth]{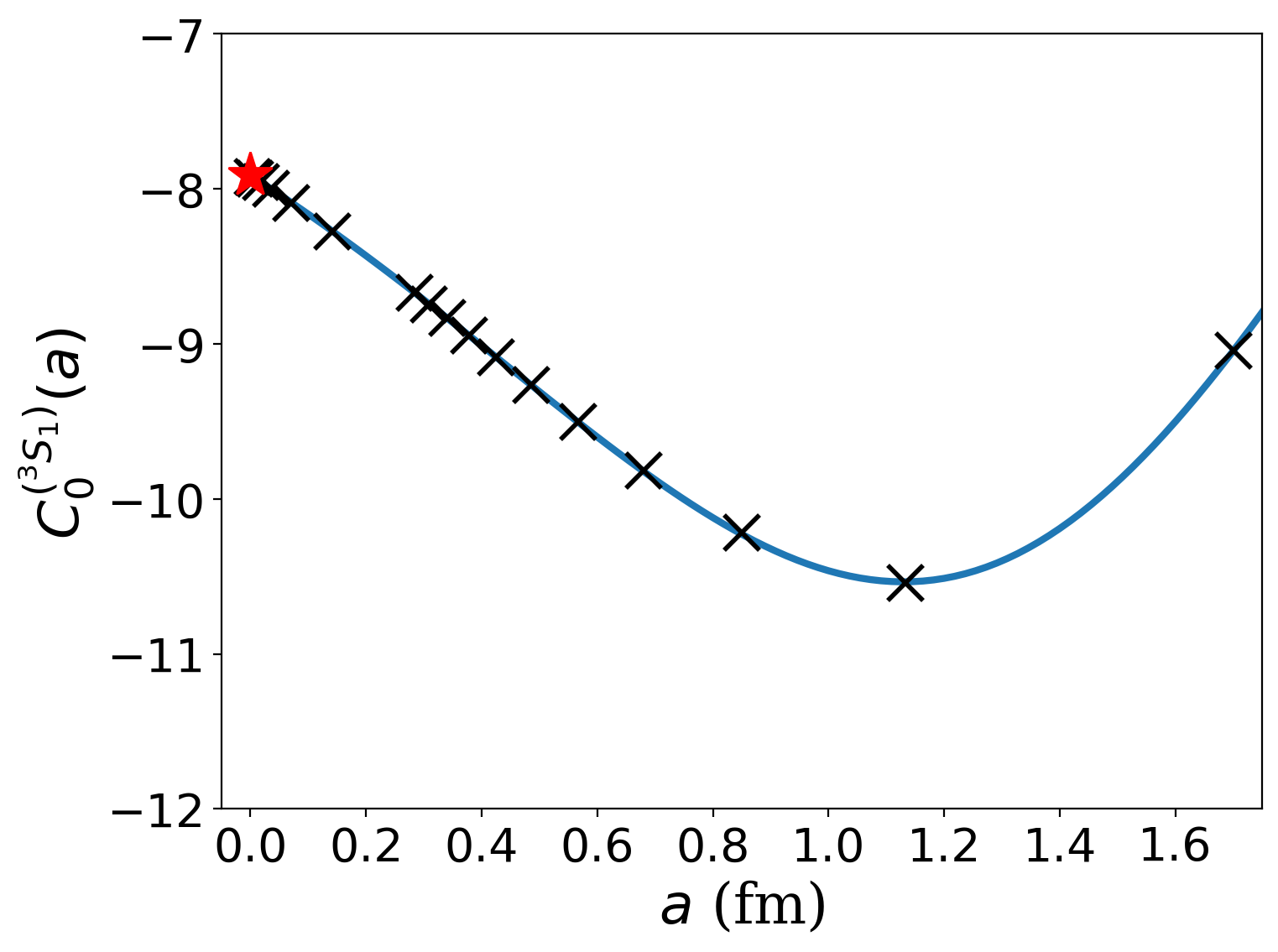}
\caption{The plot of $\Czerosing(a)$ (left plot) and $\Czerotrip(a)$ (right plot) as a function of $a$, for the physical parameters given in \cref{tab:physparams}. The lattice sizes used in the calculations include $L=2,3,..,11,12,24,48,96,192,384$ and $768$. The solid lines are given by Eq.~\eqref{eq:Ilat_dineutron} (left) and Eq.~\eqref{eq:Ilat_deuteron} (right). 
}
\label{fig:Crenorm}
\end{figure*}

We can evaluate $I_{\rm lat}$ using the eigenstates of $H_0$ which we denote as
\begin{align}
|\bn\rangle \ =\ \frac{1}{L^{3/2}}\ \sum_\bd \ e^{i(2\pi/L) \bn \cdot\bd} \ |\bd\rangle,
\end{align}
where $\bn$ is the integer vector with components $n_\alpha=0,1,2..,L-1$. It is easy to verify that the corresponding eigenvalues are $2\varepsilon {\cal E}_\bp$ where 
\begin{align}
{\cal E}_\bn = 2 \sum_\alpha\ (1-\cos(2\pi n_\alpha/L)).
\label{eq:freeE}
\end{align}
Substituting we get
\begin{align}
I_{\rm lat}(a,E)\ =\ \frac{1}{L^3} \sum_{\bn}\ \frac{1}{(E/2\varepsilon(a))-{\cal E}_\bn},
\label{eq:Ilatdef}
\end{align}
which can be numerically computed easily even for large lattice sizes. Combining this along with \cref{eq:cren} we can find the $\Czerosing(a)$ and $\Czerotrip(a)$ for large lattice sizes. In \cref{tab:lat2pparams} we tabulate these values for a variety of lattice sizes and plot them in \cref{fig:Crenorm} for lattice sizes up to $L=768$. We have verified that these results match those obtained using exact diagonalization up to $L\leq 24$.

We can compute $I_{\rm lat}(a,E)$ for small values of $a$ by replacing $2\varepsilon(a) \approx \hbar^2/\mN a^2$ and replacing $L=\Lp/a$ in \cref{eq:freeE}. It is easy to verify that 
\begin{align}
I_0 \ =\ \lim_{a\rightarrow 0}
I_{\rm lat}(a,E)\ =\ - \int_{\bp \in BZ} \frac{d^3\bp}{(2\pi)^3}\ \frac{1}{{\cal E}_\bp},
\label{eq:I0}
\end{align}
where the sum over $\bn$ is replaced by the integral over the continuous vector $\bp=2\pi \bn/L$ as $L$ becomes large. The domain of integration over $\bp$ is over the Brillouin Zone (BZ) $0\leq p_\alpha < 2\pi$. We have defined ${\cal E}_\bp = 2(3-\cos p_1 - \cos p_2 - \cos p_3)$. Up to a factor of $2$, \cref{eq:I0} is the well-known Watson's triple integral (e.g., see Ref. \cite{joyce2005evaluation}) and we get
\begin{equation}
\begin{aligned}
 I_0 & = -\frac{1}{2}\frac{\sqrt{3}-1}{96\pi^3}
    \left[\Gamma\left(\frac{1}{24}\right)\Gamma\left(\frac{11}{24}\right)\right]^2 \\
    &= -0.252731...\, .
\end{aligned}
\end{equation}
Since $I_0$ is independent of $E$ we get
\begin{align}
\lim_{a\rightarrow 0}\ \Czerosing = \lim_{a\rightarrow 0}\ 
\Czerotrip = 2/I_0 = - 7.91355...\, .  
\end{align} 
It is possible to argue that $C=-7.91355...$ is the fixed point of a renormalization group flow in the two particle channel and is referred to in the literature as the unitary limit \cite{Braaten:2004rn,Lee:2005it}. The physics of energy scales that are introduced by the interaction strength is in fact buried in the higher order terms. 

In principle we should be able to expand $I_{\rm lat}(a,E)$ in powers of $a$. Instead of finding analytic expressions for the expansion coefficients and computing them, here we obtain them by fitting our data given in \cref{tab:lat2pparams} to the form \cref{eq:cren}. We obtain
\begin{align}
    I_{\rm lat}& (a,E_0^{(nn)}) \ = I_0\  +\ 0.053148\ a  
    \nonumber \\ 
    & +\ 0.022572\ a^2 \ - 0.035412\ a^3 + ... \,,\label{eq:Ilat_dineutron}\\
    I_{\rm lat}& (a,E_0^{(np)}) \ = I_0\ +\ 0.076060\ a 
    \nonumber \\
    & +\ 0.013132\ a^2\ -\ 0.027610\ a^3 + ...\, . \label{eq:Ilat_deuteron}
\end{align}
These curves are plotted as blue curves in \cref{fig:Crenorm}.

\begin{figure*}
\includegraphics[width=0.47\textwidth]{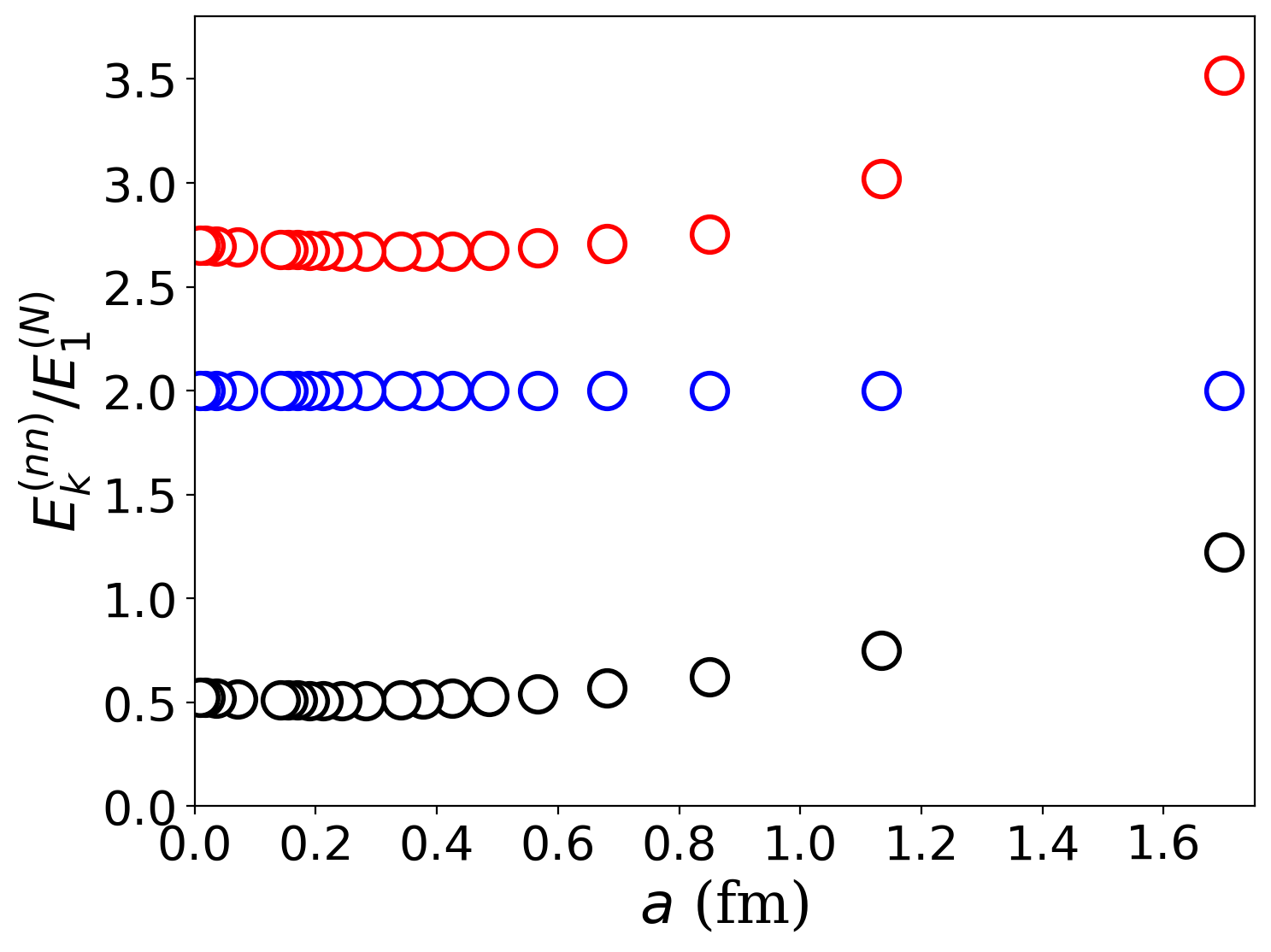}
\includegraphics[width=0.47\textwidth]{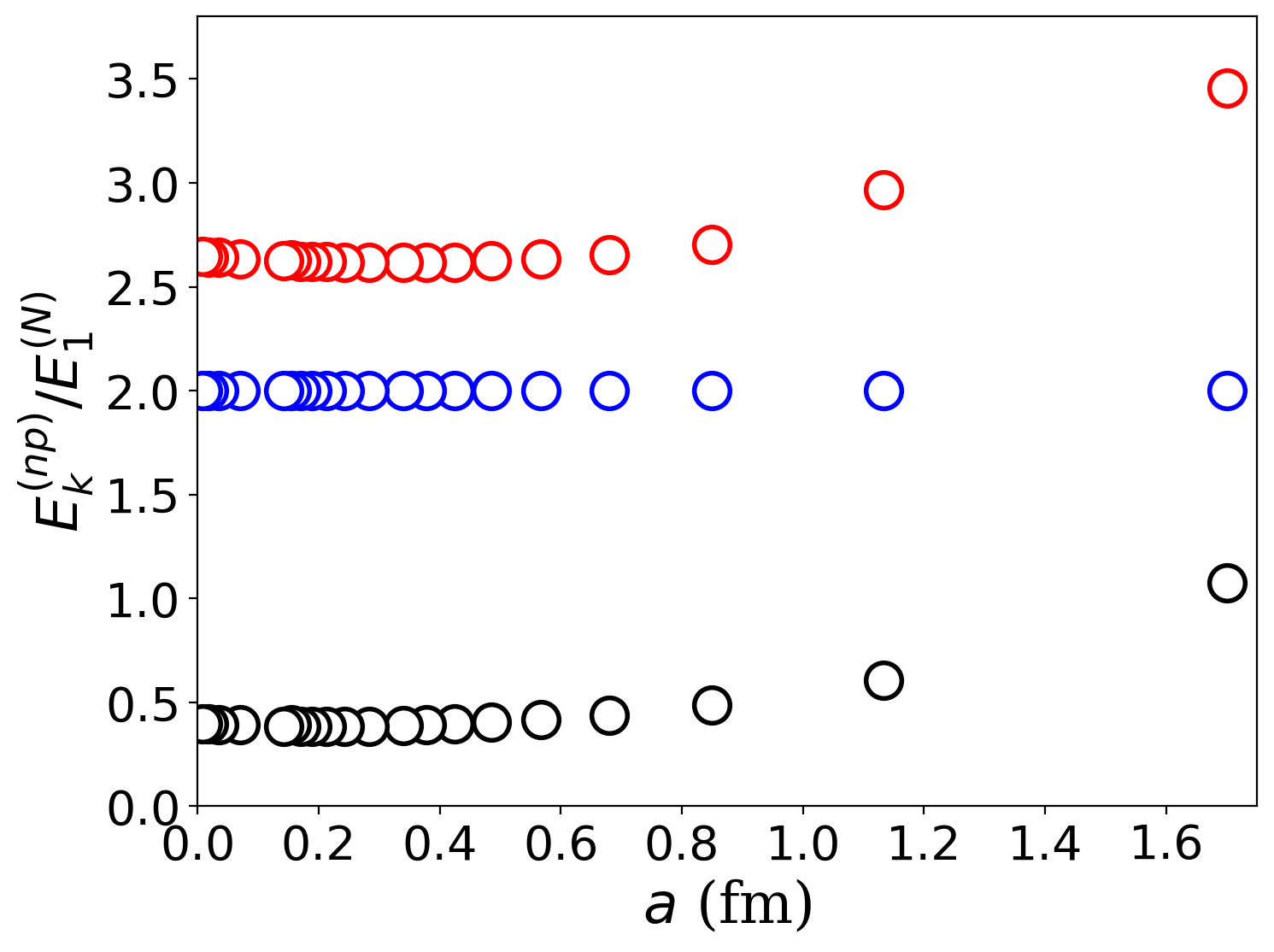}
\caption{The plot of excited energy states $E_k^{(nn)}/E_1^{(N)}$ (in the di-neutron channel, left plot) and  $E_k^{(np)}/E_1^{(N)}$ (in the deuteron channel, right plot) for the lowest three excited states $k=1,2,3$ as a function of the lattice spacing. In the continuum limit we find $E_1^{(nn)}/E_1^{(1)} = 0.524...$ (non-degenerate), $E_2^{(nn)}/E_1^{(1)} = 2$(five-fold degenerate), and $E_3^{(nn)}/E_1^{(1)} = 2.701...$ (non-degenerate), $E_1^{(np)}/E_1^{(1)} = 0.397...$(non-degenerate), $E_2^{(np)}/E_1^{(1)} = 2$(five-fold degenerate), and $E_3^{(np)}/E_1^{(1)} =  2.645...$ (non-degenerate).}
\label{fig:2pex}
\end{figure*}

Before ending this section, we briefly discuss the continuum limit of the lattice theory and the idea of renormalizability of the continuum theory. By construction, the lattice theory has a continuum limit in which the energies that were used to define the lattice parameters will be correctly reproduced. But what about the remaining energy levels in the box? These will be predictions to check. We can easily verify that in the one particle sector, all other lattice energy eigenvalues not only have a finite limit as $a\rightarrow 0$, they also match the continuum energies of free particles in a box. In the two particle sector, things are more complicated. We can use \cref{eq:cren} to compute higher energy eigenvalues once $C(a)$ is known, by finding other solutions to the equation as we vary $E$. However, this procedure only gives us the energies that are affected by the interaction, since the poles of ${\cal G}(E,a)$ miss the poles of $G(E,a)$ that remain unaffected. Combining this information with exact diagonalization on small lattices, we have found the first, the second and the third excited energies in the di-neutron and deuteron channels as function of lattice spacings. The ratio of these energies with respect to $E^{N}_1$ are plotted in \cref{fig:2pex}. Again we notice that the lattice theory has a well defined continuum limit. However, as far as we know it is unclear if the continuum theory that we obtain from our lattice approach is the same as the one that can be obtained by other continuum regularization methods.

Beyond the two body sector, things quickly get complicated and one of the goals of this work is to explore a new worldline method that can address the question of lattice renormalization in the higher particle number sector. Renormalization in the three-body sector has been studied extensively in the last two decades by other continuum regularization methods. It is widely accepted that the three-nucleon system in the spin-doublet channel requires a single three-body contact interaction with no derivatives for complete renormalization at leading order \cite{Bedaque:1998kg, Bedaque:1998km, Bedaque:1999ve}. This is the reason we introduce $\Ctb(a)$ in the lattice Hamiltonian. By studying the renormalization of this parameter and the continuum limit in the three body sector we plan to verify the widely accepted view point in a later publication.
However, we should mention the recent results which seem to suggest that a single three-body contact interaction only provides a partial renormalization of the theory. 
\cite{Epelbaum:2018zli,Epelbaum:2016ffd,Blankleider:2000vi,Gasparyan:2022isg,Epelbaum:2020maf}.
What about the four-particle sector? Do we need new parameters to obtain a sensible continuum limit of the lattice theory? This is an active area of research \cite{Platter:2004he,Bazak:2018qnu,Lin:2023zqw}, and our lattice approach could help address some of these more difficult questions. 

The continuum theory one obtains at the end is defined as renormalizable if that theory is ``universal", by which we mean other regularization and renormalization schemes also give us the same theory. In our approach, the lattice provides the regularization, and the physical parameters that we use to fix the theory is a renormalization scheme. Then the limit $a\rightarrow 0$ provides the continuum theory. If that theory is renormalizable, then we should be able to obtain the same theory via other regularizations and/or renomormalization schemes. If one needs an infinite set of tuning parameters to obtain the same continuum theory, one usually considers the theory to be non-renormalizable. While the idea of renormalizability is well developed for relativistic quantum field theories, it is still an active area of research in the case of few body physics. We already know that there is an infinite class of non-local potential models in the continuum. This means that there is an infinite class of continuum models with local interactions. The challenge would be to classify these interactions and find continuum models in each class. An interesting question is whether the pion-less effective field theory (\eftnopi) with non-derivative contact interactions is one such well defined renormalizable continuum theory. If it exists, it would describe a first approximation to nuclear physics in regimes where pions are heavy and can be integrated out.

\section{Transfer Matrix Elements}
\label{sec4}

In principle it is possible to extract the low energy spectrum of any Hamiltonian $H$ reliably if we can accurately compute all matrix elements of the transfer matrix $T(\beta) = e^{-\beta H}$ in an appropriately chosen $k$-dimensional subspace at two well chosen values of $\beta$. Let $T_k(\beta_1)$ and $T_k(\beta_2)$ be these two $k\times k$ sub-matrices of $T(\beta)$. Then, as $\beta_1$ and $\beta_2$ increase, the eigenvalues of the $k\times k$ matrix,
\begin{align}
{\cal E}_k \ =\ \frac{1}{(\beta_2-\beta_1)}\log\Big( (T_k(\beta_2))^{-1}T_k(\beta_1)\Big),
\label{eq:tmevals}
\end{align}
give a good approximation to the lowest $k$ dominant eigenvalues of $H$ whose eigenstates have non-zero overlap with the $k$ chosen basis states. Ideally, the chosen subspace should have a large overlap with these eigenstates for the method to work well.

This idea of using transfer matrix elements to compute the low energy spectrum of a Hamiltonian is well known and has been applied in lattice QCD calculations to extract excited hadron spectrum \cite{Michael:1985ne,Luscher:1990ck,Basak:2005aq,PhysRevC.105.065203}. Unfortunately, the method only works reliably if the energy eigenvalues are not very close to each other and if the matrix elements can be calculated with sufficient accuracy. If these cannot be guaranteed the method can fail and give wrong results. On the other hand the failure can be tracked by increasing the accuracy if the matrix element calculations and by increasing $\beta$. However it is important to note that such difficulties are inherent in all numerical approaches that try to compute closely spaced energy eigenvalues.

In this work we will ignore these complications and focus on constructing a worldline Monte Carlo method to compute matrix elements between a few multi-particle basis states in simple situations where the energy eigenvalues are well separated. The applicability to more realistic problems will be explored in the future and may require further innovation and refinement of our algorithm.
The basis states we work with are constructed using linear combinations of free lattice momentum eigenstates whose wavefunctions are real. We will demonstrate that we can accurately compute the low energy spectrum of the leading order pionless EFT Hamiltonian introduced in \cref{eq:Hfull} using our method on small lattices.

Let us now develop a notation to describe our multi-particle states. We first construct a complete basis of single nucleon states using the creation operators with quantum numbers $\bq = (q_x,q_y,q_z)$ where $q_x,q_y,q_z = 0,1,2,...,L-1$, using the relation
\begin{align}
\tNud{\bq,a} =  \frac{1}{L^{3/2}}\sum_{\br} u(\bq,\br) \Nud{\bq,a},
\label{eq:spmomop}
\end{align}
where $u(\bq,\br)$ are a complete basis of real position space wavefunctions as defined below. But before we define them, we divide $\bq$ into three sets ${\cal S}_1$, ${\cal S}_2$ and ${\cal S}_3$ based on the following criteria: Set ${\cal S}_1$ consists of $\bq$ such that $\sin(2\pi q_x/L) = 0$, $\sin(2\pi q_y/L) = 0$ and $\sin(2\pi q_y/L) = 0$ (note that for these values of $\bq$, $\sin(2\pi\bq\cdot\br/L)=0$), in set ${\cal S}_2$ we include those values of $\bq$ such that $\sin(2\pi q_x/L) > 1$ or $\sin(2\pi q_x/L) = 0$ and $\sin(2\pi q_y/L) > 0$ or $\sin(2\pi q_x/L) = 0$ and $\sin(2\pi q_y/L) = 0$ and $\sin(2\pi q_z/L) > 0$, in set ${\cal S}_3$ all the remaining value of $\bq$ are placed. The wavefunctions $u(\bq,\br)$ depend on which set $\bq$ belongs to and are defined as 
\begin{align}
u(\bq,\br) = 
\left\{\begin{array}{ll} 
\cos(2\pi\bq\cdot\br/L), & \bq \in {\cal S}_1\\ \\
\sqrt{2}\cos(2\pi\bq\cdot\br/L), & \bq \in {\cal S}_2 \\ \\
\sqrt{2}\sin(2\pi\bq\cdot\br/L), & \bq \in {\cal S}_3
\end{array} \right. .
\end{align}
Note that $u(\bq,\br)/L^{3/2}$ are essentially linear combination of plane-waves that give us real wavefunctions so that we can avoid complex numbers in our calculations. 

The nucleon operators $\tNud{\bq,a}$ can now be used to construct normalized multi-particle states. We will always define the basis states by creating spin-up neutrons first, followed by spin-down neutrons, and then by spin-up protons and finally by spin-down protons. Within each nucleon flavor sector we order all possible $\bq$'s and create them in that order. With this ordering in mind we assign the $i$-th nucleon the quantum number $(\bq_i,f_i)$ where $i$ labels the order of creation. Now we consider a state with $n_1$ spin-up neutrons, $n_2$ spin-down neutrons, $n_3$ spin-up protons and $n_4$ spin-down protons so that $n=n_1+n_2+n_3+n_4$. This state is represented by
\begin{align}
|\{\bq\}\rangle = \big(\tNud{\bq_{n},f_n}\tNud{\bq_{n-1},f_{n-1}}...\tNud{\bq_2,f_2}\tNud{\bq_1,f_1}\big)|0\rangle.
\label{eq:istate}
\end{align}
where $\{\bq\} = \{(\bq_1,f_1),(\bq_2,f_2)...,(\bq_n,f_n)\}$ has all the necessary information defined implicitly.

Using the notation developed above, the matrix elements we will compute in this work can be defined as
\begin{align}
\Mel_{\{\bbq\};\{\bq\}} = \langle\{\bbq\}|\ e^{-\beta H}\ |\{\bq\}\rangle.
\label{eq:matelem}
\end{align}
Our Hamiltonian allows for spin-up neutrons to be flipped to spin-down neutrons while at the same time flipping spin-down protons to spin-up protons and vice versa. This means 
while $n_1$ and $n_2$ can individually change between the initial and final states, $n_n = n_1+n_2$ (the total number of neutrons) will remain the same. Similarly, while $n_3$ and $n_4$ can change individually, $n_p = n_3+n_4$ (the total number of protons) remains the same.
Furthermore, the total spin in the $z$-direction given by $n_1-n_2+n_3-n_4$ is the same between the initial and final states. 

Since our Monte Carlo method to compute \cref{eq:matelem} works by sampling worldline configurations of hard-core bosons in position space, we introduce a complete set of position eigenstates of the nucleons
\begin{align}
|\{\br\}\rangle = \big(\Nud{\br_n,f_n} \Nud{\br_{n-1},f_{n-1}}...
\Nud{\br_2 f_2} \Nud{\br_1,f_1}\big) |0\rangle, 
\label{eq:istate-r}
\end{align}
where, in the same spirit as \cref{eq:istate}, we create spin-up neutrons first, followed by spin-down neutrons, and then by spin-up protons and finally by spin-down protons. Within each nucleon flavor sector we order the $\br$'s and create them in that order. With this ordering in mind we assign the $i$-th nucleon the quantum number $(\br_i,f_i)$ where $i$ labels the order of creation. The information about the number of nucleons of each flavor and their positions are thus implicit in $|\{\br\}\rangle$.

In the notation developed above we can then write
\begin{align}
\Mel_{\{\bbq\};\{\bq\}} \ =\ & \sum_{\{\bbr\},\{\br\}} {\langle \{\bbq\}|\{\bbr\}\rangle} \nonumber \\
& {\langle\{\bbr\}|\ e^{-\beta H}\ |\{\br\}\rangle}\ 
{\langle \{\br\}|\{\bq\}\rangle}\, ,
\label{eq:matersum}
\end{align}
where the positions of identical particles and their permutations are only counted once in the sum. Note that for a given set of $\{\bq\}$ and $\{\br\}$ we can show
\begin{align}
\langle \{\br\}|\{\bq\}\rangle\ & =\ \prod_a \frac{1}{L^{3n/2}} D_a(\{\bq\},\{\br\})\,,
\label{eq:inovlap}
\end{align}
where $D_a(\{\bq\},\{\br\})$ is the slater determinant of an $n_a\times n_a$ matrix for the nucleon flavor $a$ given by
\begin{align}
D_a(\{\bq\},\{\br\})\ =\
\begin{vmatrix}
u(\bq^{(a)}_1,\br^{(a)}_1) & ... & u(\bq^{(a)}_1,\br^{(a)}_{n_a}) \\ \\
u(\bq^{(a)}_2,\br^{(a)}_1) & ... & u(\bq^{(a)}_2,\br^{(a)}_{n_a}) \\ \\
. & ... & . \\
. & ... & . \\
. & ... & . \\ \\
u(\bq^{(a)}_{n_a},\br^{(a)}_1) & ... & u(\bq^{(a)}_{n_a},\br^{(a)}_{n_a})
\end{vmatrix}.
\label{eq:slatdet}
\end{align}
The same is true for $\langle \{\bbr\}|\{\bbq\}\rangle$, where $\{\bq\}$ and $\{\br\}$ are replaced by $\bbq$ and $\bbr$ respectively. Since our choice of wavefunctions $u(\bq,\br)$ are real, $\langle \{\bbq\}|\{\bbr\}\rangle = \langle \{\bbr\}| \{\bbq\}\rangle$.

The factor $\langle\{\bbr\}|\ e^{-\beta H}\ |\{\br\}\rangle$ is the imaginary time evolution of the position eigenstates of nucleons under the influence of the lattice Hamiltonian $H$. It is well known that such a time evolution can be written as a sum over worldline configurations $[\ell]$ of hardcore bosons, where in addition to local weights coming from fermion hopping terms, the Pauli principle can be taken into account through a permutation sign for each configuration \cite{Wiese:1992np}. This means we can write
\begin{align}
\langle\{\bbr\}|\ & e^{-\beta H}\ |\{\br\}\rangle \nonumber \\
& = \ \sum_{[\ell;\{\bbr\},\{\br\}]} \ \sgn([\ell;\{\bbr\},\{\br\}]) \ \bwt ([\ell;\{\bbr\},\{\br\}]),
\label{eq:fwlwt}
\end{align}
where $[\ell;\{\bbr\},\{\br\}]$ represents the worldline configuration of four species of bosons (one for each nucleon flavor) starting at the the positions $\{\br\}$ and ending at the positions $\{\bbr\}$. The Boltzmann weight $\bwt([\ell;\{\bbr\},\{\br\}])$ is the magnitude of the weight of the configuration, while $\sgn([\ell;\{\bbr\},\{\br\}])$ is the sign factor coming from the product of signs of local matrix elements and the fermion permutation sign. A simple way to determine $\sgn([\ell;\{\bbr\},\{\br\}])$ is to use the worldlines to find the order in which the nucleons are created at the end of the time evolution and use this ordering to determine the ordering in $\{\bbr\}$. In such a case $\sgn([\ell;\{\bbr\},\{\br\}]) = 1$, unless fermions of one flavor transform into the fermions of another flavor. With our model this can indeed occur when the neutron and proton spins flip. Then one has to reorder the nucleon creation operators before computing the matrix element. This can make $\sgn([\ell;\{\bbr\},\{\br\}])$ negative. If we substitute \cref{eq:inovlap} and \cref{eq:fwlwt} into \cref{eq:matersum} we obtain
\begin{align}
    \Mel_{\{\bbq\};\{\bq\}} \ =\ & \sum_{\{\br\},\{\bbr\}} \sum_{[\ell;\{\bbr\},\{\br\}]} \
    O(\bbq,\bq;[\ell,\{\bbr\},\{\br\}])\ \nonumber \\
& \frac{(n_n)!\ (n_p)!}{L^{3n}}\ \bwt([\ell;\{\bbr\},\{\br\}]),
\label{eq:matewlsum-0}
\end{align}
where
\begin{align}
O(\bbq,\bq;[\ell,\{\bbr\},\{\br\}]) & \ =\ \frac{\sgn([\ell;\{\bbr\},\{\br\}])}{(n_n)! (n_p)!}\
\nonumber \\
& \prod_a D_a(\{\bbq\},\{\bbr\}) D_a(\{\bq\},\{\br\}).
\label{eq:obsN}
\end{align}
The extra factors of $(n_n)!$ and $(n_p)!$ have been introduced for later convenience when we construct the Monte Carlo method. They help in implementing detailed balance more naturally.

In our discussion below we will make the notation a bit simpler, by defining worldline configurations as simply $[\ell]$ and assume that $\{\br\}$ and $\{\bbr\}$ are implicitly defined by the worldlines, since each worldline starts on a proton or a neutron with a particular spin and keeps track of the particle's position and spin as it travels in imaginary time. Note that along the worldline the spin of the nucleon could flip depending on the interaction the particle encounters. With this simplification, we can replace $O(\bbq,\bq;[\ell,\{\bbr\},\{\br\}])$ by $O(\{\bbq\},\{\bq\};[\ell])$ with the understanding that if they are not compatible with the initial and final states $|\{\bq\}\rangle$ and $|\{\bbq\}\rangle$ respectively, $O(\{\bbq\},\{\bq\};[\ell]) = 0$. We can also replace $\bwt([\ell;\{\bbr\},\{\br\}])$ with just $\bwt([\ell])$ since it does not depend on $\{\bq\}$ and $\{\bbq\}$. With these changes we rewrite \cref{eq:matewlsum-0} as
\begin{align}
    \Mel_{\{\bbq\};\{\bq\}} \ & =\ \nonumber \\
    & \sum_{[\ell]}' \ O(\{\bbq\},\{\bq\};[\ell])
\frac{(n_n)!\ (n_p)!}{L^{3n}}\ \bwt([\ell]).
\label{eq:matewlsum-1}
\end{align}
The prime symbol in the sum reminds us that we only allow configurations as long as $n_n$ and $n_p$ are less than some maximum value that we are free to set. Since the observable $O(\bbq,\bq;[\ell])$ is non-zero only in the particular sector of interest, the maximum value we choose does not affect the answer as long as the sector of interest is within the allowed sum. We will also allow the trivial sector with no particles and denote that configuration as $[0]$ and set $\bwt([0])=1$. In the next section we will design a Monte Carlo method to compute the matrix elements ${\cal M}_{\{\bbq\};\{\bq\}}$.

\begin{figure}[t]
\begin{center}
\includegraphics[width=0.5\textwidth]{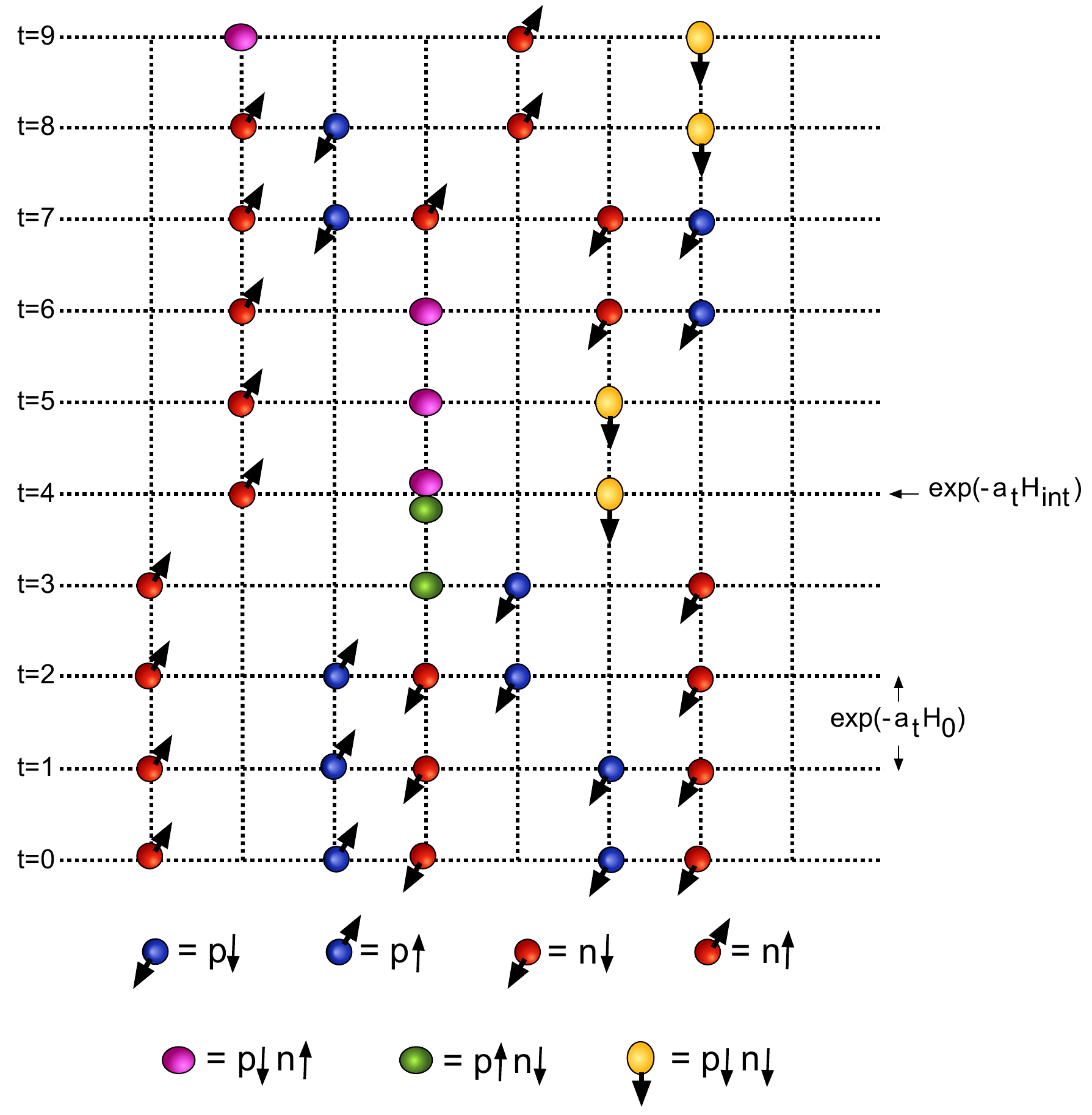}
\end{center}
\caption{\label{fig:wlconf} Illustration of a worldline configuration consisting of five nucleons moving on a space-time lattice with $L_t=9$ time slices. The circles represent neutrons while squares represent protons. The two different spins are shown with different fillings. The time evolution operator $e^{-a_t H_0}$ acts between time slices and $e^{-a_t H_{\rm int}}$ acts on a single time-slice. The latter does not change the configuration when the matrix elements are diagonal, but can change sign when an off-diagonal matrix elements are encountered. An illustration of the latter is shown in the $t=4$ time slice.}
\end{figure}

\section{The Monte Carlo Method}
\label{sec5}

In order to construct a Monte Carlo method to compute \cref{eq:matewlsum-1} we define a partition function for worldline configurations as
\begin{align}
Z \ =\ \sum_{[\ell]}'\ \frac{(n_n)!\ (n_p)!}{L^{3n}}\ \bwt([\ell])),
\label{eq:pfunc}
\end{align}
where we have just dropped the factor $O(\{\bbq\},\{\bq\};[\ell])$ in \cref{eq:matewlsum-1}. Our goal is to design a Monte Carlo method to generate the worldline configurations $[\ell]$ based on the weights given in \cref{eq:pfunc}, and then use it to compute averages of two observables
\begin{align}
\langle O(\{\bbq\},\{\bq\};[\ell]) \rangle &= \frac{1}{Z}  {\cal M}_{\{\bbq\};\{\bq\}}, 
\label{eq:obs1}
\end{align}
and 
\begin{align}
Z_0 = \langle \delta_{[\ell],[0]} \rangle &= \frac{1}{Z},
\label{eq:obs2}
\end{align}
which is the fraction of the trivial configurations in the ensemble. Using these two quantities we can compute
\begin{align}
{\cal M}_{\{\bbq\};\{\bq\}}\ =\ \frac{\langle O(\{\bbq\},\{\bq\};[\ell]) \rangle}{\langle \delta_{[\ell],[0]} \rangle}.
\label{eq:obs3}
\end{align}
Clearly, our approach will fail if ${\langle \delta_{[\ell],[0]} \rangle}$ cannot be determined accurately. This is indeed a bottleneck for the algorithm we discuss in this paper,  especially if we wish to study large number of particles. In the few body sector we seem to be able to compute ${\langle \delta_{[\ell],[0]} \rangle}$ accurately. Variants of our algorithm which can potentially overcome the bottlenecks also exist, but we have not explored them here since we consider this work as only the first of a series of papers of the subject.

\begin{table}[t]
\centering
\renewcommand{\arraystretch}{1.4}
\setlength{\tabcolsep}{4pt}
\begin{tabular}{cccc}
\TopRule
\multicolumn{4}{c}{0-particle state (Fock vacuum): $|0\rangle$} \\
\BotRule
\multicolumn{4}{c}{1-particle states} \\
\MidRule
\multicolumn{2}{c|}{$\Nud{\br,1}|0\rangle \equiv \ntd{\br,\ua}|0\rangle$} &
\multicolumn{2}{c}{$\Nud{\br,2}|0\rangle \equiv \ntd{\br,\da}|0\rangle$} \\
\multicolumn{2}{c|}{$\Nud{\br,4}|0\rangle \equiv \ptd{\br,\da}|0\rangle$} & 
\multicolumn{2}{c}{$\Nud{\br,3}|0\rangle \equiv \ptd{\br,\ua}|0\rangle$} \\
\TopRule
\multicolumn{4}{c}{2-particle states} \\
\hline
\multicolumn{2}{r|}{$I^\dagger_{\br,+}|0\rangle \ \equiv\ \ptd{\br,\da}\ptd{\br,\ua}|0\rangle$} & 
\multicolumn{2}{r}{$I^\dagger_{\br,-}|0\rangle\ \equiv\ \ntd{\br,\da}\ntd{\br,\ua}|0\rangle$} \\
\multicolumn{2}{r|}{$d^\dagger_{\br,+} |0\rangle \ \equiv\  \ptd{\br,\ua}\ntd{\br,\ua} |0\rangle$} & 
\multicolumn{2}{r}{$d^\dagger_{\br,-} |0\rangle \ \equiv\  \ptd{\br,\da}\ntd{\br,\da} |0\rangle$} \\
\multicolumn{2}{r|}{$D^\dagger_{\br,1} |0\rangle \ \equiv\ \ptd{\br,\da} \ntd{\br,\ua} |0\rangle$} &
\multicolumn{2}{r}{$D^\dagger_{\br,2} |0\rangle \ \equiv\ \ptd{\br,\ua} \ntd{\br,\da} |0\rangle$} \\
\TopRule
\multicolumn{4}{c}{3-particle states} \\
\hline
\multicolumn{2}{r|}{$t^\dagger_{\br,+} |0\rangle \equiv \ptd{\br,\ua}\ntd{\br,\da}\ntd{\br,\ua} |0\rangle$} &
\multicolumn{2}{r}{$t^\dagger_{\br,-}|0\rangle \equiv \ptd{\br,\da}\ntd{\br,\da}\ntd{\br,\ua} |0\rangle$} \\
\multicolumn{2}{r|}{$\He^\dagger_{\br,+}|0\rangle \equiv \ptd{\br,\da}\ptd{\br,\ua}\ntd{\br,\ua} |0\rangle$} &
\multicolumn{2}{r}{$\He^\dagger_{\br,-}|0\rangle \equiv \ptd{\br,\da}\ptd{\br,\ua}\ntd{\br,\da} |0\rangle$} \\
\TopRule 
\multicolumn{4}{c}{4-particle sector} \\
\MidRule
\multicolumn{4}{c}{$H |0\rangle\ \equiv \ 
\ptd{\br,\da}\ptd{\br,\ua}\ntd{\br,\da}\ntd{\br,\ua}|0\rangle$} \\
\BotRule   
\end{tabular}
\caption{\label{tab:2pOp} The table shows our notation for the sixteen dimensional local Hilbert space on each lattice site. It can either be the Fock vacuum state, one of four one particle states, one of six two-particle states, one of four three-particle states or the four-particle state. We define two, three and four particle states through creation operators defined above.}
\end{table}

Let us now discuss a worm-type algorithm to generate configurations $[\ell]$ distributed according to \cref{eq:pfunc}. We begin with \cref{eq:fwlwt} but replace the nucleons with hardcore bosons. We can then argue that 
\begin{align}
\langle\{\bbr\}|\ e^{-\beta H}\ |\{\br\}\rangle_b = \ \sum_{[\ell;\{\bbr\},\{\br\}]}' \ \bwt([\ell;\{\bbr\},\{\br\}]]),
\label{eq:bwlwt}
\end{align}
where the subscript $b$ is a reminder that the nucleons are being treated as hardcore bosons and the prime symbol means the same as the one we discussed below \cref{eq:matewlsum-1}. We introduce a finite lattice spacing in time $a_t$ and approximate
\begin{align}
e^{-\beta H}\ \approx\  \Big( e^{-a_t H_0}e^{-a_t H_{\rm int}} e^{-a_t H_0}
... e^{-a_t H_{\rm int}}e^{-a_t H_0} \Big),
\label{eq:trotter}
\end{align}
where we have introduced $L_t-1$ factors of $e^{-a_t H_{\rm int}}e^{-a_t H_0}$ and one final factor of $e^{-a_t H_0}$ at the end to make the operator on the right hand side Hermitian. Here we assume $a_t L_t=\beta$ and to obtain the matrix elements accurately we will need to take the limit $a_t \rightarrow 0$ and $L_t \rightarrow \infty$ while keeping $\beta$ fixed. In this work we fix $a_t=0.001/\varepsilon$ or $a_t=0.0005/\varepsilon$ in our calculations. In a few cases we also study extrapolation to small $a_t$ to demonstrate that our answers agree with exact calculations. From the point of effective field theories, finite $a_t$ errors can also be treated as a renormalization of the Hamiltonian and may not play an important role as long as the symmetries of the theory are maintained and we are in the scaling regime. If smaller values of $a_t$ are necessary, they can be reached with further computational cost.

Within the approximation introduced above we can compute the matrix elements $\langle\{\bbr\}|\ e^{-\beta H}\ |\{\br\}\rangle_b$ by inserting a complete basis of nucleon occupation number states after every time evolution operator $e^{-a_t H_{\rm int}}e^{-a_t H_0}$. We label each of these states with a Euclidean time index $t=1,2,...,L_t-1$. The initial and final states are assumed to be at $t=0$ and $t=L_t$ respectively. Due to particle number conservation every non-zero matrix element obtained with a given choice of the basis states defines worldlines of particles $[\ell]$ moving on a space-time lattice. An illustration of such a configuration in one spatial dimension is shown in \cref{fig:wlconf}. The state on each space-time lattice site is given by one of sixteen possible states as defined in \cref{tab:2pOp}, where we have also introduced new creation operators for multi-particle states for later convenience. In order to find the Boltzmann weight $\bwt([\ell])$ for each configuration we need to compute the matrix elements of $e^{-a_t H_0}$ and $e^{-a_t H_{\rm int}}$ in the occupation number basis using $H_0$ and $H_{\rm int}$ as defined in \cref{eq:Hfree} and \cref{eq:Hint}. When particles hop due to $H_0$, we will make a further approximation and define
\begin{align}
e^{-a_t H_0} \ \approx\ & \prod_{\br}\ \Big(
(1- 6 a_t \varepsilon \Nud{\br}\Nu{\br}) \nonumber \\
& + a_t \varepsilon \sum_{\hat{\alpha}}\ (\Nud{\br+\hat{\alpha}}\Nu{\br} + \Nud{\br-\hat{\alpha}}\Nu{\br}),
\Big).
\end{align}
where we assume nucleons can only hop to the nearest neighbor sites or remain on the same site after each time step. An useful feature of this approximation is that the weight of a nucleon hop to the neighboring site is given by $\omega_h = a_t\varepsilon$ while the weight to remain on the same site is $\omega_t = (1-6 \omega_h)$. When $a_t$ is small these weights can be used as probabilities during the worm update.

\begin{table*}[t]
\centering
\renewcommand{\arraystretch}{2.0}
\setlength{\tabcolsep}{4pt}
\begin{tabular}{llll}
\TopRule
$\omega_1$ & =\quad
$\langle 0|D_{\br,2} \ T_{\rm int}(\br)\  D^\dagger_{\br,1}|0\rangle$ & =\quad  
$\langle 0|D_{\br,1} \ T_{\rm int}(\br)\  D^\dagger_{\br,2}|0\rangle$ & =\quad  
$e^{-a_t \varepsilon (\Czerotrip + \Czerosing)/2}\  
\sinh(a_t \varepsilon (\Czerosing - \Czerotrip)/2)$ \\
$\omega_2$ & =\quad
$\langle 0|D_{\br,1} \ T_{\rm int}(\br)\ D^\dagger_{\br,1}|0\rangle$ & =\quad 
$\langle 0|D_{\br,2} \ T_{\rm int}(\br)\ D^\dagger_{\br,2}|0\rangle$ & =\quad
$e^{-a_t \varepsilon (\Czerotrip + \Czerosing)/2}\ 
\cosh(a_t \varepsilon (\Czerosing - \Czerotrip)/2)$ \\
$\omega_3$ & =\quad
$\langle 0|I_{\br,+} \ T_{\rm int}(\br)\ I^\dagger_{\br,+}|0\rangle$ & =\quad 
$\langle 0|I_{\br,-} \ T_{\rm int}(\br)\ I^\dagger_{\br,-}|0\rangle$ & =\quad
$e^{-a_t \varepsilon \Czerosing}$ \\
$\omega_4$ & =\quad
$\langle 0|d_{\br,+} \ T_{\rm int}(\br)\ d^\dagger_{\br,+}|0\rangle$ & =\quad 
$\langle 0|d_{\br,-} \ T_{\rm int}(\br)\ d^\dagger_{\br,-}|0\rangle$ & =\quad
$e^{-a_t \varepsilon \Czerotrip}$ \\
$\omega_5$ & =\quad
$\langle 0|t_{\br,+} \ T_{\rm int}(\br)\ t^\dagger_{\br,+}|0\rangle$ & =\quad 
$\langle 0|t_{\br,-} \ T_{\rm int}(\br)\ t^\dagger_{\br,-}|0\rangle$ & =\quad
$e^{-a_t \varepsilon (\Ctb + 3(\Czerotrip+\Czerosing)/2)}$ \\
$\omega_6$ & =\quad
$\langle 0|He_{\br,+} \ T_{\rm int}(\br)\ He^\dagger_{\br,+}|0\rangle$ & =\quad 
$\langle 0|He_{\br,-} \ T_{\rm int}(\br)\ He^\dagger_{\br,-}|0\rangle$ & =\quad
$e^{-a_t \varepsilon (\Ctb + 3(\Czerotrip+\Czerosing)/2)}$ \\
$\omega_7$ & =\quad
$\langle 0|H{\br} \ T_{\rm int}(\br)\ H^\dagger_{\br}|0\rangle$ & =\quad
$e^{-a_t \varepsilon (4\Ctb + 3(\Czerotrip+\Czerosing))}$ & \\
\BotRule
\end{tabular}
\caption{\label{tab:intme} The non-zero matrix elements of the local interaction transfer matrix $T_{\rm int}(\br) = e^{-a_t (H_{\rm int}^{\rm d}(\br)+ H_{\rm int}^{\rm o}(\br))}$ between states with at least two particles. In addition to these the other non-zero matrix elements are $\langle 0|\ T_{\rm int}(\br)\ |0\rangle = \langle 0|\Nu{\br,a} \ T_{\rm int}(\br)\ \Nud{\br,a}|0\rangle\ =\ 1$.}
\end{table*}

In contrast to $H_0$, the interaction term $H_{\rm int}$ is a single site operator which has both a diagonal part and an offdiagonal part. To make this explicit we write 
\begin{align}
e^{-a_t H_{\rm int}}\ =\ \prod_\br e^{-a_t (H_{\rm int}^{\rm d}(\br)+ H_{\rm int}^{\rm o}(\br))}.
\end{align}
In order to identify $H_{\rm int}^d(\br)$ and $H_{\rm int}^o(\br)$ explicitly we substitute $\ProjS{a}$ and $\ProjT{a}$, defined in \cref{eq:projectors}, into \cref{eq:Hint}. The nucleon bilinear terms that are spin-singlets and isospin-triplets are given by
\begin{align}
\NuT{\br} \ProjS{1} \Nu{\br} &= \frac{1}{\sqrt{2}}\big(-\pt{\br,\ua}\pt{\br,\da} + \nt{\br,\ua}\nt{\br,\da}\big),\\
\NuT{\br} \ProjS{2} \Nu{\br} &= -\frac{i}{\sqrt{2}}\big(\pt{\br,\ua}\pt{\br,\da} + \nt{\br,\ua}\nt{\br,\da}\big),\\
\NuT{\br} \ProjS{3} \Nu{\br} &= \frac{1}{\sqrt{2}}\big(n_{\br,\ua}p_{\br,\da} - n_{\br,\da}p_{\br,\ua}\big),
\end{align}
while those that are spin-triplets and isospin-singlets are given by
\begin{align}
\NuT{\br} \ProjT{1} \Nu{\br} &= \frac{1}{\sqrt{2}}\big(n_{\br,\ua}p_{\br,\ua} - n_{\br,\da}p_{\br,\da}\big),
\\ 
\NuT{\br} \ProjT{2} \Nu{\br} &= \frac{i}{\sqrt{2}}\big(n_{\br,\ua}p_{\br,\ua} + n_{\br,\da}p_{\br,\da}\big),
\\ 
\NuT{\br} \ProjT{3} \Nu{\br} &=- \frac{1}{\sqrt{2}}\big(n_{\br,\ua}p_{\br,\da} + n_{\br,\da}p_{\br,\ua}\big).
\end{align}
Using these we can identify the diagonal interaction term to be
\begin{align}
H_{\rm int}^d(\br) \ & =\ \varepsilon
\Bigg\{\ \Czerosing\big(\ I^\dagger_{\br,+}I_{\br,+} + I^\dagger_{\br,-}I_{\br,-} \big)
\nonumber \\
& + \Czerotrip \ \big( d^\dagger_{\br,+}d_{\br,+} + d^\dagger_{\br,-}d_{\br,-} \big)
\nonumber \\
& + \ C_{3B} \ \Big[
\Big(t^\dagger_{\br,+}t_{\br,+} + 
t^\dagger_{\br,-}t_{\br,-} 
\nonumber \\
& \qquad + \Big(\He^\dagger_{\br,+}\He_{\br,+} + 
\He^\dagger_{\br,-}\He_{\br,-}\Big) \Big]
\nonumber \\
& \frac{1}{2} (\Czerotrip+ \Czerosing)
\big( D^\dagger_{\br,1}D_{\br,1} + D^\dagger_{\br,2}D_{\br,2} \big)
\Bigg\}.
\end{align}
while the off-diagonal term is given by
\begin{align}
H_{\rm int}^o(\br) \ =\ & 
\frac{\varepsilon}{2}(\Czerotrip - \Czerosing)  ( D^\dagger_{\br,1}D_{\br,2} + D^\dagger_{\br,2}D_{\br,1}).
\label{eq:intH}
\end{align}
We can now compute the matrix elements of $T_{\rm int}(\br) = e^{-a_t (H_{\rm int}^{\rm d}(\br)+ H_{\rm int}^{\rm o}(\br))}$, between the sixteen states shown in \cref{tab:2pOp}. The non-zero matrix elements are given in table \cref{tab:intme}. Notice that there are only two non-zero off-diagonal matrix elements with weights $\omega_1$ which will be negative if $\Czerotrip > \Czerosing$. These negative signs will be ignored during the worm update and but taken into account through $\sgn([\ell;\{\bbr\},\{\br\}])$. Thus, every worldline configuration $[\ell]$ can be assigned a unique Boltzmann weight $\bwt([\ell])$. 

We use two types of worm algorithms to update the worldline configurations. The first is the \ac{PNA} and the second is the \ac{SFA}. In the \ac{PNA} we propose to add or remove a nucleon of a random spin. When both nuetrons and protons are present, we perform two \ac{PNA} algorithms, one for protons \ac{PNA}-p and one for neutrons \ac{PNA}-n sequentially. This algorithm can change the number of nucleons and their worldlines. In our algorithm the \ac{PNA} cannot create or remove off-diagonal interactions that appear with weight $\omega_1$ in \cref{tab:2pOp}. To overcome this limitation we introduce the second update, the \ac{SFA}, in which the worldlines of the nucleons are frozen and but their spins are flipped. The \ac{SFA} can create and remove 
off-diagonal interactions. If $\omega_1=0$ the \ac{SFA} is not necessary, but may still help in reducing autocorrelation times.

The \ac{PNA} starts by either adding a nucleon of a random spin on a random spatial site chosen on the time-slice $t=0$ or by removing an existing nucleon of either spin on the time slice $t=L_t$. This natural first step introduces the extra weight $(n_n)!(n_p)!/{L^{3n}}$ that we introduced in \cref{eq:matewlsum-0}. We allow particles to be added until a maximum number is reached beyond which the algorithm forbids any new additions. Once the algorithm begins, it allows the particle worldlines to either grow or shrink like a worm whose tail is always on the $t=0$ slice but the head is allowed to move. Each local move satisfies detailed balance using the well established procedures \cite{PhysRevD.99.074511}. If the growth proposal is rejected then the worm begins to shrink and vice versa. The update can end in three possible ways: (1) by either adding a new particle worldline, (2) by removing an existing particle worldline, (3) keeping the number of particles the same but possibly changing the worldline. During the growth step, when the worm head at lattice site $(\br,t)$ moves to one of the six neighboring sites $(\br\pm \hat{\alpha},t+1)$, it does so with probability $\omega_h$. With the remaining probability $\omega_t = 1 - 6\omega_h$ the site worm head moves to $(\br,t+1)$. Since these weights are also weights of the local worldlines these local growth steps satisfy detailed balance and are always accepted. Similarly, during the shrink step when the worm head at lattice site $(\br,t)$ moves to the unique site $(\br',t-1)$ connected by the worldline, it does so always 
and still satisfies detailed balance. Thus, unless the worm head encounters another particle, the forward or backward motion moves without an accept-reject, which makes it very efficient to create free particles far from each other. 

When the worm head encounters another particle, then the interactions given in \cref{tab:2pOp} play a role in determining if the worm head moves forward or backwards. Here we use an accept-reject step based on the ratio of the interaction weights after and before the worm move. These weights of course depend on all the other particles that exist at the location of the head. There are again three possible scenarios: First, if the site contained an off-diagonal configuration with weight $\omega_1$ before the worm head reached the site, the worm is forbidden to enter it and so the worm move always bounces (i.e., if the worm was growing it would begin to shrink and vice-versa). Second, if the worm head encounters another identical nucleon of the same spin as it grows, then the head and the tail are switched to the worldline of the other identical nucleon and the worm begins to shrink. Finally, if the worm encounters the site with one of the allowed diagonal interactions with weights $\omega_i, i = 2,3,..7$ given in \cref{tab:2pOp}, then it creates another of these diagonal interactions when the new nucleon is created on the site. If $\omega_I$ and $\omega_F$ are the initial and final interaction weights, a Metropolis accept-reject step is performed based on the ratio $(\omega_F/\omega_I)$ to either continue with the worm move or perform a bounce.

The \ac{PNA} is not ergodic by itself when $\omega_1 \neq 0$ and it must be supplemented with the \ac{SFA}, which can again be of two types based on how the algorithm begins. In first type of algorithm (\ac{SFA}-1) begins on a nucleon at the $t=0$ or $t=L_t$ time slices. For this type of algorithm the worm update also ends at one of these time slices. In particular it can begin at $t=0$ and end at $t=L_t$ or vice-versa. The second type of algorithm (\ac{SFA}-2), begins with a nucleon at an arbitry site $(\br,t)$ where $0 < t < L_t$. In this case the worm is always a loop algorithm. This is because the worm always bounces at $t=0$ or $t=L_t$. Construction of \ac{SFA} agorithms are based on local detailed balance and their procedures are well established \cite{PhysRevE.66.046701}. Combining \ac{PNA} and \ac{SFA} we can construct an algorithm that creates worldline configurations $[\ell]$ ditributed according to the partition function \cref{eq:pfunc}. For each configuration computing the observables given in \cref{eq:obs1}, \cref{eq:obs2} are straight forward. From these two we can compute the transfer matrix element using \cref{eq:obs3}.

\section{Testing the Algorithm}
\label{sec6}

In this section we test our MC algorithm by comparing results obtained by it with results obtained using exact diagonalization on the $L=2$ lattice. We fix our lattice parameters using the physical parameters given in \cref{tab:physparams}. For example, substituting $L=2$ in \cref{eq:varepsilon} we obtain
\begin{align}
\varepsilon = \pi^2 \hbar^2/(2m_N \Lpsq) = 10.14\ \mbox{MeV} ,
\end{align}
where we have approximated $1 \mbox{fm} = (197 \mbox{MeV})^{-1}$ in natural units. Using the above value of $\varepsilon$, we impose that the exact diagonalization must give us $E^{(nn)}_0 \ =\ -1.755\ \varepsilon$, $E^{(pn)}_0\ =\ -2.505\ \varepsilon$ and $E^{(pnn)}_0\ =\ -6.469\ \varepsilon$. This fixes $\Czerosing=-7.373$, $\Czerotrip=-9.044$ and $\Ctb=5.109$. We summarize these parameters in \cref{tab:L2latparams}.

\begin{table}[!htb]
\centering
\renewcommand{\arraystretch}{1.4}
\setlength{\tabcolsep}{4pt}
\begin{tabular}{c|c|c|c|c|c}
\TopRule
 $L$ & $a$ (fm) & $\varepsilon$ (MeV) & $\Czerosing$ & $\Czerotrip$ & $\Ctb$ \\
\MidRule
 2 & 1.70 & 10.14 & $-7.373$ & $-9.044 $ & $5.109$ \\
\BotRule 
\end{tabular}
\caption{Lattice parameters for the physical parameters given in \cref{tab:physparams}.\label{tab:L2latparams} 
}
\end{table}

Since our current implementation of the Monte Carlo method works at a non-zero $a_t$, there is always a systematic error associated with that. While we can in principle eliminate it by extrapolating $a_t$ to zero, for this first study we will try to minimize that error by working at a small value of $a_t$. We will provide some evidence that the finite temporal lattice spacing errors can be eliminated through extrapolations when necessary.

For convenience in defining the initial and final states of the matrix elements we compute, we introduce eight single nucleon quantum numbers $\bq$ as defined in \cref{sec5}. We label them as $\bk_i = (k_{x,i},k_{y,i},k_{z,i})$, $i = 1,2...,8$ and define them in \cref{tab:qnums}. 
\begin{table}[!htb]
\centering
\renewcommand{\arraystretch}{1.4}
\setlength{\tabcolsep}{4pt}
\begin{tabular}{l|l|l|l}
\TopRule
$\bk_1 = (0,0,0)$ & $\bk_2 = (1,0,0)$ & $\bk_3 = (0,1,0)$ & $\bk_4 = (0,0,1)$ \\
$\bk_5 = (1,1,0)$ & $\bk_6 = (1,0,1)$ & $\bk_7 = (0,1,1)$ & $\bk_8 = (1,1,1)$ \\
\BotRule 
\end{tabular}
\caption{\label{tab:qnums} Eight different quantum numbers $\bq$ as defined in \cref{sec5}.}
\end{table}
Multi-nucleon states $|\{q\}\rangle$ are defined by simply choosing each nucleon in one of these eight states in an order as discussed in \cref{sec5}. In the subsections below we will discuss our results starting with free particles and then the di-neutron, deuteron, triton and Helium sectors. 

\subsection{Free Particles}

While the physics of free fermions is trivial from an analytic perspective, to obtain the same results using a Monte Carlo sampling of hard-core worldlines in position space is difficult, since it involves cancellations due to fermion permutation signs \cite{Wiese:1992np}. For example to accurately compute the ground state energy of several free fermions using the usual formula
\begin{align}
E_F = \lim_{\beta \rightarrow \infty} \frac{\mathrm{Tr}(H_0 e^{-\beta H_0})}{\mathrm{Tr}(e^{-\beta H_0})},
\end{align}
where the right hand side is computed using a Monte Carlo method that samples position space fermion worldlines, like we plan to do, is known to be extremely difficult. This difficulty can be circumvented in our approach since we can compute the diagonal matrix elements
\begin{align}
{\cal M}_{\{\bq\};\{\bq\}} = \langle\{\bq\}|\ e^{-\beta H_0}\ |\{\bq\}\rangle = e^{-\beta E_F},
\label{eq:frme}
\end{align}
where $|\{\bq\}\rangle$ is the eigenstate of $H_0$ and $E_F$ is the corresponding eigenvalue. For example, the ground state energy with $N$ free nucleons is given by the Fermi energy
\begin{align}
E_F = 2\varepsilon\sum_{i=1}^N \sum_\alpha \big(1 - \cos(2\pi k_{\alpha,i}/L)\big),
\end{align}
where the state with $\{\bq\} = \{\bk_1,\bk_2,...,\bk_N\}$ containing $N$ nucleons of the same flavor is the ground state on a $2^3$ lattice. Although computing the matrix element in \cref{eq:frme} at large values of $\beta$ will continue to be difficult with our method, we can compute it efficiently at small $\beta$ and since $E_F = -\log(e^{-\beta E_F})/\beta$ is valid for all values of $\beta$ we can accurately determine $E_f$. We will argue below that a similar method can also work in the interacting case at least for few body physics in a box.

In \cref{tab:freef} we compare the exact results with those obtained using our Monte Carlo method at $\beta=0.1/\varepsilon$ with the choice $a_t = 0.001/\varepsilon$ and $L_t=100$. 
\begin{table}[!htb]
\centering
\renewcommand{\arraystretch}{1.4}
\setlength{\tabcolsep}{4pt}
\begin{tabular}{c|c|c|c}
\TopRule
 $N$ & $E_F/\varepsilon$ & $e^{-\beta E_F}$ & MC \\
\MidRule
 2 & 4 & 0.67032... & 0.6699(4)\\
 3 & 8 & 0.44932... & 0.4491(4)\\
 4 & 12 & 0.30119... & 0.2999(4)\\
 5 & 20 & 0.13533... & 0.1347(5)\\
 6 & 28 & 0.06081... & 0.0604(3)\\
 7 & 36 & 0.02732... & 0.0268(3)\\
 8 & 48 & 0.00822... & 0.0081(3)\\
\BotRule 
\end{tabular}
\caption{We compare the exact results for the ground state matrix elements given in \cref{eq:frme} with those obtained with our Monte Carlo method. We use $\beta=0.1/\varepsilon$, obtained using $a_t=0.001/\varepsilon$ and $L_t=100$, for these calculations. \label{tab:freef}}
\end{table}
Note that as the particle number increases we have to work harder even at a fixed value of $\beta$ since $E_F$ increases and the signal degrades. However, in this case we can reduce $\beta$ if necessary to get a better signal. Although the temperature is very high in these calculations, knowing the exact eigenstate solves the sign problem. While this is not surprising, it proves that the physics of a degenerate Fermi gas can emerge by sampling hard core boson worldlines. Since adding interactions is straight forward in this framework, it would be interesting to explore if the method can  continue to be useful in cases where the solution to the sign problem is difficult. We will postpone such a study to later publications.

\subsection{The di-neutron channel}

Next we test our algorithm in the di-neutron channel. For this we construct the $5\times 5$ transfer matrix $T_5(\beta)$ with matrix elements 
\begin{align}
\Mel_{ij}(\beta)\ =\ \langle \psi_i|e^{-\beta H}|\psi_j\rangle,
\label{eq:dinme}
\end{align}
where the five di-neutron states we choose are defined as
\begin{align}
|\psi_i\rangle \ =\ \tNud{\bk_i,2}\tNud{\bk_i,1}  |0\rangle\,, i=1,2,3,4,5
\end{align}
at $\beta=0.4/\varepsilon$ and $0.5/\varepsilon$. The choice of $\beta$ is somewhat arbitrary, but we are guided by the ability to extract the ground state energy reliably by 
minimizing the contamination from higher excited states while at the same time keeping it sufficiently small to make sure we can compute the matrix elements accurately. In \cref{tab:di-neutron-test} we tabulate the exact answers and the corresponding values obtained using our Monte Carlo method for $a_t=0.0005/\varepsilon$. 

\begin{table}[!htb]
\centering
\renewcommand{\arraystretch}{1.4}
\setlength{\tabcolsep}{4pt}
\begin{tabular}{c|c|c|c|c}
\TopRule
& \multicolumn{2}{|c|}{$\Mel_{ij}(\beta=0.4/\varepsilon)$} & \multicolumn{2}{|c}{$\Mel_{ij}(\beta=0.5/\varepsilon)$} \\
\MidRule
 $(i,j)$ & exact & MC & exact & MC\\
\MidRule
(1,1) &  1.799597... & 1.7978(8) &  2.135911... & 2.1352(11) \\
(1 2) &  0.296652... & 0.2949(5) &  0.367658... & 0.3659(6) \\
(1 3) &  0.296652... & 0.2949(5) &  0.367658... & 0.3659(6) \\
(1 4) &  0.296652... & 0.2949(5) &  0.367658... & 0.3659(6) \\
(1 5) &  0.169404... & 0.1678(5) &  0.205997... & 0.2044(6) \\
(2 2) &  0.123554... & 0.1222(4) &  0.104679... & 0.1041(7) \\
(2 3) &  0.082792... & 0.0820(5) &  0.086363... & 0.0853(6) \\
(2 4) &  0.082792... & 0.0820(5) &  0.086363... & 0.0853(6) \\
(2 5) &  0.042156... & 0.0420(4) &  0.044335... & 0.0441(6) \\
(3 3) &  0.123554... & 0.1222(4) &  0.104679... & 0.1041(7) \\
(3 4) &  0.082792... & 0.0820(5) &  0.086363... & 0.0853(6) \\
(3 5) &  0.042156... & 0.0420(4) &  0.044335... & 0.0441(6) \\
(4 4) &  0.123554... & 0.1222(4) &  0.104679... & 0.1041(7) \\
(4 5) &  0.042156... & 0.0420(4) &  0.044335... & 0.0441(6) \\
(5 5) &  0.022624... & 0.0224(5) &  0.023111... & 0.0225(6) \\
\BotRule
\end{tabular}
\caption{Comparison of di-neutron matrix elements defined in \cref{eq:dinme} between exact diagonalization results and worldline MC. The MC calculations are done at $a_t=0.0005/\varepsilon$. Since the matrix is symmetric, we only give the fifteen independent elements. Several matrix elements are identical due to rotational symmetry. In the Monte Carlo calculations we do not compute them independently.\label{tab:di-neutron-test}.}
\end{table}

\begin{figure}[htb]
\includegraphics[width=0.48\textwidth]{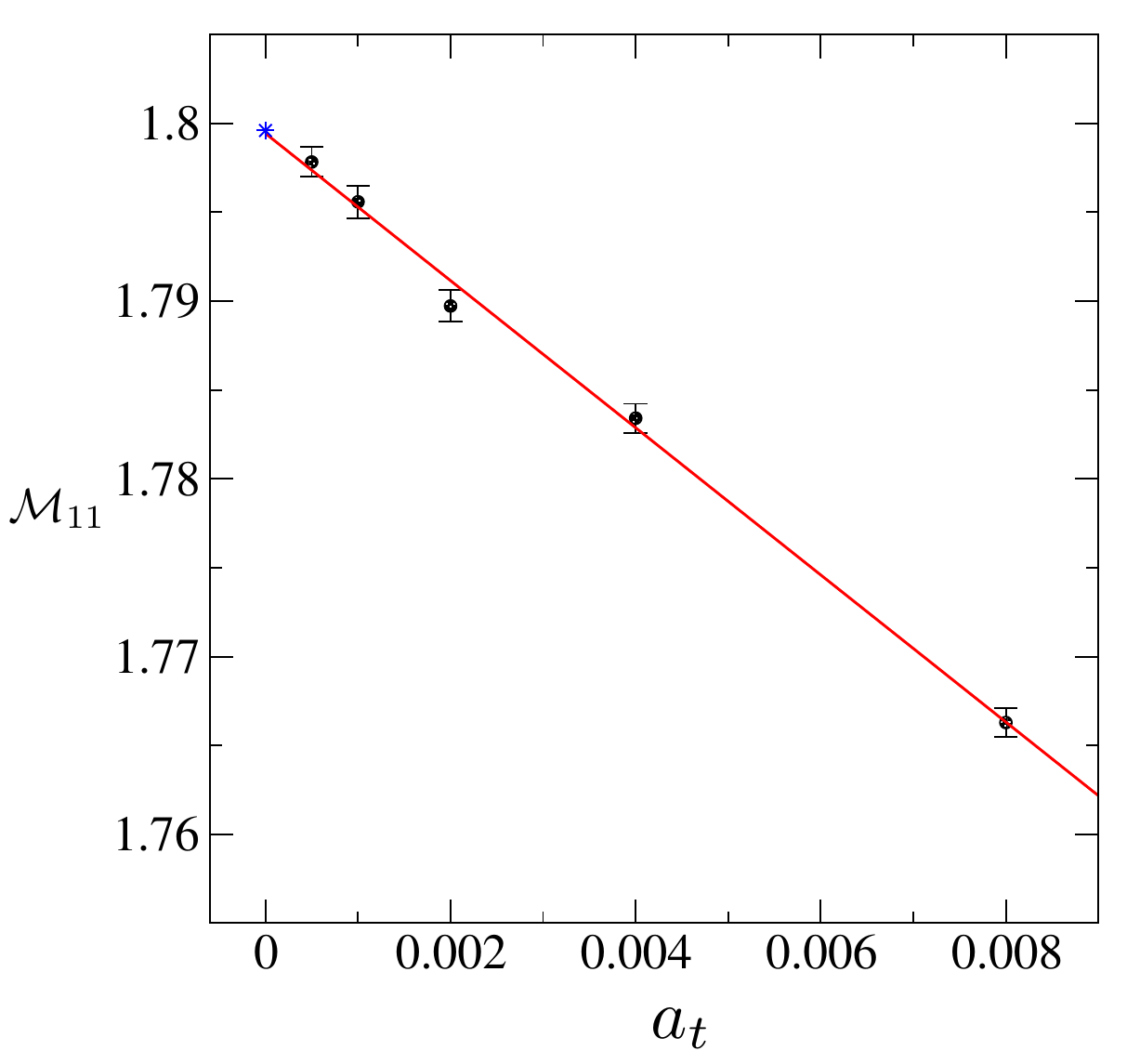}
\caption{A demonstration that the systematic deviation of the MC result for $\Mel_{11}$ in the di-neutron channel at $\beta=0.4/\varepsilon$ in \cref{tab:di-neutron-test} as compared to the exact result is due to finite $a_t$ errors. The solid line is a linear fit to the data. The data shown at $a_t=0$ indicates the exact result $\Mel_{11}=1.799597$ and the fit gives us $\Mel=1.7994(6)$. \label{fig:aterr}}
\end{figure}

While our MC results are close to the exact results, they seem to be systematically lower by about 3-4 $\sigma$ in many cases. For example the exact matrix element $\Mel_{11} = 1.799597...$ is about 3-$\sigma$ away from the MC result of $1.7978(8)$. While this is of course statistically possible in a few rare cases, many more matrix elements show a similar trend. Since the only source of error in our calculations is due to the finite value of $a_t$, we performed a small $a_t$ extrapolation to make sure we can eliminate it if needed. In \cref{fig:aterr} we show our MC results at several values of $a_t$ along with a linear fit that helps extrapolate our data to $a_t=0$. After the extrapolation we get $\Mel_{11} = 1.7994(6)$ which is closer to the exact result.

\begin{table}[!htb]
\centering
\renewcommand{\arraystretch}{1.4}
\setlength{\tabcolsep}{4pt}
\begin{tabular}{c|c|c|c|c|c}
\TopRule
\multirow{2}{*}{$H$} & \multicolumn{5}{c}{${\cal E}_5$} \\
\cline{2-6}
& exact & $10^{-2}$ & $10^{-3}$ & $10^{-4}$ & MC\\
\MidRule
-1.7552... & -1.7550... & -1.79 & -1.753 & -1.7549 & -1.764(1)  \\
4.8967... & 4.9024... & 0.00 & 5.031 & 4.9023  & 5.07(2) \\
8.0000... & 8.0000... & 13.86 & 7.691 & 8.0178  & 7.61(3)\\
8.0000... & 8.0000... & 13.86 & 7.691 & 8.0178  & 7.61(3) \\
14.064... & 14.9075... & --- & 15.445 & 14.4461 & --- \\
\BotRule
\end{tabular}
\caption{ Comparison of the lowest five dominant eigenvalues of $H$ that contribute to the di-neutron matrix elements in \cref{tab:di-neutron-test} with the eigenvalues of ${\cal E}_5$ defined in \cref{eq:tmevals}. The first column gives the eigenvalues of $H$, while the second column gives the eigenvalues of ${\cal E}_5$ for $\beta_1=0.4/\varepsilon$ and $\beta_2=0.5/\varepsilon$ using the exact matrix elements. The third column shows the results of the second column, assuming the matrix elements are determined up to an accuracy of $10^{-3}$ in a hypothetical Monte Carlo calculation. All numbers shown are in units of $\varepsilon$.
\label{tab:di-neutron-e}}
\end{table}

\begin{table}[!htb]
\centering
\renewcommand{\arraystretch}{1.4}
\setlength{\tabcolsep}{4pt}
\begin{tabular}{c|c|c|c|c}
\TopRule
& \multicolumn{2}{|c|}{$M_{ij}(\beta=0.4/\varepsilon)$} & \multicolumn{2}{|c}{$M_{ij}(\beta=0.5/\varepsilon)$} \\
\MidRule
 $(i,j)$ & exact & MC & exact & MC\\
\MidRule
(1,1) &  2.239217... & 2.2348(9) &  2.856016... &  2.8504(13)\\
(1 2) &  0.488407... & 0.4860(6) &  0.651407... &  0.6480(8)\\
(1 3) &  0.488407... & 0.4860(6) &  0.651407... &  0.6480(8)\\
(1 4) &  0.488407... & 0.4860(6) &  0.651407... &  0.6480(8)\\
(1 5) &  0.286565... & 0.2849(6) &  0.375937... &  0.3725(7)\\
(2 2) &  0.196787... & 0.1947(6) &  0.202076... &  0.2001(8) \\
(2 3) &  0.156025... & 0.1546(6) &  0.183761... &  0.1816(8)\\
(2 4) &  0.156025... & 0.1546(6) &  0.183761... &  0.1816(8)\\
(2 5) &  0.084723... & 0.0835(6) &  0.100467... &  0.0989(8)\\
(3 3) &  0.196787... & 0.1947(6) &  0.202076... & 0.2001(8) \\
(3 4) &  0.156025... & 0.1546(6) &  0.183761... &  0.1816(8) \\
(3 5) &  0.084723... & 0.0835(6) &  0.100467... &  0.0989(8) \\
(4 4) &  0.196787... & 0.1947(6) &  0.202076... &  0.2001(8) \\
(4 5) &  0.084723... & 0.0835(6) &  0.100467... &  0.0989(8) \\
(5 5) &  0.047091... & 0.0470(7) &  0.055331... &  0.0548(7)\\
\BotRule
\end{tabular}
\caption{Comparison of deuteron matrix elements between exact diagonalization results and worldline MC calculations with $a_t=0.0005/\varepsilon$. Only the independent elements of a symmetric matrix are shown and  rotational symmetry is assumed \label{tab:deuteron-test}.}
\end{table}

We can use the matrix elements in \cref{tab:di-neutron-test} to construct $T_5(\beta_1=0.4/\varepsilon)$ and $T_5(\beta_2=0.5/\varepsilon)$ and then using \cref{eq:tmevals} we can evaluate the five eigenvalues of the matrix ${\cal E}_5$. We can then compare it with the lowest five exact energy eigenvalues of $H$ that have a dominant overlap with that the five dimensional sub-space that we are projecting out through $T_5(\beta)$. In \cref{tab:di-neutron-e} we show these results. The first column shows the exact eigenvalues of $H$ that contribute to the matrix elements. The next four columns give eigenvalues of ${\cal E}_5$ assuming that we can compute the the matrix elements $T_5(\beta_1=0.4/\varepsilon)$ and $T_5(\beta_2=0.5/\varepsilon)$ at various levels of accuracy after extrapolating to $a_t=0$. The column labeled exact assumes double precision accuracy of the matrix elements, while the other three columns labeled $10^{-p}$ assume the elements are evaluated accurately up to $p$ decimal numbers. The last column marked as MC uses the numbers for the matrix elements given in \cref{tab:di-neutron-test} without $a_t$ extrapolation. The central value and error are estimated based by a stochastic method which involves randomly adding or subtracting 1-sigma errors to the central values. The five columns under ${\cal E}_5$ give a sense of Monte Carlo errors that can come from both finite $a_t$ errors and statistical errors. Based on the results of the extrapolation in \cref{fig:aterr} if we can assume that in a hypothetical MC calculation with $a_t$ extrapolations we can compute all matrix elements with an accuracy of $10^{-3}$, from \cref{tab:di-neutron-e} we see that the energy levels can be extracted reasonably well. In this example, since the various energy levels are well separated, this is not surprising.

\begin{table}[!htb]
\centering
\renewcommand{\arraystretch}{1.4}
\setlength{\tabcolsep}{4pt}
\begin{tabular}{c|c|c|c|c|c}
\TopRule
\multirow{2}{*}{$H$} & \multicolumn{5}{c}{${\cal E}_5$} \\
\cline{2-6}
& exact & $10^{-2}$ & $10^{-3}$ & $10^{-4}$ & MC \\
\MidRule
-2.5053... & -2.5050.. & -2.51 & -2.506 & -2.5051 & -2.5046(4)\\
4.2971... & 4.3028... & 2.69 & 4.211 & 4.2890 & 4.223(41)\\
8.0000... & 8.0000... & 6.93 & 8.232 & 8.0178 & 7.511(11)\\
8.0000... & 8.0000... & 6.93 & 8.232 & 8.0178 & 7.511(11)\\
13.817... & 14.6926... & 8.30 & 13.482 & 15.5810 & 11.17(75)\\
\BotRule
\end{tabular}
\caption{ Comparison of the lowest five dominant eigenvalues of $H$ that contribute to the matrix elements in \cref{tab:deuteron-test} with the eigenvalues of ${\cal E}_5$ defined in \cref{eq:tmevals}. All numbers shown are in units of $\varepsilon$. The meanings of the three columns are the same as \cref{tab:di-neutron-e}.
\label{tab:deutron-e}}
\end{table}

\subsection{The deuteron channel}

We now repeat the above calculations in the deuteron channel. We focus on the five states similar to the di-neuteron channel, except that the spin-down neuteron is replaced by a spin-up proton. Thus, the five states now take the form
\begin{align}
|\psi_i\rangle \ =\ \tNud{\bk_i,3}\tNud{\bk_i,1}  |0\rangle, i=1,2,3,4,5.
\end{align}
We again compute the corresponding $5\times 5$ matrix defined in \cref{eq:dinme} at $\beta_1=0.4/\varepsilon$ and $\beta_2=0.5/\varepsilon$ assuming $a_t=0.0005/\varepsilon$. Our results for the matrix elements are tabulated in \cref{tab:deuteron-test} in the same format as in the di-neutron case.

Using the exact matrix elements in \cref{tab:deuteron-test} we can again construct $T_5(\beta=0.4/\varepsilon)$ and $T_5(\beta=0.5/\varepsilon)$ and the matrix ${\cal E}_5$ using \cref{eq:tmevals}. We again compare the lowest five exact energy eigenvalues of $H$ that have a dominant overlap with the five dimensional sub-space with the five eigenvalues of ${\cal E}_5$. The results of this comparison are shown in \cref{tab:deutron-e}. In this comparison we again assume the matrix elements of $T_5(\beta=0.4/\varepsilon)$ and $T_5(\beta=0.5/\varepsilon)$ are obtained with various percisions as explained during the discussion of the results in \cref{tab:di-neutron-e}. We again notice that the accuracy with which the matrix elements are computed greatly affects the extracted eigenvalues. We observe that if we can extract the matrix elements with a precision of $10^{-3}$ the lowest two energies can be extracted with about two percent errors. However notice that the energies are well separated.

\begin{table}[!htb]
\centering
\renewcommand{\arraystretch}{1.4}
\setlength{\tabcolsep}{4pt}
\begin{tabular}{r|r|c|r|c}
\TopRule
& \multicolumn{2}{|c|}{$M_{ij}(\beta=1.0/\varepsilon)$} & \multicolumn{2}{|c}{$M_{ij}(\beta=1.1/\varepsilon)$} \\
\MidRule
 $(i,j)$ & exact & MC & exact & MC\\
\MidRule
(1,1) &  21.96367... & 21.54(19) &  41.84479... & 40.90(47) \\
(1 2) &  8.06498... &  7.72(03) &  15.35780... & 14.65(20) \\
(1 3) &  11.4549... &  11.05(10) &  21.85445... & 20.95(23) \\
(1 4) &  7.50681... &  7.10(09) &  14.32638... & 13.90(19) \\
(1 5) &  4.54160... &  4.30(06) &  8.65113... & 8.20(13) \\
(2 2) &  2.98205... & 2.84(08) &  5.65721... & 5.69(18)\\
(2 3) &  4.20247... & 4.08(10) &  8.01511... & 7.55(20) \\
(2 4) &  2.75182... & 2.61(07) &  5.25175... & 5.11(17) \\
(2 5) &  1.67677... & 1.53(06) &  3.18442... & 3.02(12) \\
(3 3) &  5.98551... & 5.71(09) &  11.42462... & 10.73(21) \\
(3 4) &  3.92403... & 3.71(07) &  7.49092... & 6.96(17) \\
(3 5) &  2.36777... & 2.18(06) &  4.51619... & 4.11(12) \\
(4 4) &  2.57295... & 2.46(08)&  4.91209... & 4.81(15) \\
(4 5) &  1.55078... & 1.44(07)&  2.95948... & 2.78(12) \\
(5 5) &  0.94473... & 0.95(06) &  1.79401... & 1.59(11) \\
\BotRule
\end{tabular}
\caption{Comparison of triton matrix elements between exact diagonalization results and worldline MC calculations with $a_t=0.001/\varepsilon$. Only the independent elements of a symmetric matrix are shown. Since rotational symmetry does not help here, the matrix elements are all different. \label{tab:triton-test}}
\end{table}

\subsection{The triton channel}

In the triton channel we focus on the five basis states defined as
\begin{align}
|\psi_1\rangle \ &=\ \tNud{\bk_2,3}\tNud{\bk_2,2}\tNud{\bk_1,1} |0\rangle
\nonumber \\ 
|\psi_2\rangle \ &=\ \tNud{\bk_5,3}\tNud{\bk_3,2}\tNud{\bk_2,1} |0\rangle
\nonumber \\ 
|\psi_3\rangle \ &=\ \tNud{\bk_5,3}\tNud{\bk_5,2}\tNud{\bk_1,1} |0\rangle
\nonumber \\ 
|\psi_4\rangle \ &=\ \tNud{\bk_8,3}\tNud{\bk_8,2}\tNud{\bk_1,1} |0\rangle
\nonumber \\ 
|\psi_5\rangle \ &=\ \tNud{\bk_8,3}\tNud{\bk_4,2}\tNud{\bk_5,1} |0\rangle,
\end{align}
and as before, compute the $5\times 5$ matrix defined in \cref{eq:dinme}. Here we choose states that are not related by by rotations in order to learn how the lack of symmetries affects our ability to extract the low energy spectrum. Since the higher energy levels are more closely packed in the triton case we choose slightly higher values of $\beta$. We choose $\beta=1.0/\varepsilon$ and $\beta = 1.1/\varepsilon$. The Monte Carlo results are also obtained at a slightly higher temporal lattice spacing of $a_t=0.001/\varepsilon$. In \cref{tab:triton-test} we tabulate our results for the matrix elements following the same format as \cref{tab:di-neutron-test} and \cref{tab:deuteron-test}. We again notice that the Monte Carlo results are close to the exact answers but clearly systematically off due to $a_t$ errors. We also notice larger fluctuations in the Monte Carlo data, which means we need more statistics to obtain accurate values of the matrix elements. The matrix elements we compute are an order of magnitude larger than those in the two body sector, essentially due to the larger triton binding energy and the higher value of $\beta$ used in the calculations. Reducing the errors by a factor of ten than what is shown in \cref{tab:triton-test} is in principle feasible.

\begin{figure}[hbt]
\includegraphics[width=0.46\textwidth]{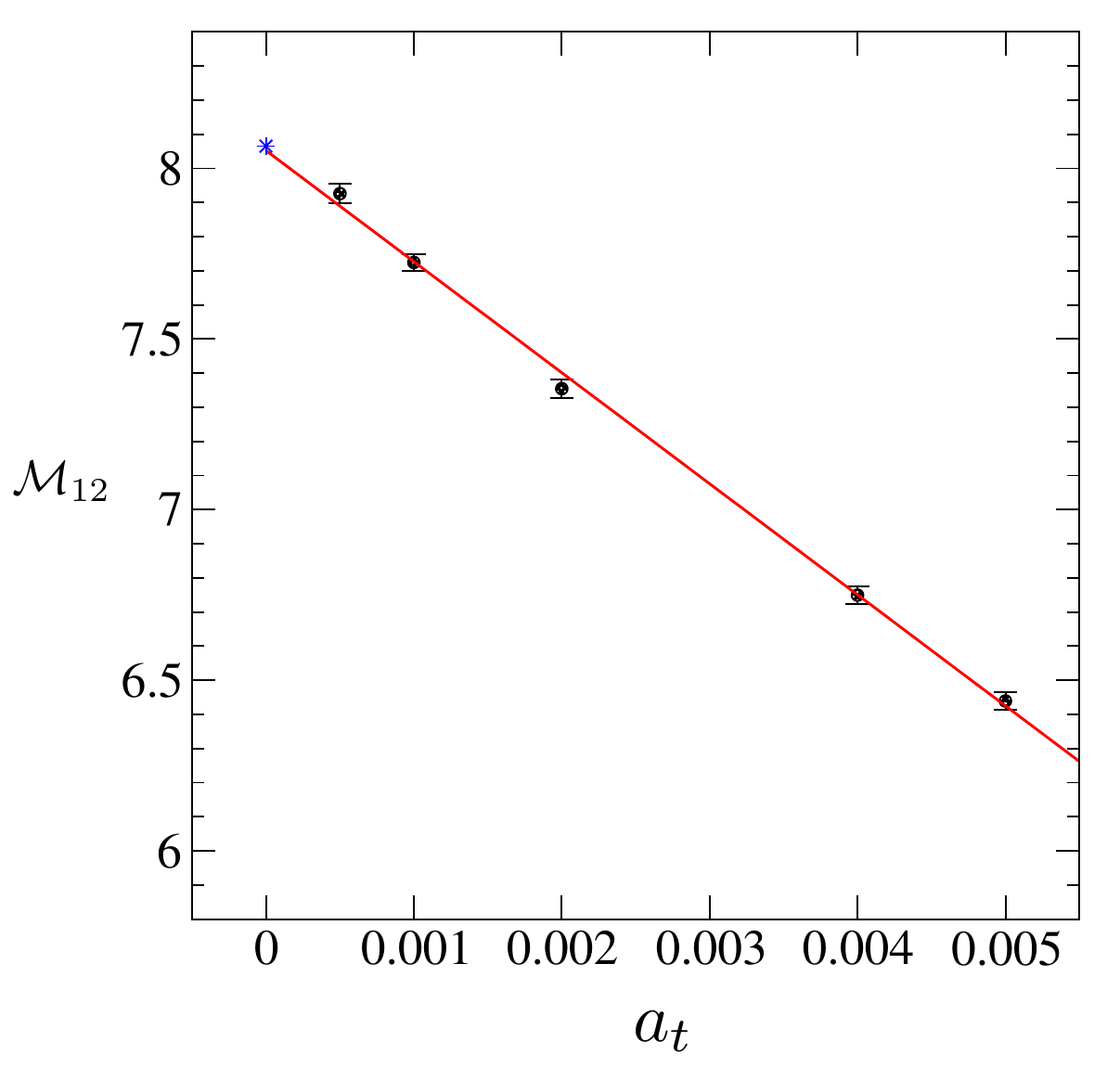}
\caption{A demonstration that the deviation of the MC result for $\Mel_{12}$ at $\beta=1.0/\varepsilon$ using $a_t=0.001/\varepsilon$ as compared to the exact result, shown in \cref{tab:triton-test}, is due to finite $a_t$ errors. We show data at several values of $a_t$ and fit it to a linear function of $a_t$ (solid line). The data shown at $a_t=0$ indicates the exact result $\Mel_{12}=8.06498...$ and the fit gives us $\Mel=8.05(2)$. \label{fig:aterr-trit}}
\end{figure}

\begin{table}[!htb]
\centering
\renewcommand{\arraystretch}{1.4}
\setlength{\tabcolsep}{4pt}
\begin{tabular}{c|c|c|c|c|c}
\TopRule
\multirow{2}{*}{$H$} & \multicolumn{5}{c}{${\cal E}_5$} \\
\cline{2-6}
& exact & $10^{-1}$ & $10^{-2}$ & $10^{-4}$ & MC\\
\MidRule
-6.4692... &  -6.4681... & 6.1 & -15.08 & -6.4674 & -6.321(49) \\
-1.0711... & -1.0340... & 22.3 & -6.55 &  -0.9972 & ---\\
1.7049... &  2.4491... & 331.5 & -1.53 &  2.4680 &--- \\
2.3679... &  2.6395... & --- & 5.74 &  1.9510 & ---\\
2.3679... & 5.1663... & --- & --- &  7.0213 & ---\\
\TopRule
\multirow{2}{*}{$H$} & \multicolumn{5}{c}{${\cal E}_3$} \\
\cline{2-6}
& exact & $10^{-1}$ & $10^{-2}$ &  $10^{-4}$ & MC \\
\MidRule
-6.4692... &  -6.4665... & -10.8 & -6.45 & -6.4665 & -6.542(69) \\
-1.0711... & -0.8537... & -6.4 & 2.23 &  -0.8501 & --- \\
1.7049... &  2.6144... & -5.0 & 7.61 & 2.7679 & ---\\
\TopRule
\multirow{2}{*}{$H$} & \multicolumn{5}{c}{${\cal E}_2$} \\
\cline{2-6}
& exact & $10^{-1}$ & $10^{-2}$ & $10^{-4}$ & MC \\
\MidRule
-6.4692... &  -6.4664... & -6.4 & -6.45 & -6.4465 & -6.385(11)\\
-1.0711... & -0.0016... & -3.9 & 0.27 & 0.0244 & ---\\
\BotRule
\end{tabular}
\caption{ Comparison of the lowest five dominant eigenvalues of $H$ that contribute to the triton matrix elements with the eigenvalues of ${\cal E}_5$ assuming the matrix elements in \cref{tab:triton-test} are known at various levels of accuracy. All numbers shown are in units of $\varepsilon$. We notice that the energy eigenvalues of ${\cal E}_5$ are highly sensitive to the accuracy with which matrix elements are computed. In some cases, when the accuracy is low, the eigenvalues of $e^{{\cal E}_5}$ become negative. These are shown as dashes in the table.  
\label{tab:triton-e}}
\end{table}

The finite $a_t$ errors of the matrix elements can again be eliminated using an extrapolation to small $a_t$. To demonstrate this we have computed the matrix element $\Mel_{1,2}$ at several values of $a_t$. Our results for the extrapolation are plotted in \cref{fig:aterr-trit}. While the exact answer is $\Mel_{11}=8.06498$ the fit of our data gives us $\Mel=8.034(38)$. Assuming we can compute all the matrix elements with this accuracy, how well can we extract the dominant eigenvalues of $H$ from  ${\cal E}_5$ as we did for the di-neutron and deuteron cases. Our results are tabulated in \cref{tab:triton-e}. We again notice several new features. First, we notice that with accuracy of $10^{-2}$ we get a spurious energy level with an anomalously large binding energy of $-15.08\varepsilon$. This bound state disappears when the precision is increased to $10^{-3}$. Even at that precision, one of the eigenvalues of $e^{{\cal E}_5}$ turns out to be negative. Finally, although the lowest two energy eigenvalues of $H$ are well separated, the exact energies of the higher states are more closely packed. 

Where did the deeply bound spurious state appear from? In order to address this puzzle let us look at ${\cal E}_2$, where we form the transfer matrices using $|\psi_1\rangle$ and $|\psi_2\rangle$, and on ${\cal E}_3$ by including the additional state $|\psi_3\rangle$. The results from these smaller subspaces are shown in the lower rows of \cref{tab:triton-e}. We notice that the deeply bound state has disappeared in the ${\cal E}_2$, while it appears in ${\cal E}_3$ at a different value when the precision is reduced to $10^{-1}$. However, in all the cases the lowest stable bound state eigenvalue is approximately $-6.5(1)$, which agrees with the exact eigenvalue of $H$. Thus, we learn that without sufficient accuracy we can in principle obtain spurious energy levels in the spectrum of ${\cal E}_k$. But by varying $k$ and increasing the precision we can identify the stable physical eigenvalues.

\begin{table}[!t]
\centering
\renewcommand{\arraystretch}{1.4}
\setlength{\tabcolsep}{4pt}
\begin{tabular}{r|r|c|r|c}
\TopRule
& \multicolumn{2}{|c|}{$M_{ij}(\beta=0.8/\varepsilon)$} & \multicolumn{2}{|c}{$M_{ij}(\beta=0.9/\varepsilon)$} \\
\MidRule
 $(i,j)$ & exact & MC & exact & MC\\
\MidRule
(1,1) &  202.4837... & 193.4(14) &  623.5825... &  596(5) \\
(1 2) &  59.6957... &  56.3(05) &  185.2066... & 182(2) \\
(1 3) &  97.8796... &  93.3(07) & 303.7617... &  289(3) \\
(1 4) &  61.2623... &  58.2(02) &  190.2936... &  179(2) \\
(1 5) &  27.4097... &  26.1(3) &  85.2486... & 77.3(12) \\
(2 2) &  18.5980... & 17.7(03) &  56.5716... & 53.2(12) \\
(2 3) &  28.9417... & 27.0(04) &  90.2168... & 85.6(16) \\
(2 4) &  18.0076... & 17.3(03) &  56.3303... & 54.5(11) \\
(2 5) &  8.3833... & 7.8(03) &  25.8156... & 25.0(10) \\
(3 3) &  47.9952... & 44.9(05) &  148.9880 & 140(02) \\
(3 4) &  30.0876... & 28.3(03) &  93.4354... & 87.3(14) \\
(3 5) &  13.3538... &  12.4(04) &  41.6357... &  38.6(14)\\
(4 4) &  18.8981... & 17.5(03) &  58.6521... & 55.2(12) \\
(4 5) &  8.3464... & 7.4(03) &  26.0566... & 24.6(9)\\
(5 5) &  3.9308... & 4.0(03) & 11.9996... & 11.5(9) \\
\BotRule
\end{tabular}
\caption{Comparison of matrix elements in the helium channel between exact diagonalization results and worldline MC calculations with $a_t=0.001/\varepsilon$. Only the independent elements of a symmetric matrix are shown. Since rotational symmetry does not help, all the independent matrix elements are all different. \label{tab:helium-test}}
\end{table}

\begin{figure}[!b]
\includegraphics[width=0.48\textwidth]{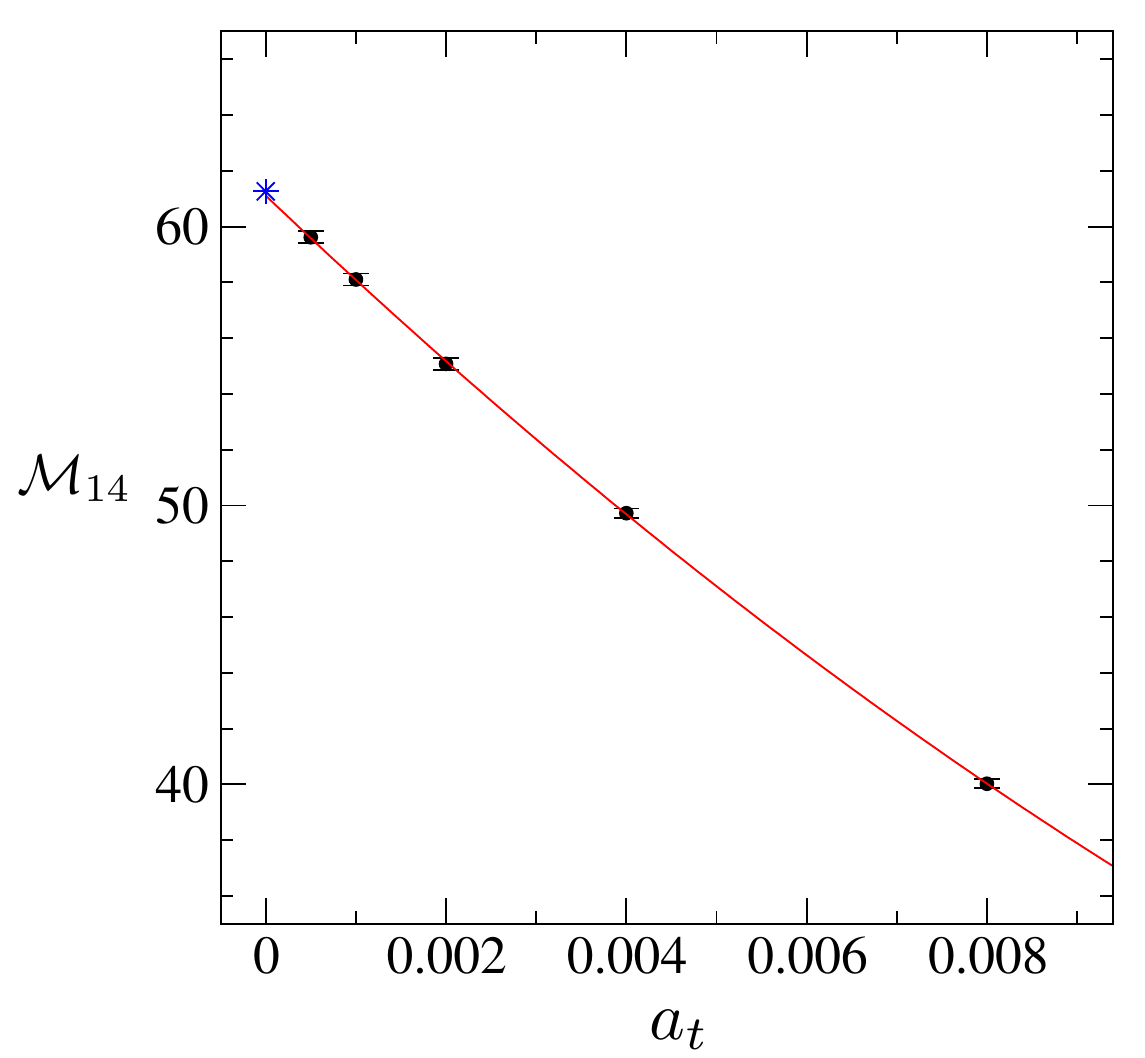}
\caption{The systematic deviation of the MC result for $\Mel_{14}$ at $\beta=0.8$ in \cref{tab:helium-test} as compared to the exact result is shown to be due to finite $a_t$ errors. The solid line is a quadratic fit to the data. The star symbol shown at $a_t=0$ indicates the exact result $\Mel_{14}=61.2623...$ and the fit gives us $\Mel=61.1(2)$. A linear fit has a bad $\chi^2$/DOF. \label{fig:aterr-hel}}
\end{figure}

\begin{table}[!t]
\centering
\renewcommand{\arraystretch}{1.4}
\setlength{\tabcolsep}{4pt}
\begin{tabular}{c|c|c|c|c|c}
\TopRule
\multirow{2}{*}{$H$} & \multicolumn{5}{c}{${\cal E}_5$} \\
\cline{2-6}
& exact & $10^{-1}$ & $10^{-2}$ & $10^{-3}$ & MC \\
\MidRule
-11.3836... & -11.3450... & -11.3 & -11.34 & -11.345 & -11.441(34) \\
-6.2516... & -5.8384... & -5.4 & -5.93 & -5.827 & ---\\
-4.3515... &  -3.8719... & -3.5 & -3.72 & -3.953 & ---- \\
-4.3515... &  -3.0434... & 6.2 &  -2.73 & -3.026 & --- \\
-4.0528... & -0.6487... & --- & 5.17 & -1.028 & --- \\
\TopRule
\multirow{2}{*}{$H$} & \multicolumn{5}{c}{${\cal E}_3$} \\
\cline{2-6}
& exact & $10^{-1}$ & $10^{-2}$ &  $10^{-4}$ & MC \\
\MidRule
-11.3836... &  -11.3346... & -11.3 & -11.33 & -11.3347 &  -11.398(6)\\
-6.2516... & -5.3802... & -5.1 & -5.29 &  -5.3789 & ---\\
-4.3515... &  -3.1589... & -3.5 & -3.20 & -3.1592 & ---\\
\TopRule
\multirow{2}{*}{$H$} & \multicolumn{5}{c}{${\cal E}_2$} \\
\cline{2-6}
& exact & $10^{-1}$ & $10^{-2}$ & $10^{-4}$ & MC \\
\MidRule
-11.3836... &  -11.2678... & -11.3 & -11.27 & -11.2679 &  -11.451(5)\\
-6.2516... & -4.4689... & -4.7 & -4.45 & -4.4682 & ---\\\BotRule
\end{tabular}
\caption{ Comparison of the lowest five dominant eigenvalues of $H$ that contribute to the helium matrix elements with the eigenvalues of ${\cal E}_5$ assuming the matrix elements are known at various levels of accuracy. There are several higher eigenvalues of $H$ (like $-3.9971...$ and $-3.9354...$) which also have large overlap with some of the chosen states, but are not shown here.\label{tab:helium-e}}
\end{table}

\subsection{The Helium Channel}

We next consider calculations in the four-body channel by focusing on the five basis states defined as
\begin{align}
|\psi_1\rangle \ &=\ 
\tNud{\bk_2,4}\tNud{\bk_1,3}\tNud{\bk_2,2}\tNud{\bk_1,1} |0\rangle
\nonumber \\ 
|\psi_2\rangle \ &=\ \tNud{\bk_5,4}\tNud{\bk_1,3}\tNud{\bk_3,2}\tNud{\bk_2,1} |0\rangle
\nonumber \\ 
|\psi_3\rangle \ &=\ \tNud{\bk_5,4}\tNud{\bk_1,3}\tNud{\bk_5,2}\tNud{\bk_1,1} |0\rangle
\nonumber \\ 
|\psi_4\rangle \ &=\ \tNud{\bk_8,4}\tNud{\bk_1,3}\tNud{\bk_8,2}\tNud{\bk_1,1} |0\rangle
\nonumber \\ 
|\psi_5\rangle \ &=\ \tNud{\bk_8,4}\tNud{\bk_1,3}\tNud{\bk_4,2}\tNud{\bk_5,1} |0\rangle.
\end{align}
We again compute the $5\times 5$ matrix defined in \cref{eq:dinme} at $\beta=0.8/\varepsilon$ and $0.9/\varepsilon$ with $a_t=0.001/\varepsilon$ and compare with results from the exact diagonalization. The results are shown in \cref{tab:helium-test}. Again the matrix elements computed using the Monte Carlo method are close to the exact values but remain systematically off due to finite $a_t$ errors. In \cref{fig:aterr-hel} we show how these errors can be removed by performing extrapolations using the example of $\Mel_{14}$. Since the data at various values of $a_t$ show some curvature, the extrapolation involves a quadratic function in $a_t$. The extrapolated result is again in excellent agreement with the exact result. A linear extrapolation gives a large $\chi^2$/DOF.

We can again extract the lowest five energy levels from ${\cal E}_5$ and compare them with the exact eigenvalues of $H$. Our results are tabulated in \cref{tab:helium-e} where the various columns have the same meaning as already discussed in the two and three body channels. An important difference from the pervious cases is that there are several closely packed bound states. This clearly affects our ability to extract the higher energy states. Since the lowest energy is deeply bound, it can be extracted reliably.
As in the triton case, if we are only interested in the lowest energy state even ${\cal E}_2$ is sufficient. The lowest energy is given by $E_0^{He} = -11.3 \varepsilon = -114.6$ MeV which agrees well with the previous lattice QCD result in the hypothetical physical system we are studying \cite{Eliyahu:2019nkz}.

\begin{table*}[!htb]
\centering
\renewcommand{\arraystretch}{1.4}
\setlength{\tabcolsep}{4pt}
\begin{tabular}{r|r|r|r|r|r|r}
\TopRule
& \multicolumn{2}{c|}{$M_{ij}(L=4)$} & \multicolumn{2}{c|}{$M_{ij}(L=8)$} & \multicolumn{2}{c}{$M_{ij}(L=16)$} \\
\MidRule
$(i,j)$ & 
$(\beta=1.0/\varepsilon)$ & 
$(\beta=1.1/\varepsilon)$ &
$(\beta=1.0/\varepsilon)$ & 
$(\beta=1.1/\varepsilon)$ &
$(\beta=1.0/\varepsilon)$ & 
$(\beta=1.1/\varepsilon)$ \\
\MidRule
(1,1) & 11.00(15) & 22.49(35) & 0.452(2) & 0.459(3) & 0.7577(4) & 0.7383(4)\\
(1,2) & 0.004(03) & 0.008(5) & 0.019(1)& 0.0331(14) & 0.0001(2) & 0.0005(3)\\
(1,3) & 0.751(20) & 1.818(62) & 0.0326(7)& 0.0471(11) & 0.0034(2) & 0.0036(3) \\
(1,4) & 3.695(49) & 8.09(16) & 0.0325(7) & 0.0503(11) & 0.0016(3) & 0.0021(3) \\
(1,5) & 0.281(13) & 0.769(34) & 0.0001(4) & 0.0000(5) & 0.0003(3) & 0.0002(3) \\
(2,2) & 0.007(2) & 0.003(4) & 0.0187(13) & 0.0271(21)& 0.1435(3) & 0.1183(3)\\
(2,3) & -0.022(38) & -0.03(11)&0.0066(8) & 0.0138(12) & 0.0010(2) & 0.0010(3)\\
(2,4) & -0.004(27) & 0.008(85)& 0.0008(4) & 0.0012(7) & -0.0001(2) & 0.0002(2)\\
(2,5) & 0.013(25) & -0.055(61) & -0.0001(3) & -0.0002(4) & 0.0001(3) & -0.0001(3)\\
(3,3) & 10.66(13) & 23.02(32) & 0.0121(6) & 0.0179(11)& 0.088(3)& 0.0690(3)\\
(3,4) & -0.003(3) & 0.003(6) & 0.0079(9) & 0.0139(13) & 0.0009(3)& 0.0009(3)\\
(3,5) & 3.25(16) & 9.1(6) & 0.0136(7) & 0.0186(9) & 0.0005(2) & 0.0006(3) \\
(4,4) & 4.215(67) & 8.85(18) & 0.4533(18) & 0.4636(27) & 0.0005(3) & 0.0009(3)\\
(4,5) & 2.894(61) & 6.39(18) & -0.0001(4)& 0.0003(5) & 0.0001(3) & 0.0000(3)\\
(5,5) & 3.755(87) & 8.16(25) & 0.0036(4) & 0.0036(5) & 0.0009(3) & 0.0004(3)\\
\BotRule
\end{tabular}
\caption{Matrix elements in the triton channel on larger lattices \label{tab:lltriton}.}
\end{table*}

\section{Challenges on larger lattices}
\label{sec7}

Our Monte Carlo method continues to work well on larger lattices as long as $\beta \varepsilon$ is fixed, which makes sense if we wish to study our lattice theory at a fixed lattice spacing but on larger physical volumes. However, if we wish to hold the physical volume fixed and approach the continuum limit, we will need to renormalize our theory and change $\varepsilon$ according to \cref{tab:lat2pparams} as we make our lattice sizes larger. Since $\varepsilon$ is the lattice cutoff energy scale, we observe that it grows as we approach the continuum limit. On the other hand we will need to fix the scale $\beta$ to some physical value to be able to extract the low energy physics. This means we will need to explore larger values of $\beta\varepsilon$. As expected, in this limit our algorithm encounters new challenges and needs some further refinement. We discuss these issues in this section. 

First, we wish to give some sense of how our algorithm behaves at larger lattices. To do this we will change $\Czerosing$ and $\Czerotrip$ according to the values given in \cref{tab:lat2pparams} as we go to larger lattices so as to keep the lattice spacing fixed.
Ideally we would have liked to change $\Ctb$ with the lattice spacing, but this is beyond the scope of what we could study in our current work, since it requires a lot of systematic Monte Carlo work or analytic work in the three body sector on the lattice. We postpone this study to a future publication. Here we fix $\Ctb=5.109$, which was the result at $L=2$. Thus, our results for triton and helium that we discuss below for larger lattices, must be viewed as an exploration of how our algorithm performs.

In \cref{tab:lltriton} we tabulate the matrix elements in the triton channel in the same five dimensional subspace that we introduced in \cref{sec6}, but now on larger lattices of $L=4$, $8$ and $16$. As mentioned above we fix $\Ctb=5.109$ while changing $\Czerosing$ and $\Czerotrip$ change according to \cref{tab:lat2pparams}. We again choose $\beta=1.0/\varepsilon$ and $\beta=1.1/\varepsilon$ so as to learn how the Monte Carlo algorithm performs as the lattice size increases. In \cref{tab:lltriton} we notice several matrix elements are close to zero, but there also a few that are not small and can be determined well. It is important to recognize that the matrix element ${\cal M}_{ij}$ between the states $|\psi_i\rangle$ and $|\psi_j\rangle$ can be small for two reasons: (1) the overlap $\langle E_k|\psi_i\rangle$ or $\langle E_k|\psi_j\rangle$ with the energy eigenstates $|E_k\rangle$ in the low energy subspace is small,(2) all Botlzmann weights $e^{-\beta E_k}$ in the low energy subspace are small. If there is a bound state the latter cannot be small unless the former is small. 

\begin{table}[tb]
\centering
\renewcommand{\arraystretch}{1.4}
\setlength{\tabcolsep}{4pt}
\begin{tabular}{r|r|r}
\TopRule
$\beta$ & ${\cal M}_{11}$ & \% error \\
\MidRule
$0.8/\varepsilon$ & 0.5991(09) & 0.15 \\
$0.9/\varepsilon$ & 0.6300(19) & 0.30 \\
$1.2/\varepsilon$ & 1.918(27) & 1.4 \\
$1.5/\varepsilon$ &  32.52(65) & 2.0 \\
$1.8/\varepsilon$ & 846(34) & 4.1 \\
$2.0/\varepsilon$ & 8816(801) & 9.1 \\
\BotRule
\end{tabular}
\caption{ The Helium matrix element ${\cal M}_{11}$ on an $L=8$ lattice as a function of $\beta$. The errors are obtained using the same number of Monte Carlo updates. We notice that the error has increased by a factor of 60 when $\beta$ increases from $0.8/\varepsilon$ to $2.0/\varepsilon$. \label{tab:llbhel}.}
\end{table}

Using the data in \cref{tab:lltriton} for $L=4$ we find evidence of at least two bound states. One is comparatively more stable and has a value of roughly $E_0^{nnp}/\varepsilon \approx -7.15(2)$, while the other is unstable and can fluctuate between $-4.07(5)$ and $-8.68(1)$ depending on the matrix elements chosen to analyze it. Substituting $\varepsilon = 20.28$ MeV at $L=4$ from \cref{tab:lat2pparams} we see that the triton bound state we find is $E_0^{nnp} \approx -145$ MeV compared to the $-65.6$ MeV from \cref{tab:physparams}. This strong binding suggests that $\Ctb$ needs to be more positive to increase repulsion. The presence of multiple bound states could also be related to the well known Efimov effect \cite{Braaten:2004rn,Naidon:2016dpf, Kievsky:2021ghz}. When we use $L=8$ and $L=16$ notice that the large non-zero diagonal matrix elements like ${\cal M}_{11}$ and ${\cal M}_{44}$ don't seem to change much between $\beta=1.0/\varepsilon$ and $1.1/\varepsilon$ suggesting that the matrix elements are not yet sensitive to the bound states. Indeed we find that the bound state extracted from the data are very unstable implying that we will need to explore larger values of $\beta \varepsilon$. While this will require more computing resources, our algorithm will continue to work as we explain below.

\begin{figure*}[!t]
\includegraphics[width=0.49\textwidth]{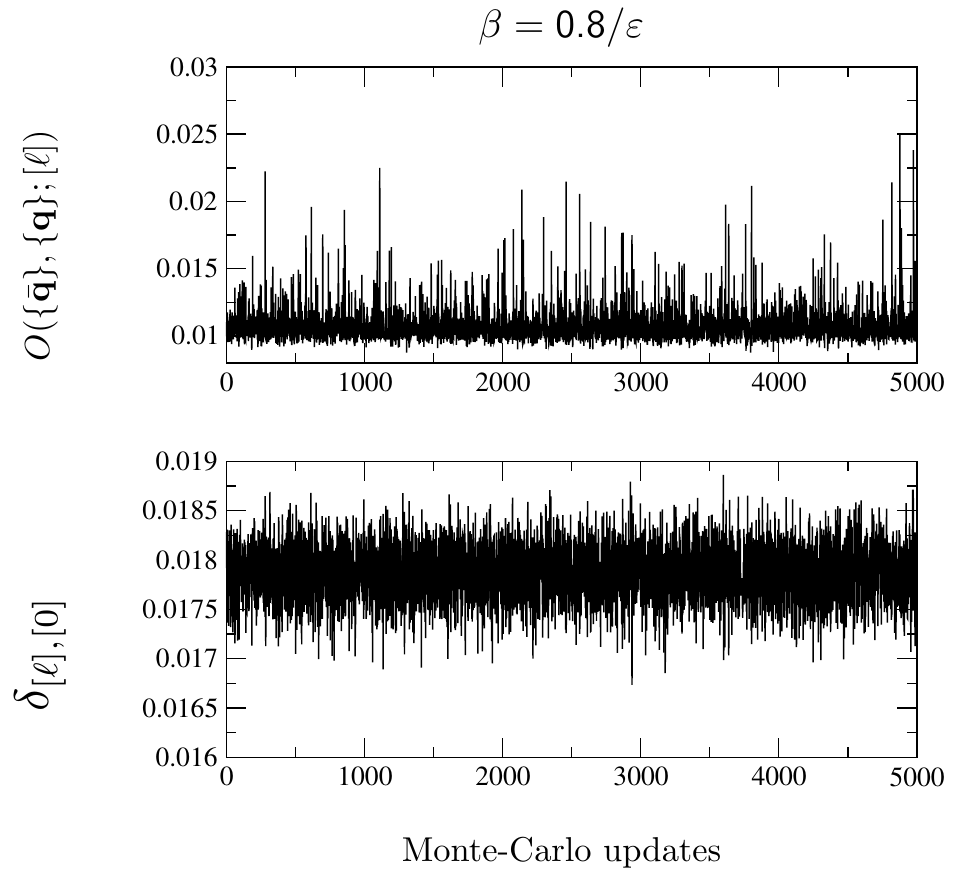}
\includegraphics[width=0.48\textwidth]{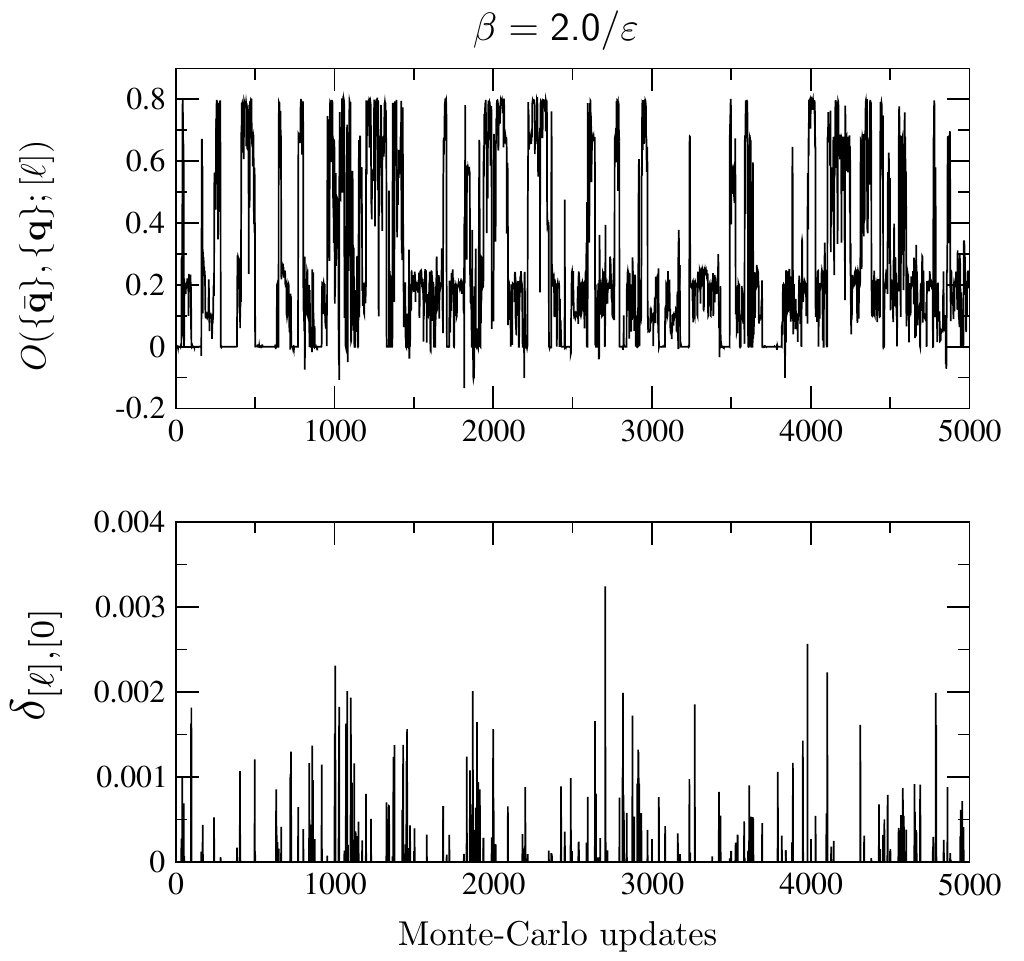}
\caption{Monte Carlo fluctuations in the calculation of ${\cal M}_{11}$ in the helium channel based on \cref{eq:obs3}. The top graphs show the fluctuations of the numerator at $\beta=0.8/\varepsilon$ (left) and $\beta=2.0/\varepsilon$ (right), while the bottom graph shows the fluctuations of the denominators for the same values of $\beta$.
\label{fig:llbz}}
\end{figure*}

In order to study renormalization of the lattice theory we need to explore physics at the same physical volume but smaller lattice spacing. In this case the larger lattices help reduce the lattice spacing through renormalization. As explained earlier, $\varepsilon$ increases according to the values given in \cref{tab:lat2pparams}. On the other hand, in order to explore the same low energy subspace, we will need to hold $\beta$ fixed in physical units, which means $\beta \varepsilon$ will need to grow. While we postpone a systematic study of this situation to a later publication, here we explain some of the challenges we will encounter in such a study. To understand how our algorithm performs when $\beta \varepsilon$ increases, in \cref{tab:llbhel} we tabulate ${\cal M}_{11}$, defined in \cref{sec6} while discussing the helium channel, as a function of $\beta \varepsilon$. For these calculations we fix $L=8$, $\Czerosing=-8.742$, $\Czerotrip=-9.085$. While $\Ctb$ must also change, for the present study we still choose $\Ctb=5.109$.
We notice that the errors in our calculation with the same number of Monte Carlo sweeps increases by about 60-fold when we go from $\beta = 0.8/\varepsilon$ to $\beta = 2.0/\varepsilon$. Let us understand this increase.

There are two reasons why statistical errors can increase as we increase $\beta \varepsilon$. Note that the matrix element is computed using \cref{eq:obs3}, which involves the ratio of two Monte Carlo observables: the numerator given by $\langle O(\{\bar{\bq}\},\{\bq\};[\ell])\rangle$ and the denominator by $\langle \delta_{[\ell],[0]}\rangle$. The fluctuations in the numerator $\langle O(\{\bar{\bq}\},\{\bq\};[\ell])\rangle$ can come from two sources. First we need to generate a worldline with the correct number of nucleons and spins in the initial and the final states. Let us refer to these as the number or N-fluctuations. In addition to these fluctuations, the numerator, as defined in \cref{eq:obsN}, contains the fermion permutation signs along with determinantal factors that can fluctuate. Let us refer to these as S-fluctuations since it is related to the sign problem. In contrast, the denominator $\langle \delta_{[\ell],[0]}\rangle$ only contains N-flucutaions and is free of S-fluctuations.

In our algorithm we currently sample all quantum number sectors up to a maximum particle number without special reweighting techniques. This means, the algorithm can get stuck in nucleon number sectors that contain deep bound states. This increases the N-fluctuations in both the numerator and the denominator as $\beta \varepsilon$ increases. As an illustration of this problem, in \cref{fig:llbz} we plot the Monte Carlo fluctuations of the numerator and the denominator in the calculation of ${\cal M}_{11}$ for $L=8$ in the helium channel. We notice large fluctuations and autocorrelation times in both the numerator and the denominator at $\beta=2.0/\varepsilon$ as compared to $\beta=0.8/\varepsilon$. While there should in principle be S-fluctuations hidden in the numerator at both values of $\beta$, they seem to be almost absent suggesting that the sign problem is mild. These results suggest that the main challenge for going to larger values of $\beta\varepsilon$ is to tame the N-fluctuations. We are currently investigating a refinement of our algorithm, that does not require us to sample all nucleon number sectors to compute the low energy spectrum.

Clearly the N-fluctuations will also increase when the number of particles increase. Our current algorithm samples all particle number sectors and waits for these specific sectors to emerge during the sampling process. When the particle numbers are small this is not a problem, but as the particle numbers increase it will become exponentially difficult to efficiently sample both the numerator and the denominator of \cref{eq:obsN}. In this case one can build particle numbers in steps. Assume $Z_N$ is the partition function in the $N$ particle sector. Although $Z_N/Z_0$ scales exponentially, we can compute it as a product using the relation
\begin{align}
\frac{Z_N}{Z_0}\ =\  \frac{Z_1}{Z_0}\frac{Z_2}{Z_1}...\frac{Z_N}{Z_{N-1}}.
\end{align}
This approach has been used in the past to compute exponentially small quantities \cite{deForcrand:2004jt}. As we approach larger values $\beta\varepsilon$ we also expect S-fluctuations to increase when the particle numbers are large. How far can we push our approach would ultimately depend on the physics of the problem.

\section{Conclusions}
\label{sec8}

In this paper we have introduced a new fermionic Monte Carlo method for quantum mechanical systems which treats fermions as hard core bosons and absorbs sign problems into observables. This approach does not solve the sign problem but can be used for computing the transfer matrix elements $\exp(-\beta H)$ at values of $\beta$ that help us extract the lowest bound states with a small number of particles. The method is expected to encounter problems when $\beta$ or the number of particles become large.

One of the applications of our method we have explored in this work is to pionless nuclear effective field theory. We explored a simple Hamiltonian lattice regularization of the leading order theory and discussed a scheme to renormalize the lattice parameters as we change the lattice spacing. Taking a hypothetical nuclear system we computed the lattice parameters in the one and two-body sectors for all lattice spacings. The three body coupling was only determined for the smallest lattice using exact diagonalization methods. Using these lattice parameters we showed that we can recover exact lattice results using our algorithm on the smallest lattice size with up to four-particles. We also showed that the method scales well on large system sizes as long as $\beta$ is not very large.

Sign problems in Fermi systems are known to be notoriously difficult even on small system sizes. The fact that we could extract the matrix elements accurately, we believe indicates the potential of our method. We believe we can make refinements to our approach to study renormalizability of nuclear effective field theories, which as far as we know is an open problem when more particles are introduced. For example, renormalizabilty in the four-body sector is currently an exciting area of research where we believe our method could help \cite{Platter:2004he,Bazak:2018qnu,Lin:2023zqw}.

Another feature of our Monte Carlo method, which we did not discuss much in this work, is that it treats all types of interactions on an equal footing. In particular the sign problem affects both attractive and repulsive interactions equally. We believe it may be worthwhile to explore models with severe sign problems to see how our algorithm performs in these more difficult situations.

\section*{Acknowledgments}

We thank R.~P.~Springer, X.~Lin, and Sonia Bacca for helpful discussions. This work is supported in part by the U.S. Department of Energy, Office of Science, Nuclear Physics program under Award No. DE-FG02-05ER41368.
This work was supported in part by the Deutsche Forschungsgemeinschaft (DFG) through the Cluster of Excellence ``Precision Physics, Fundamental Interactions, and Structure of Matter'' (PRISMA${}^+$ EXC 2118/1) funded by the DFG within the German Excellence Strategy (Project ID 390831469).

\bibliography{qmc,eft-renorm,ref}

\begin{thebibliography}{100}%
\makeatletter
\providecommand \@ifxundefined [1]{%
 \@ifx{#1\undefined}
}%
\providecommand \@ifnum [1]{%
 \ifnum #1\expandafter \@firstoftwo
 \else \expandafter \@secondoftwo
 \fi
}%
\providecommand \@ifx [1]{%
 \ifx #1\expandafter \@firstoftwo
 \else \expandafter \@secondoftwo
 \fi
}%
\providecommand \natexlab [1]{#1}%
\providecommand \enquote  [1]{``#1''}%
\providecommand \bibnamefont  [1]{#1}%
\providecommand \bibfnamefont [1]{#1}%
\providecommand \citenamefont [1]{#1}%
\providecommand \href@noop [0]{\@secondoftwo}%
\providecommand \href [0]{\begingroup \@sanitize@url \@href}%
\providecommand \@href[1]{\@@startlink{#1}\@@href}%
\providecommand \@@href[1]{\endgroup#1\@@endlink}%
\providecommand \@sanitize@url [0]{\catcode `\\12\catcode `\$12\catcode `\&12\catcode `\#12\catcode `\^12\catcode `\_12\catcode `\%12\relax}%
\providecommand \@@startlink[1]{}%
\providecommand \@@endlink[0]{}%
\providecommand \url  [0]{\begingroup\@sanitize@url \@url }%
\providecommand \@url [1]{\endgroup\@href {#1}{\urlprefix }}%
\providecommand \urlprefix  [0]{URL }%
\providecommand \Eprint [0]{\href }%
\providecommand \doibase [0]{https://doi.org/}%
\providecommand \selectlanguage [0]{\@gobble}%
\providecommand \bibinfo  [0]{\@secondoftwo}%
\providecommand \bibfield  [0]{\@secondoftwo}%
\providecommand \translation [1]{[#1]}%
\providecommand \BibitemOpen [0]{}%
\providecommand \bibitemStop [0]{}%
\providecommand \bibitemNoStop [0]{.\EOS\space}%
\providecommand \EOS [0]{\spacefactor3000\relax}%
\providecommand \BibitemShut  [1]{\csname bibitem#1\endcsname}%
\let\auto@bib@innerbib\@empty
\bibitem [{\citenamefont {Ceperley}\ \emph {et~al.}(1977)\citenamefont {Ceperley}, \citenamefont {Chester},\ and\ \citenamefont {Kalos}}]{PhysRevB.16.3081}%
  \BibitemOpen
  \bibfield  {author} {\bibinfo {author} {\bibfnamefont {D.}~\bibnamefont {Ceperley}}, \bibinfo {author} {\bibfnamefont {G.~V.}\ \bibnamefont {Chester}},\ and\ \bibinfo {author} {\bibfnamefont {M.~H.}\ \bibnamefont {Kalos}},\ }\bibfield  {title} {\bibinfo {title} {Monte carlo simulation of a many-fermion study},\ }\href {https://doi.org/10.1103/PhysRevB.16.3081} {\bibfield  {journal} {\bibinfo  {journal} {Phys. Rev. B}\ }\textbf {\bibinfo {volume} {16}},\ \bibinfo {pages} {3081} (\bibinfo {year} {1977})}\BibitemShut {NoStop}%
\bibitem [{\citenamefont {Troyer}\ and\ \citenamefont {Wiese}(2005)}]{PhysRevLett.94.170201}%
  \BibitemOpen
  \bibfield  {author} {\bibinfo {author} {\bibfnamefont {M.}~\bibnamefont {Troyer}}\ and\ \bibinfo {author} {\bibfnamefont {U.-J.}\ \bibnamefont {Wiese}},\ }\bibfield  {title} {\bibinfo {title} {Computational complexity and fundamental limitations to fermionic quantum monte carlo simulations},\ }\href {https://doi.org/10.1103/PhysRevLett.94.170201} {\bibfield  {journal} {\bibinfo  {journal} {Phys. Rev. Lett.}\ }\textbf {\bibinfo {volume} {94}},\ \bibinfo {pages} {170201} (\bibinfo {year} {2005})}\BibitemShut {NoStop}%
\bibitem [{\citenamefont {Wiese}(1993)}]{Wiese:1992np}%
  \BibitemOpen
  \bibfield  {author} {\bibinfo {author} {\bibfnamefont {U.~J.}\ \bibnamefont {Wiese}},\ }\bibfield  {title} {\bibinfo {title} {{Bosonization and cluster updating of lattice fermions}},\ }\href {https://doi.org/10.1016/0370-2693(93)90561-U} {\bibfield  {journal} {\bibinfo  {journal} {Phys. Lett. B}\ }\textbf {\bibinfo {volume} {311}},\ \bibinfo {pages} {235} (\bibinfo {year} {1993})},\ \Eprint {https://arxiv.org/abs/hep-lat/9210019} {arXiv:hep-lat/9210019} \BibitemShut {NoStop}%
\bibitem [{\citenamefont {Singh}\ and\ \citenamefont {Chandrasekharan}(2019)}]{PhysRevD.99.074511}%
  \BibitemOpen
  \bibfield  {author} {\bibinfo {author} {\bibfnamefont {H.}~\bibnamefont {Singh}}\ and\ \bibinfo {author} {\bibfnamefont {S.}~\bibnamefont {Chandrasekharan}},\ }\bibfield  {title} {\bibinfo {title} {Few-body physics on a spacetime lattice in the worldline approach},\ }\href {https://doi.org/10.1103/PhysRevD.99.074511} {\bibfield  {journal} {\bibinfo  {journal} {Phys. Rev. D}\ }\textbf {\bibinfo {volume} {99}},\ \bibinfo {pages} {074511} (\bibinfo {year} {2019})}\BibitemShut {NoStop}%
\bibitem [{\citenamefont {Zhang}\ \emph {et~al.}(1997)\citenamefont {Zhang}, \citenamefont {Carlson},\ and\ \citenamefont {Gubernatis}}]{Zhang:1996us}%
  \BibitemOpen
  \bibfield  {author} {\bibinfo {author} {\bibfnamefont {S.}~\bibnamefont {Zhang}}, \bibinfo {author} {\bibfnamefont {J.}~\bibnamefont {Carlson}},\ and\ \bibinfo {author} {\bibfnamefont {J.~E.}\ \bibnamefont {Gubernatis}},\ }\bibfield  {title} {\bibinfo {title} {{A Constrained path Monte Carlo method for fermion ground states}},\ }\href {https://doi.org/10.1103/PhysRevB.55.7464} {\bibfield  {journal} {\bibinfo  {journal} {Phys. Rev. B}\ }\textbf {\bibinfo {volume} {55}},\ \bibinfo {pages} {7464} (\bibinfo {year} {1997})},\ \Eprint {https://arxiv.org/abs/cond-mat/9607062} {arXiv:cond-mat/9607062} \BibitemShut {NoStop}%
\bibitem [{\citenamefont {Zhang}(2013)}]{zhang2013}%
  \BibitemOpen
  \bibfield  {author} {\bibinfo {author} {\bibfnamefont {S.}~\bibnamefont {Zhang}},\ }\bibfield  {title} {\bibinfo {title} {Auxiliary-field quantum monte carlo for correlated electron systems},\ }in\ \href@noop {} {\emph {\bibinfo {booktitle} {Emergent Phenomena in Correlated Matter}}},\ \bibinfo {editor} {edited by\ \bibinfo {editor} {\bibfnamefont {E.}~\bibnamefont {Pavarini}}, \bibinfo {editor} {\bibfnamefont {E.}~\bibnamefont {Koch}},\ and\ \bibinfo {editor} {\bibfnamefont {U.}~\bibnamefont {Schollw\"{o}ck}}}\ (\bibinfo  {publisher} {Verlag des Forschungszentrum J\"{u}lich},\ \bibinfo {year} {2013})\BibitemShut {NoStop}%
\bibitem [{\citenamefont {Carlson}\ \emph {et~al.}(2015)\citenamefont {Carlson}, \citenamefont {Gandolfi}, \citenamefont {Pederiva}, \citenamefont {Pieper}, \citenamefont {Schiavilla}, \citenamefont {Schmidt},\ and\ \citenamefont {Wiringa}}]{RevModPhys.87.1067}%
  \BibitemOpen
  \bibfield  {author} {\bibinfo {author} {\bibfnamefont {J.}~\bibnamefont {Carlson}}, \bibinfo {author} {\bibfnamefont {S.}~\bibnamefont {Gandolfi}}, \bibinfo {author} {\bibfnamefont {F.}~\bibnamefont {Pederiva}}, \bibinfo {author} {\bibfnamefont {S.~C.}\ \bibnamefont {Pieper}}, \bibinfo {author} {\bibfnamefont {R.}~\bibnamefont {Schiavilla}}, \bibinfo {author} {\bibfnamefont {K.~E.}\ \bibnamefont {Schmidt}},\ and\ \bibinfo {author} {\bibfnamefont {R.~B.}\ \bibnamefont {Wiringa}},\ }\bibfield  {title} {\bibinfo {title} {Quantum monte carlo methods for nuclear physics},\ }\href {https://doi.org/10.1103/RevModPhys.87.1067} {\bibfield  {journal} {\bibinfo  {journal} {Rev. Mod. Phys.}\ }\textbf {\bibinfo {volume} {87}},\ \bibinfo {pages} {1067} (\bibinfo {year} {2015})}\BibitemShut {NoStop}%
\bibitem [{\citenamefont {Curry}\ \emph {et~al.}(2023)\citenamefont {Curry}, \citenamefont {Dissanayake}, \citenamefont {Gandolfi},\ and\ \citenamefont {Gezerlis}}]{Curry:2023sxh}%
  \BibitemOpen
  \bibfield  {author} {\bibinfo {author} {\bibfnamefont {R.}~\bibnamefont {Curry}}, \bibinfo {author} {\bibfnamefont {J.}~\bibnamefont {Dissanayake}}, \bibinfo {author} {\bibfnamefont {S.}~\bibnamefont {Gandolfi}},\ and\ \bibinfo {author} {\bibfnamefont {A.}~\bibnamefont {Gezerlis}},\ }\bibfield  {title} {\bibinfo {title} {{Auxiliary Field Quantum Monte Carlo for Nuclear Physics on the Lattice}},\ }\href@noop {} {\bibfield  {journal} {\bibinfo  {journal} {arXiv:2310.01504}\ } (\bibinfo {year} {2023})}\BibitemShut {NoStop}%
\bibitem [{\citenamefont {Chandrasekharan}\ \emph {et~al.}(2003)\citenamefont {Chandrasekharan}, \citenamefont {Pepe}, \citenamefont {Steffen},\ and\ \citenamefont {Wiese}}]{Chandrasekharan:2003wy}%
  \BibitemOpen
  \bibfield  {author} {\bibinfo {author} {\bibfnamefont {S.}~\bibnamefont {Chandrasekharan}}, \bibinfo {author} {\bibfnamefont {M.}~\bibnamefont {Pepe}}, \bibinfo {author} {\bibfnamefont {F.~D.}\ \bibnamefont {Steffen}},\ and\ \bibinfo {author} {\bibfnamefont {U.~J.}\ \bibnamefont {Wiese}},\ }\bibfield  {title} {\bibinfo {title} {{Nonlinear realization of chiral symmetry on the lattice}},\ }\href {https://doi.org/10.1088/1126-6708/2003/12/035} {\bibfield  {journal} {\bibinfo  {journal} {JHEP}\ }\textbf {\bibinfo {volume} {12}},\ \bibinfo {pages} {035}},\ \Eprint {https://arxiv.org/abs/hep-lat/0306020} {arXiv:hep-lat/0306020} \BibitemShut {NoStop}%
\bibitem [{\citenamefont {Blankenbecler}\ \emph {et~al.}(1981)\citenamefont {Blankenbecler}, \citenamefont {Scalapino},\ and\ \citenamefont {Sugar}}]{PhysRevD.24.2278}%
  \BibitemOpen
  \bibfield  {author} {\bibinfo {author} {\bibfnamefont {R.}~\bibnamefont {Blankenbecler}}, \bibinfo {author} {\bibfnamefont {D.~J.}\ \bibnamefont {Scalapino}},\ and\ \bibinfo {author} {\bibfnamefont {R.~L.}\ \bibnamefont {Sugar}},\ }\bibfield  {title} {\bibinfo {title} {Monte carlo calculations of coupled boson-fermion systems. i},\ }\href {https://doi.org/10.1103/PhysRevD.24.2278} {\bibfield  {journal} {\bibinfo  {journal} {Phys. Rev. D}\ }\textbf {\bibinfo {volume} {24}},\ \bibinfo {pages} {2278} (\bibinfo {year} {1981})}\BibitemShut {NoStop}%
\bibitem [{\citenamefont {Scalettar}\ \emph {et~al.}(1986)\citenamefont {Scalettar}, \citenamefont {Scalapino},\ and\ \citenamefont {Sugar}}]{PhysRevB.34.7911}%
  \BibitemOpen
  \bibfield  {author} {\bibinfo {author} {\bibfnamefont {R.~T.}\ \bibnamefont {Scalettar}}, \bibinfo {author} {\bibfnamefont {D.~J.}\ \bibnamefont {Scalapino}},\ and\ \bibinfo {author} {\bibfnamefont {R.~L.}\ \bibnamefont {Sugar}},\ }\bibfield  {title} {\bibinfo {title} {New algorithm for the numerical simulation of fermions},\ }\href@noop {} {\bibfield  {journal} {\bibinfo  {journal} {Phys. Rev. B}\ }\textbf {\bibinfo {volume} {34}},\ \bibinfo {pages} {7911} (\bibinfo {year} {1986})}\BibitemShut {NoStop}%
\bibitem [{\citenamefont {Scalettar}\ \emph {et~al.}(1987)\citenamefont {Scalettar}, \citenamefont {Scalapino}, \citenamefont {Sugar},\ and\ \citenamefont {Toussaint}}]{PhysRevB.36.8632}%
  \BibitemOpen
  \bibfield  {author} {\bibinfo {author} {\bibfnamefont {R.~T.}\ \bibnamefont {Scalettar}}, \bibinfo {author} {\bibfnamefont {D.~J.}\ \bibnamefont {Scalapino}}, \bibinfo {author} {\bibfnamefont {R.~L.}\ \bibnamefont {Sugar}},\ and\ \bibinfo {author} {\bibfnamefont {D.}~\bibnamefont {Toussaint}},\ }\bibfield  {title} {\bibinfo {title} {Hybrid molecular-dynamics algorithm for the numerical simulation of many-electron systems},\ }\href {https://doi.org/10.1103/PhysRevB.36.8632} {\bibfield  {journal} {\bibinfo  {journal} {Phys. Rev. B}\ }\textbf {\bibinfo {volume} {36}},\ \bibinfo {pages} {8632} (\bibinfo {year} {1987})}\BibitemShut {NoStop}%
\bibitem [{\citenamefont {Duane}\ \emph {et~al.}(1987)\citenamefont {Duane}, \citenamefont {Kennedy}, \citenamefont {Pendleton},\ and\ \citenamefont {Roweth}}]{Duane1987216}%
  \BibitemOpen
  \bibfield  {author} {\bibinfo {author} {\bibfnamefont {S.}~\bibnamefont {Duane}}, \bibinfo {author} {\bibfnamefont {A.}~\bibnamefont {Kennedy}}, \bibinfo {author} {\bibfnamefont {B.~J.}\ \bibnamefont {Pendleton}},\ and\ \bibinfo {author} {\bibfnamefont {D.}~\bibnamefont {Roweth}},\ }\bibfield  {title} {\bibinfo {title} {Hybrid monte carlo},\ }\href {https://doi.org/https://doi.org/10.1016/0370-2693(87)91197-X} {\bibfield  {journal} {\bibinfo  {journal} {Physics Letters B}\ }\textbf {\bibinfo {volume} {195}},\ \bibinfo {pages} {216} (\bibinfo {year} {1987})}\BibitemShut {NoStop}%
\bibitem [{\citenamefont {Sorella}\ \emph {et~al.}(1989)\citenamefont {Sorella}, \citenamefont {Baroni}, \citenamefont {Car},\ and\ \citenamefont {Parrinello}}]{Sorella1989}%
  \BibitemOpen
  \bibfield  {author} {\bibinfo {author} {\bibfnamefont {S.}~\bibnamefont {Sorella}}, \bibinfo {author} {\bibfnamefont {S.}~\bibnamefont {Baroni}}, \bibinfo {author} {\bibfnamefont {R.}~\bibnamefont {Car}},\ and\ \bibinfo {author} {\bibfnamefont {M.}~\bibnamefont {Parrinello}},\ }\bibfield  {title} {\bibinfo {title} {A novel technique for the simulation of interacting fermion systems},\ }\href {https://doi.org/10.1209/0295-5075/8/7/014} {\bibfield  {journal} {\bibinfo  {journal} {Europhysics Letters}\ }\textbf {\bibinfo {volume} {8}},\ \bibinfo {pages} {663} (\bibinfo {year} {1989})}\BibitemShut {NoStop}%
\bibitem [{\citenamefont {Kennedy}(2005)}]{Kennedy:2004ae}%
  \BibitemOpen
  \bibfield  {author} {\bibinfo {author} {\bibfnamefont {A.~D.}\ \bibnamefont {Kennedy}},\ }\bibfield  {title} {\bibinfo {title} {{Algorithms for lattice QCD with dynamical fermions}},\ }\href {https://doi.org/10.1016/j.nuclphysbps.2004.11.243} {\bibfield  {journal} {\bibinfo  {journal} {Nucl. Phys. B Proc. Suppl.}\ }\textbf {\bibinfo {volume} {140}},\ \bibinfo {pages} {190} (\bibinfo {year} {2005})},\ \Eprint {https://arxiv.org/abs/hep-lat/0409167} {arXiv:hep-lat/0409167} \BibitemShut {NoStop}%
\bibitem [{\citenamefont {Assaad}\ and\ \citenamefont {Evertz}(2008)}]{Assaad2008}%
  \BibitemOpen
  \bibfield  {author} {\bibinfo {author} {\bibfnamefont {F.}~\bibnamefont {Assaad}}\ and\ \bibinfo {author} {\bibfnamefont {H.}~\bibnamefont {Evertz}},\ }\bibinfo {title} {World-line and determinantal quantum monte carlo methods for spins, phonons and electrons},\ in\ \href {https://doi.org/10.1007/978-3-540-74686-7_10} {\emph {\bibinfo {booktitle} {Computational Many-Particle Physics}}},\ \bibinfo {editor} {edited by\ \bibinfo {editor} {\bibfnamefont {H.}~\bibnamefont {Fehske}}, \bibinfo {editor} {\bibfnamefont {R.}~\bibnamefont {Schneider}},\ and\ \bibinfo {editor} {\bibfnamefont {A.}~\bibnamefont {Wei{\ss}e}}}\ (\bibinfo  {publisher} {Springer Berlin Heidelberg},\ \bibinfo {address} {Berlin, Heidelberg},\ \bibinfo {year} {2008})\ pp.\ \bibinfo {pages} {277--356}\BibitemShut {NoStop}%
\bibitem [{\citenamefont {Lee}(2009)}]{Lee:2008fa}%
  \BibitemOpen
  \bibfield  {author} {\bibinfo {author} {\bibfnamefont {D.}~\bibnamefont {Lee}},\ }\bibfield  {title} {\bibinfo {title} {{Lattice simulations for few- and many-body systems}},\ }\href {https://doi.org/10.1016/j.ppnp.2008.12.001} {\bibfield  {journal} {\bibinfo  {journal} {Prog. Part. Nucl. Phys.}\ }\textbf {\bibinfo {volume} {63}},\ \bibinfo {pages} {117} (\bibinfo {year} {2009})},\ \Eprint {https://arxiv.org/abs/0804.3501} {arXiv:0804.3501 [nucl-th]} \BibitemShut {NoStop}%
\bibitem [{\citenamefont {Drut}\ and\ \citenamefont {Nicholson}(2013)}]{Drut_2013}%
  \BibitemOpen
  \bibfield  {author} {\bibinfo {author} {\bibfnamefont {J.~E.}\ \bibnamefont {Drut}}\ and\ \bibinfo {author} {\bibfnamefont {A.~N.}\ \bibnamefont {Nicholson}},\ }\bibfield  {title} {\bibinfo {title} {Lattice methods for strongly interacting many-body systems},\ }\href {https://doi.org/10.1088/0954-3899/40/4/043101} {\bibfield  {journal} {\bibinfo  {journal} {Journal of Physics G: Nuclear and Particle Physics}\ }\textbf {\bibinfo {volume} {40}},\ \bibinfo {pages} {043101} (\bibinfo {year} {2013})}\BibitemShut {NoStop}%
\bibitem [{\citenamefont {Huffman}\ and\ \citenamefont {Chandrasekharan}(2014)}]{PhysRevB.89.111101}%
  \BibitemOpen
  \bibfield  {author} {\bibinfo {author} {\bibfnamefont {E.~F.}\ \bibnamefont {Huffman}}\ and\ \bibinfo {author} {\bibfnamefont {S.}~\bibnamefont {Chandrasekharan}},\ }\bibfield  {title} {\bibinfo {title} {Solution to sign problems in half-filled spin-polarized electronic systems},\ }\href {https://doi.org/10.1103/PhysRevB.89.111101} {\bibfield  {journal} {\bibinfo  {journal} {Phys. Rev. B}\ }\textbf {\bibinfo {volume} {89}},\ \bibinfo {pages} {111101} (\bibinfo {year} {2014})}\BibitemShut {NoStop}%
\bibitem [{\citenamefont {Li}\ \emph {et~al.}(2015)\citenamefont {Li}, \citenamefont {Jiang},\ and\ \citenamefont {Yao}}]{PhysRevB.91.241117}%
  \BibitemOpen
  \bibfield  {author} {\bibinfo {author} {\bibfnamefont {Z.-X.}\ \bibnamefont {Li}}, \bibinfo {author} {\bibfnamefont {Y.-F.}\ \bibnamefont {Jiang}},\ and\ \bibinfo {author} {\bibfnamefont {H.}~\bibnamefont {Yao}},\ }\bibfield  {title} {\bibinfo {title} {Solving the fermion sign problem in quantum monte carlo simulations by majorana representation},\ }\href {https://doi.org/10.1103/PhysRevB.91.241117} {\bibfield  {journal} {\bibinfo  {journal} {Phys. Rev. B}\ }\textbf {\bibinfo {volume} {91}},\ \bibinfo {pages} {241117} (\bibinfo {year} {2015})}\BibitemShut {NoStop}%
\bibitem [{\citenamefont {Li}\ \emph {et~al.}(2016)\citenamefont {Li}, \citenamefont {Jiang},\ and\ \citenamefont {Yao}}]{PhysRevLett.117.267002}%
  \BibitemOpen
  \bibfield  {author} {\bibinfo {author} {\bibfnamefont {Z.-X.}\ \bibnamefont {Li}}, \bibinfo {author} {\bibfnamefont {Y.-F.}\ \bibnamefont {Jiang}},\ and\ \bibinfo {author} {\bibfnamefont {H.}~\bibnamefont {Yao}},\ }\bibfield  {title} {\bibinfo {title} {Majorana-time-reversal symmetries: A fundamental principle for sign-problem-free quantum monte carlo simulations},\ }\href {https://doi.org/10.1103/PhysRevLett.117.267002} {\bibfield  {journal} {\bibinfo  {journal} {Phys. Rev. Lett.}\ }\textbf {\bibinfo {volume} {117}},\ \bibinfo {pages} {267002} (\bibinfo {year} {2016})}\BibitemShut {NoStop}%
\bibitem [{\citenamefont {Wei}\ \emph {et~al.}(2016)\citenamefont {Wei}, \citenamefont {Wu}, \citenamefont {Li}, \citenamefont {Zhang},\ and\ \citenamefont {Xiang}}]{PhysRevLett.116.250601}%
  \BibitemOpen
  \bibfield  {author} {\bibinfo {author} {\bibfnamefont {Z.~C.}\ \bibnamefont {Wei}}, \bibinfo {author} {\bibfnamefont {C.}~\bibnamefont {Wu}}, \bibinfo {author} {\bibfnamefont {Y.}~\bibnamefont {Li}}, \bibinfo {author} {\bibfnamefont {S.}~\bibnamefont {Zhang}},\ and\ \bibinfo {author} {\bibfnamefont {T.}~\bibnamefont {Xiang}},\ }\bibfield  {title} {\bibinfo {title} {Majorana positivity and the fermion sign problem of quantum monte carlo simulations},\ }\href {https://doi.org/10.1103/PhysRevLett.116.250601} {\bibfield  {journal} {\bibinfo  {journal} {Phys. Rev. Lett.}\ }\textbf {\bibinfo {volume} {116}},\ \bibinfo {pages} {250601} (\bibinfo {year} {2016})}\BibitemShut {NoStop}%
\bibitem [{\citenamefont {Berger}\ \emph {et~al.}(2021)\citenamefont {Berger}, \citenamefont {Rammelm\"uller}, \citenamefont {Loheac}, \citenamefont {Ehmann}, \citenamefont {Braun},\ and\ \citenamefont {Drut}}]{Berger:2019odf}%
  \BibitemOpen
  \bibfield  {author} {\bibinfo {author} {\bibfnamefont {C.~E.}\ \bibnamefont {Berger}}, \bibinfo {author} {\bibfnamefont {L.}~\bibnamefont {Rammelm\"uller}}, \bibinfo {author} {\bibfnamefont {A.~C.}\ \bibnamefont {Loheac}}, \bibinfo {author} {\bibfnamefont {F.}~\bibnamefont {Ehmann}}, \bibinfo {author} {\bibfnamefont {J.}~\bibnamefont {Braun}},\ and\ \bibinfo {author} {\bibfnamefont {J.~E.}\ \bibnamefont {Drut}},\ }\bibfield  {title} {\bibinfo {title} {{Complex Langevin and other approaches to the sign problem in quantum many-body physics}},\ }\href {https://doi.org/10.1016/j.physrep.2020.09.002} {\bibfield  {journal} {\bibinfo  {journal} {Phys. Rept.}\ }\textbf {\bibinfo {volume} {892}},\ \bibinfo {pages} {1} (\bibinfo {year} {2021})},\ \Eprint {https://arxiv.org/abs/1907.10183} {arXiv:1907.10183 [cond-mat.quant-gas]} \BibitemShut {NoStop}%
\bibitem [{\citenamefont {Alexandru}\ \emph {et~al.}(2022)\citenamefont {Alexandru}, \citenamefont {Basar}, \citenamefont {Bedaque},\ and\ \citenamefont {Warrington}}]{Alexandru:2020wrj}%
  \BibitemOpen
  \bibfield  {author} {\bibinfo {author} {\bibfnamefont {A.}~\bibnamefont {Alexandru}}, \bibinfo {author} {\bibfnamefont {G.}~\bibnamefont {Basar}}, \bibinfo {author} {\bibfnamefont {P.~F.}\ \bibnamefont {Bedaque}},\ and\ \bibinfo {author} {\bibfnamefont {N.~C.}\ \bibnamefont {Warrington}},\ }\bibfield  {title} {\bibinfo {title} {{Complex paths around the sign problem}},\ }\href {https://doi.org/10.1103/RevModPhys.94.015006} {\bibfield  {journal} {\bibinfo  {journal} {Rev. Mod. Phys.}\ }\textbf {\bibinfo {volume} {94}},\ \bibinfo {pages} {015006} (\bibinfo {year} {2022})},\ \Eprint {https://arxiv.org/abs/2007.05436} {arXiv:2007.05436 [hep-lat]} \BibitemShut {NoStop}%
\bibitem [{\citenamefont {Assaad}\ \emph {et~al.}(2022)\citenamefont {Assaad}, \citenamefont {Bercx}, \citenamefont {Goth}, \citenamefont {G\"otz}, \citenamefont {Hofmann}, \citenamefont {Huffman}, \citenamefont {Liu}, \citenamefont {Parisen~Toldin}, \citenamefont {Portela},\ and\ \citenamefont {Schwab}}]{ALF:2020tyi}%
  \BibitemOpen
  \bibfield  {author} {\bibinfo {author} {\bibfnamefont {F.~F.}\ \bibnamefont {Assaad}}, \bibinfo {author} {\bibfnamefont {M.}~\bibnamefont {Bercx}}, \bibinfo {author} {\bibfnamefont {F.}~\bibnamefont {Goth}}, \bibinfo {author} {\bibfnamefont {A.}~\bibnamefont {G\"otz}}, \bibinfo {author} {\bibfnamefont {J.~S.}\ \bibnamefont {Hofmann}}, \bibinfo {author} {\bibfnamefont {E.}~\bibnamefont {Huffman}}, \bibinfo {author} {\bibfnamefont {Z.}~\bibnamefont {Liu}}, \bibinfo {author} {\bibfnamefont {F.}~\bibnamefont {Parisen~Toldin}}, \bibinfo {author} {\bibfnamefont {J.~S.~E.}\ \bibnamefont {Portela}},\ and\ \bibinfo {author} {\bibfnamefont {J.}~\bibnamefont {Schwab}} (\bibinfo {collaboration} {ALF}),\ }\bibfield  {title} {\bibinfo {title} {{The ALF (Algorithms for Lattice Fermions) project release 2.4. Documentation for the auxiliary-field quantum Monte Carlo code}},\ }\href {https://doi.org/10.21468/SciPostPhysCodeb.1} {\bibfield  {journal} {\bibinfo  {journal} {SciPost Phys. Codeb.}\ }\textbf {\bibinfo {volume}
  {2022}},\ \bibinfo {pages} {1} (\bibinfo {year} {2022})},\ \Eprint {https://arxiv.org/abs/2012.11914} {arXiv:2012.11914 [cond-mat.str-el]} \BibitemShut {NoStop}%
\bibitem [{\citenamefont {Ceperley}(1995)}]{RevModPhys.67.279}%
  \BibitemOpen
  \bibfield  {author} {\bibinfo {author} {\bibfnamefont {D.~M.}\ \bibnamefont {Ceperley}},\ }\bibfield  {title} {\bibinfo {title} {Path integrals in the theory of condensed helium},\ }\href {https://doi.org/10.1103/RevModPhys.67.279} {\bibfield  {journal} {\bibinfo  {journal} {Rev. Mod. Phys.}\ }\textbf {\bibinfo {volume} {67}},\ \bibinfo {pages} {279} (\bibinfo {year} {1995})}\BibitemShut {NoStop}%
\bibitem [{\citenamefont {Prokof'ev}\ and\ \citenamefont {Svistunov}(2001)}]{PhysRevLett.87.160601}%
  \BibitemOpen
  \bibfield  {author} {\bibinfo {author} {\bibfnamefont {N.}~\bibnamefont {Prokof'ev}}\ and\ \bibinfo {author} {\bibfnamefont {B.}~\bibnamefont {Svistunov}},\ }\bibfield  {title} {\bibinfo {title} {Worm algorithms for classical statistical models},\ }\href {https://doi.org/10.1103/PhysRevLett.87.160601} {\bibfield  {journal} {\bibinfo  {journal} {Phys. Rev. Lett.}\ }\textbf {\bibinfo {volume} {87}},\ \bibinfo {pages} {160601} (\bibinfo {year} {2001})}\BibitemShut {NoStop}%
\bibitem [{\citenamefont {Sylju\aa{}sen}\ and\ \citenamefont {Sandvik}(2002)}]{PhysRevE.66.046701}%
  \BibitemOpen
  \bibfield  {author} {\bibinfo {author} {\bibfnamefont {O.~F.}\ \bibnamefont {Sylju\aa{}sen}}\ and\ \bibinfo {author} {\bibfnamefont {A.~W.}\ \bibnamefont {Sandvik}},\ }\bibfield  {title} {\bibinfo {title} {Quantum monte carlo with directed loops},\ }\href {https://doi.org/10.1103/PhysRevE.66.046701} {\bibfield  {journal} {\bibinfo  {journal} {Phys. Rev. E}\ }\textbf {\bibinfo {volume} {66}},\ \bibinfo {pages} {046701} (\bibinfo {year} {2002})}\BibitemShut {NoStop}%
\bibitem [{\citenamefont {Adams}\ and\ \citenamefont {Chandrasekharan}(2003)}]{Adams:2003cc}%
  \BibitemOpen
  \bibfield  {author} {\bibinfo {author} {\bibfnamefont {D.~H.}\ \bibnamefont {Adams}}\ and\ \bibinfo {author} {\bibfnamefont {S.}~\bibnamefont {Chandrasekharan}},\ }\bibfield  {title} {\bibinfo {title} {Chiral limit of strongly coupled lattice gauge theories},\ }\href@noop {} {\bibfield  {journal} {\bibinfo  {journal} {Nucl. Phys.}\ }\textbf {\bibinfo {volume} {B662}},\ \bibinfo {pages} {220} (\bibinfo {year} {2003})}\BibitemShut {NoStop}%
\bibitem [{\citenamefont {Frank}\ \emph {et~al.}(2020)\citenamefont {Frank}, \citenamefont {Huffman},\ and\ \citenamefont {Chandrasekharan}}]{Frank:2019jzv}%
  \BibitemOpen
  \bibfield  {author} {\bibinfo {author} {\bibfnamefont {J.}~\bibnamefont {Frank}}, \bibinfo {author} {\bibfnamefont {E.}~\bibnamefont {Huffman}},\ and\ \bibinfo {author} {\bibfnamefont {S.}~\bibnamefont {Chandrasekharan}},\ }\bibfield  {title} {\bibinfo {title} {{Emergence of Gauss' law in a $Z_2$ lattice gauge theory in 1 + 1 dimensions}},\ }\href {https://doi.org/10.1016/j.physletb.2020.135484} {\bibfield  {journal} {\bibinfo  {journal} {Phys. Lett. B}\ }\textbf {\bibinfo {volume} {806}},\ \bibinfo {pages} {135484} (\bibinfo {year} {2020})},\ \Eprint {https://arxiv.org/abs/1904.05414} {arXiv:1904.05414 [cond-mat.str-el]} \BibitemShut {NoStop}%
\bibitem [{\citenamefont {Elhatisari}\ \emph {et~al.}(2017)\citenamefont {Elhatisari}, \citenamefont {Epelbaum}, \citenamefont {Krebs}, \citenamefont {L\"ahde}, \citenamefont {Lee}, \citenamefont {Li}, \citenamefont {Lu}, \citenamefont {Mei\ss{}ner},\ and\ \citenamefont {Rupak}}]{PhysRevLett.119.222505}%
  \BibitemOpen
  \bibfield  {author} {\bibinfo {author} {\bibfnamefont {S.}~\bibnamefont {Elhatisari}}, \bibinfo {author} {\bibfnamefont {E.}~\bibnamefont {Epelbaum}}, \bibinfo {author} {\bibfnamefont {H.}~\bibnamefont {Krebs}}, \bibinfo {author} {\bibfnamefont {T.~A.}\ \bibnamefont {L\"ahde}}, \bibinfo {author} {\bibfnamefont {D.}~\bibnamefont {Lee}}, \bibinfo {author} {\bibfnamefont {N.}~\bibnamefont {Li}}, \bibinfo {author} {\bibfnamefont {B.-n.}\ \bibnamefont {Lu}}, \bibinfo {author} {\bibfnamefont {U.-G.}\ \bibnamefont {Mei\ss{}ner}},\ and\ \bibinfo {author} {\bibfnamefont {G.}~\bibnamefont {Rupak}},\ }\bibfield  {title} {\bibinfo {title} {Ab initio calculations of the isotopic dependence of nuclear clustering},\ }\href {https://doi.org/10.1103/PhysRevLett.119.222505} {\bibfield  {journal} {\bibinfo  {journal} {Phys. Rev. Lett.}\ }\textbf {\bibinfo {volume} {119}},\ \bibinfo {pages} {222505} (\bibinfo {year} {2017})}\BibitemShut {NoStop}%
\bibitem [{\citenamefont {Lu}\ \emph {et~al.}(2020{\natexlab{a}})\citenamefont {Lu}, \citenamefont {Li}, \citenamefont {Elhatisari}, \citenamefont {Lee}, \citenamefont {Drut}, \citenamefont {L\"ahde}, \citenamefont {Epelbaum},\ and\ \citenamefont {Mei\ss{}ner}}]{PhysRevLett.125.192502}%
  \BibitemOpen
  \bibfield  {author} {\bibinfo {author} {\bibfnamefont {B.-N.}\ \bibnamefont {Lu}}, \bibinfo {author} {\bibfnamefont {N.}~\bibnamefont {Li}}, \bibinfo {author} {\bibfnamefont {S.}~\bibnamefont {Elhatisari}}, \bibinfo {author} {\bibfnamefont {D.}~\bibnamefont {Lee}}, \bibinfo {author} {\bibfnamefont {J.~E.}\ \bibnamefont {Drut}}, \bibinfo {author} {\bibfnamefont {T.~A.}\ \bibnamefont {L\"ahde}}, \bibinfo {author} {\bibfnamefont {E.}~\bibnamefont {Epelbaum}},\ and\ \bibinfo {author} {\bibfnamefont {U.-G.}\ \bibnamefont {Mei\ss{}ner}},\ }\bibfield  {title} {\bibinfo {title} {Ab initio nuclear thermodynamics},\ }\href {https://doi.org/10.1103/PhysRevLett.125.192502} {\bibfield  {journal} {\bibinfo  {journal} {Phys. Rev. Lett.}\ }\textbf {\bibinfo {volume} {125}},\ \bibinfo {pages} {192502} (\bibinfo {year} {2020}{\natexlab{a}})}\BibitemShut {NoStop}%
\bibitem [{\citenamefont {Chandrasekharan}\ and\ \citenamefont {Wiese}(1999)}]{PhysRevLett.83.3116}%
  \BibitemOpen
  \bibfield  {author} {\bibinfo {author} {\bibfnamefont {S.}~\bibnamefont {Chandrasekharan}}\ and\ \bibinfo {author} {\bibfnamefont {U.-J.}\ \bibnamefont {Wiese}},\ }\bibfield  {title} {\bibinfo {title} {Meron-cluster solution of fermion sign problems},\ }\href {https://doi.org/10.1103/PhysRevLett.83.3116} {\bibfield  {journal} {\bibinfo  {journal} {Phys. Rev. Lett.}\ }\textbf {\bibinfo {volume} {83}},\ \bibinfo {pages} {3116} (\bibinfo {year} {1999})}\BibitemShut {NoStop}%
\bibitem [{\citenamefont {Chandrasekharan}(2010)}]{PhysRevD.82.025007}%
  \BibitemOpen
  \bibfield  {author} {\bibinfo {author} {\bibfnamefont {S.}~\bibnamefont {Chandrasekharan}},\ }\bibfield  {title} {\bibinfo {title} {Fermion bag approach to lattice field theories},\ }\href {https://doi.org/10.1103/PhysRevD.82.025007} {\bibfield  {journal} {\bibinfo  {journal} {Phys. Rev. D}\ }\textbf {\bibinfo {volume} {82}},\ \bibinfo {pages} {025007} (\bibinfo {year} {2010})}\BibitemShut {NoStop}%
\bibitem [{\citenamefont {Huffman}\ and\ \citenamefont {Chandrasekharan}(2020)}]{PhysRevD.101.074501}%
  \BibitemOpen
  \bibfield  {author} {\bibinfo {author} {\bibfnamefont {E.}~\bibnamefont {Huffman}}\ and\ \bibinfo {author} {\bibfnamefont {S.}~\bibnamefont {Chandrasekharan}},\ }\bibfield  {title} {\bibinfo {title} {Fermion-bag inspired hamiltonian lattice field theory for fermionic quantum criticality},\ }\href {https://doi.org/10.1103/PhysRevD.101.074501} {\bibfield  {journal} {\bibinfo  {journal} {Phys. Rev. D}\ }\textbf {\bibinfo {volume} {101}},\ \bibinfo {pages} {074501} (\bibinfo {year} {2020})}\BibitemShut {NoStop}%
\bibitem [{\citenamefont {Lee}(2002)}]{Lee:2002sy}%
  \BibitemOpen
  \bibfield  {author} {\bibinfo {author} {\bibfnamefont {D.}~\bibnamefont {Lee}},\ }\bibfield  {title} {\bibinfo {title} {{Permutation zones and the fermion sign problem}},\ }\href@noop {} {\bibfield  {journal} {\bibinfo  {journal} {arXiv:cond-mat/0202283}\ } (\bibinfo {year} {2002})}\BibitemShut {NoStop}%
\bibitem [{\citenamefont {Kaplan}\ \emph {et~al.}(1998)\citenamefont {Kaplan}, \citenamefont {Savage},\ and\ \citenamefont {Wise}}]{Kaplan:1998tg}%
  \BibitemOpen
  \bibfield  {author} {\bibinfo {author} {\bibfnamefont {D.~B.}\ \bibnamefont {Kaplan}}, \bibinfo {author} {\bibfnamefont {M.~J.}\ \bibnamefont {Savage}},\ and\ \bibinfo {author} {\bibfnamefont {M.~B.}\ \bibnamefont {Wise}},\ }\bibfield  {title} {\bibinfo {title} {{A New expansion for nucleon-nucleon interactions}},\ }\href {https://doi.org/10.1016/S0370-2693(98)00210-X} {\bibfield  {journal} {\bibinfo  {journal} {Phys. Lett. B}\ }\textbf {\bibinfo {volume} {424}},\ \bibinfo {pages} {390} (\bibinfo {year} {1998})},\ \Eprint {https://arxiv.org/abs/nucl-th/9801034} {arXiv:nucl-th/9801034} \BibitemShut {NoStop}%
\bibitem [{\citenamefont {Kaplan}\ \emph {et~al.}(1996)\citenamefont {Kaplan}, \citenamefont {Savage},\ and\ \citenamefont {Wise}}]{Kaplan:1996xu}%
  \BibitemOpen
  \bibfield  {author} {\bibinfo {author} {\bibfnamefont {D.~B.}\ \bibnamefont {Kaplan}}, \bibinfo {author} {\bibfnamefont {M.~J.}\ \bibnamefont {Savage}},\ and\ \bibinfo {author} {\bibfnamefont {M.~B.}\ \bibnamefont {Wise}},\ }\bibfield  {title} {\bibinfo {title} {{Nucleon - nucleon scattering from effective field theory}},\ }\href {https://doi.org/10.1016/0550-3213(96)00357-4} {\bibfield  {journal} {\bibinfo  {journal} {Nucl. Phys. B}\ }\textbf {\bibinfo {volume} {478}},\ \bibinfo {pages} {629} (\bibinfo {year} {1996})},\ \Eprint {https://arxiv.org/abs/nucl-th/9605002} {arXiv:nucl-th/9605002} \BibitemShut {NoStop}%
\bibitem [{\citenamefont {Bedaque}\ \emph {et~al.}(1999{\natexlab{a}})\citenamefont {Bedaque}, \citenamefont {Hammer},\ and\ \citenamefont {van Kolck}}]{Bedaque:1998kg}%
  \BibitemOpen
  \bibfield  {author} {\bibinfo {author} {\bibfnamefont {P.~F.}\ \bibnamefont {Bedaque}}, \bibinfo {author} {\bibfnamefont {H.~W.}\ \bibnamefont {Hammer}},\ and\ \bibinfo {author} {\bibfnamefont {U.}~\bibnamefont {van Kolck}},\ }\bibfield  {title} {\bibinfo {title} {{Renormalization of the three-body system with short range interactions}},\ }\href {https://doi.org/10.1103/PhysRevLett.82.463} {\bibfield  {journal} {\bibinfo  {journal} {Phys. Rev. Lett.}\ }\textbf {\bibinfo {volume} {82}},\ \bibinfo {pages} {463} (\bibinfo {year} {1999}{\natexlab{a}})},\ \Eprint {https://arxiv.org/abs/nucl-th/9809025} {arXiv:nucl-th/9809025} \BibitemShut {NoStop}%
\bibitem [{\citenamefont {Bedaque}\ \emph {et~al.}(1999{\natexlab{b}})\citenamefont {Bedaque}, \citenamefont {Hammer},\ and\ \citenamefont {van Kolck}}]{Bedaque:1998km}%
  \BibitemOpen
  \bibfield  {author} {\bibinfo {author} {\bibfnamefont {P.~F.}\ \bibnamefont {Bedaque}}, \bibinfo {author} {\bibfnamefont {H.~W.}\ \bibnamefont {Hammer}},\ and\ \bibinfo {author} {\bibfnamefont {U.}~\bibnamefont {van Kolck}},\ }\bibfield  {title} {\bibinfo {title} {{The Three boson system with short range interactions}},\ }\href {https://doi.org/10.1016/S0375-9474(98)00650-2} {\bibfield  {journal} {\bibinfo  {journal} {Nucl. Phys. A}\ }\textbf {\bibinfo {volume} {646}},\ \bibinfo {pages} {444} (\bibinfo {year} {1999}{\natexlab{b}})},\ \Eprint {https://arxiv.org/abs/nucl-th/9811046} {arXiv:nucl-th/9811046} \BibitemShut {NoStop}%
\bibitem [{\citenamefont {Bedaque}\ \emph {et~al.}(2000)\citenamefont {Bedaque}, \citenamefont {Hammer},\ and\ \citenamefont {van Kolck}}]{Bedaque:1999ve}%
  \BibitemOpen
  \bibfield  {author} {\bibinfo {author} {\bibfnamefont {P.~F.}\ \bibnamefont {Bedaque}}, \bibinfo {author} {\bibfnamefont {H.~W.}\ \bibnamefont {Hammer}},\ and\ \bibinfo {author} {\bibfnamefont {U.}~\bibnamefont {van Kolck}},\ }\bibfield  {title} {\bibinfo {title} {{Effective theory of the triton}},\ }\href {https://doi.org/10.1016/S0375-9474(00)00205-0} {\bibfield  {journal} {\bibinfo  {journal} {Nucl. Phys. A}\ }\textbf {\bibinfo {volume} {676}},\ \bibinfo {pages} {357} (\bibinfo {year} {2000})},\ \Eprint {https://arxiv.org/abs/nucl-th/9906032} {arXiv:nucl-th/9906032} \BibitemShut {NoStop}%
\bibitem [{\citenamefont {Beane}\ \emph {et~al.}(2004)\citenamefont {Beane}, \citenamefont {Bedaque}, \citenamefont {Parreno},\ and\ \citenamefont {Savage}}]{Beane:2003da}%
  \BibitemOpen
  \bibfield  {author} {\bibinfo {author} {\bibfnamefont {S.~R.}\ \bibnamefont {Beane}}, \bibinfo {author} {\bibfnamefont {P.~F.}\ \bibnamefont {Bedaque}}, \bibinfo {author} {\bibfnamefont {A.}~\bibnamefont {Parreno}},\ and\ \bibinfo {author} {\bibfnamefont {M.~J.}\ \bibnamefont {Savage}},\ }\bibfield  {title} {\bibinfo {title} {{Two nucleons on a lattice}},\ }\href {https://doi.org/10.1016/j.physletb.2004.02.007} {\bibfield  {journal} {\bibinfo  {journal} {Phys. Lett. B}\ }\textbf {\bibinfo {volume} {585}},\ \bibinfo {pages} {106} (\bibinfo {year} {2004})},\ \Eprint {https://arxiv.org/abs/hep-lat/0312004} {arXiv:hep-lat/0312004} \BibitemShut {NoStop}%
\bibitem [{\citenamefont {Epelbaum}\ \emph {et~al.}(2009{\natexlab{a}})\citenamefont {Epelbaum}, \citenamefont {Hammer},\ and\ \citenamefont {Meissner}}]{Epelbaum:2008ga}%
  \BibitemOpen
  \bibfield  {author} {\bibinfo {author} {\bibfnamefont {E.}~\bibnamefont {Epelbaum}}, \bibinfo {author} {\bibfnamefont {H.-W.}\ \bibnamefont {Hammer}},\ and\ \bibinfo {author} {\bibfnamefont {U.-G.}\ \bibnamefont {Meissner}},\ }\bibfield  {title} {\bibinfo {title} {{Modern Theory of Nuclear Forces}},\ }\href {https://doi.org/10.1103/RevModPhys.81.1773} {\bibfield  {journal} {\bibinfo  {journal} {Rev. Mod. Phys.}\ }\textbf {\bibinfo {volume} {81}},\ \bibinfo {pages} {1773} (\bibinfo {year} {2009}{\natexlab{a}})},\ \Eprint {https://arxiv.org/abs/0811.1338} {arXiv:0811.1338 [nucl-th]} \BibitemShut {NoStop}%
\bibitem [{\citenamefont {Hammer}\ \emph {et~al.}(2020)\citenamefont {Hammer}, \citenamefont {K\"onig},\ and\ \citenamefont {van Kolck}}]{Hammer:2019poc}%
  \BibitemOpen
  \bibfield  {author} {\bibinfo {author} {\bibfnamefont {H.~W.}\ \bibnamefont {Hammer}}, \bibinfo {author} {\bibfnamefont {S.}~\bibnamefont {K\"onig}},\ and\ \bibinfo {author} {\bibfnamefont {U.}~\bibnamefont {van Kolck}},\ }\bibfield  {title} {\bibinfo {title} {{Nuclear effective field theory: status and perspectives}},\ }\href {https://doi.org/10.1103/RevModPhys.92.025004} {\bibfield  {journal} {\bibinfo  {journal} {Rev. Mod. Phys.}\ }\textbf {\bibinfo {volume} {92}},\ \bibinfo {pages} {025004} (\bibinfo {year} {2020})},\ \Eprint {https://arxiv.org/abs/1906.12122} {arXiv:1906.12122 [nucl-th]} \BibitemShut {NoStop}%
\bibitem [{\citenamefont {Epelbaum}\ \emph {et~al.}(2020{\natexlab{a}})\citenamefont {Epelbaum}, \citenamefont {Krebs},\ and\ \citenamefont {Reinert}}]{Epelbaum:2019kcf}%
  \BibitemOpen
  \bibfield  {author} {\bibinfo {author} {\bibfnamefont {E.}~\bibnamefont {Epelbaum}}, \bibinfo {author} {\bibfnamefont {H.}~\bibnamefont {Krebs}},\ and\ \bibinfo {author} {\bibfnamefont {P.}~\bibnamefont {Reinert}},\ }\bibfield  {title} {\bibinfo {title} {{High-precision nuclear forces from chiral EFT: State-of-the-art, challenges and outlook}},\ }\href {https://doi.org/10.3389/fphy.2020.00098} {\bibfield  {journal} {\bibinfo  {journal} {Front. in Phys.}\ }\textbf {\bibinfo {volume} {8}},\ \bibinfo {pages} {98} (\bibinfo {year} {2020}{\natexlab{a}})},\ \Eprint {https://arxiv.org/abs/1911.11875} {arXiv:1911.11875 [nucl-th]} \BibitemShut {NoStop}%
\bibitem [{\citenamefont {Epelbaum}\ \emph {et~al.}(2017)\citenamefont {Epelbaum}, \citenamefont {Gegelia}, \citenamefont {Mei\ss{}ner},\ and\ \citenamefont {Yao}}]{Epelbaum:2016ffd}%
  \BibitemOpen
  \bibfield  {author} {\bibinfo {author} {\bibfnamefont {E.}~\bibnamefont {Epelbaum}}, \bibinfo {author} {\bibfnamefont {J.}~\bibnamefont {Gegelia}}, \bibinfo {author} {\bibfnamefont {U.-G.}\ \bibnamefont {Mei\ss{}ner}},\ and\ \bibinfo {author} {\bibfnamefont {D.-L.}\ \bibnamefont {Yao}},\ }\bibfield  {title} {\bibinfo {title} {{Renormalization of the three-boson system with short-range interactions revisited}},\ }\href {https://doi.org/10.1140/epja/i2017-12288-3} {\bibfield  {journal} {\bibinfo  {journal} {Eur. Phys. J. A}\ }\textbf {\bibinfo {volume} {53}},\ \bibinfo {pages} {98} (\bibinfo {year} {2017})},\ \Eprint {https://arxiv.org/abs/1611.06040} {arXiv:1611.06040 [nucl-th]} \BibitemShut {NoStop}%
\bibitem [{\citenamefont {Epelbaum}\ \emph {et~al.}(2018)\citenamefont {Epelbaum}, \citenamefont {Gasparyan}, \citenamefont {Gegelia},\ and\ \citenamefont {Mei\ss{}ner}}]{Epelbaum:2018zli}%
  \BibitemOpen
  \bibfield  {author} {\bibinfo {author} {\bibfnamefont {E.}~\bibnamefont {Epelbaum}}, \bibinfo {author} {\bibfnamefont {A.~M.}\ \bibnamefont {Gasparyan}}, \bibinfo {author} {\bibfnamefont {J.}~\bibnamefont {Gegelia}},\ and\ \bibinfo {author} {\bibfnamefont {U.-G.}\ \bibnamefont {Mei\ss{}ner}},\ }\bibfield  {title} {\bibinfo {title} {{How (not) to renormalize integral equations with singular potentials in effective field theory}},\ }\href {https://doi.org/10.1140/epja/i2018-12632-1} {\bibfield  {journal} {\bibinfo  {journal} {Eur. Phys. J. A}\ }\textbf {\bibinfo {volume} {54}},\ \bibinfo {pages} {186} (\bibinfo {year} {2018})},\ \Eprint {https://arxiv.org/abs/1810.02646} {arXiv:1810.02646 [nucl-th]} \BibitemShut {NoStop}%
\bibitem [{\citenamefont {Epelbaum}\ \emph {et~al.}(2020{\natexlab{b}})\citenamefont {Epelbaum}, \citenamefont {Gasparyan}, \citenamefont {Gegelia}, \citenamefont {Mei\ss{}ner},\ and\ \citenamefont {Ren}}]{Epelbaum:2020maf}%
  \BibitemOpen
  \bibfield  {author} {\bibinfo {author} {\bibfnamefont {E.}~\bibnamefont {Epelbaum}}, \bibinfo {author} {\bibfnamefont {A.~M.}\ \bibnamefont {Gasparyan}}, \bibinfo {author} {\bibfnamefont {J.}~\bibnamefont {Gegelia}}, \bibinfo {author} {\bibfnamefont {U.-G.}\ \bibnamefont {Mei\ss{}ner}},\ and\ \bibinfo {author} {\bibfnamefont {X.~L.}\ \bibnamefont {Ren}},\ }\bibfield  {title} {\bibinfo {title} {{How to renormalize integral equations with singular potentials in effective field theory}},\ }\href {https://doi.org/10.1140/epja/s10050-020-00162-4} {\bibfield  {journal} {\bibinfo  {journal} {Eur. Phys. J. A}\ }\textbf {\bibinfo {volume} {56}},\ \bibinfo {pages} {152} (\bibinfo {year} {2020}{\natexlab{b}})},\ \Eprint {https://arxiv.org/abs/2001.07040} {arXiv:2001.07040 [nucl-th]} \BibitemShut {NoStop}%
\bibitem [{\citenamefont {Gasparyan}\ and\ \citenamefont {Epelbaum}(2023)}]{Gasparyan:2022isg}%
  \BibitemOpen
  \bibfield  {author} {\bibinfo {author} {\bibfnamefont {A.~M.}\ \bibnamefont {Gasparyan}}\ and\ \bibinfo {author} {\bibfnamefont {E.}~\bibnamefont {Epelbaum}},\ }\bibfield  {title} {\bibinfo {title} {{\textquotedblleft{}Renormalization-group-invariant effective field theory\textquotedblright{} for few-nucleon systems is cutoff dependent}},\ }\href {https://doi.org/10.1103/PhysRevC.107.034001} {\bibfield  {journal} {\bibinfo  {journal} {Phys. Rev. C}\ }\textbf {\bibinfo {volume} {107}},\ \bibinfo {pages} {034001} (\bibinfo {year} {2023})},\ \Eprint {https://arxiv.org/abs/2210.16225} {arXiv:2210.16225 [nucl-th]} \BibitemShut {NoStop}%
\bibitem [{\citenamefont {K\"orber}\ \emph {et~al.}(2019)\citenamefont {K\"orber}, \citenamefont {Berkowitz},\ and\ \citenamefont {Luu}}]{Korber:2019cuq}%
  \BibitemOpen
  \bibfield  {author} {\bibinfo {author} {\bibfnamefont {C.}~\bibnamefont {K\"orber}}, \bibinfo {author} {\bibfnamefont {E.}~\bibnamefont {Berkowitz}},\ and\ \bibinfo {author} {\bibfnamefont {T.}~\bibnamefont {Luu}},\ }\bibfield  {title} {\bibinfo {title} {{Renormalization of a Contact Interaction on a Lattice}},\ }\href@noop {} {\bibfield  {journal} {\bibinfo  {journal} {arXiv:1912.04425}\ } (\bibinfo {year} {2019})}\BibitemShut {NoStop}%
\bibitem [{\citenamefont {Lee}\ and\ \citenamefont {Sch\"afer}(2005)}]{Lee:2004qd}%
  \BibitemOpen
  \bibfield  {author} {\bibinfo {author} {\bibfnamefont {D.}~\bibnamefont {Lee}}\ and\ \bibinfo {author} {\bibfnamefont {T.}~\bibnamefont {Sch\"afer}},\ }\bibfield  {title} {\bibinfo {title} {{Neutron matter on the lattice with pionless effective field theory}},\ }\href {https://doi.org/10.1103/PhysRevC.72.024006} {\bibfield  {journal} {\bibinfo  {journal} {Phys. Rev. C}\ }\textbf {\bibinfo {volume} {72}},\ \bibinfo {pages} {024006} (\bibinfo {year} {2005})},\ \Eprint {https://arxiv.org/abs/nucl-th/0412002} {arXiv:nucl-th/0412002} \BibitemShut {NoStop}%
\bibitem [{\citenamefont {Lee}\ \emph {et~al.}(2004)\citenamefont {Lee}, \citenamefont {Borasoy},\ and\ \citenamefont {Sch\"afer}}]{Lee:2004si}%
  \BibitemOpen
  \bibfield  {author} {\bibinfo {author} {\bibfnamefont {D.}~\bibnamefont {Lee}}, \bibinfo {author} {\bibfnamefont {B.}~\bibnamefont {Borasoy}},\ and\ \bibinfo {author} {\bibfnamefont {T.}~\bibnamefont {Sch\"afer}},\ }\bibfield  {title} {\bibinfo {title} {{Nuclear lattice simulations with chiral effective field theory}},\ }\href {https://doi.org/10.1103/PhysRevC.70.014007} {\bibfield  {journal} {\bibinfo  {journal} {Phys. Rev. C}\ }\textbf {\bibinfo {volume} {70}},\ \bibinfo {pages} {014007} (\bibinfo {year} {2004})},\ \Eprint {https://arxiv.org/abs/nucl-th/0402072} {arXiv:nucl-th/0402072} \BibitemShut {NoStop}%
\bibitem [{\citenamefont {Borasoy}\ \emph {et~al.}(2006)\citenamefont {Borasoy}, \citenamefont {Krebs}, \citenamefont {Lee},\ and\ \citenamefont {Meissner}}]{Borasoy:2005yc}%
  \BibitemOpen
  \bibfield  {author} {\bibinfo {author} {\bibfnamefont {B.}~\bibnamefont {Borasoy}}, \bibinfo {author} {\bibfnamefont {H.}~\bibnamefont {Krebs}}, \bibinfo {author} {\bibfnamefont {D.}~\bibnamefont {Lee}},\ and\ \bibinfo {author} {\bibfnamefont {U.~G.}\ \bibnamefont {Meissner}},\ }\bibfield  {title} {\bibinfo {title} {{The Triton and three-nucleon force in nuclear lattice simulations}},\ }\href {https://doi.org/10.1016/j.nuclphysa.2006.01.009} {\bibfield  {journal} {\bibinfo  {journal} {Nucl. Phys. A}\ }\textbf {\bibinfo {volume} {768}},\ \bibinfo {pages} {179} (\bibinfo {year} {2006})},\ \Eprint {https://arxiv.org/abs/nucl-th/0510047} {arXiv:nucl-th/0510047} \BibitemShut {NoStop}%
\bibitem [{\citenamefont {Borasoy}\ \emph {et~al.}(2007{\natexlab{a}})\citenamefont {Borasoy}, \citenamefont {Epelbaum}, \citenamefont {Krebs}, \citenamefont {Lee},\ and\ \citenamefont {Meissner}}]{Borasoy:2006qn}%
  \BibitemOpen
  \bibfield  {author} {\bibinfo {author} {\bibfnamefont {B.}~\bibnamefont {Borasoy}}, \bibinfo {author} {\bibfnamefont {E.}~\bibnamefont {Epelbaum}}, \bibinfo {author} {\bibfnamefont {H.}~\bibnamefont {Krebs}}, \bibinfo {author} {\bibfnamefont {D.}~\bibnamefont {Lee}},\ and\ \bibinfo {author} {\bibfnamefont {U.-G.}\ \bibnamefont {Meissner}},\ }\bibfield  {title} {\bibinfo {title} {{Lattice Simulations for Light Nuclei: Chiral Effective Field Theory at Leading Order}},\ }\href {https://doi.org/10.1140/epja/i2006-10154-1} {\bibfield  {journal} {\bibinfo  {journal} {Eur. Phys. J. A}\ }\textbf {\bibinfo {volume} {31}},\ \bibinfo {pages} {105} (\bibinfo {year} {2007}{\natexlab{a}})},\ \Eprint {https://arxiv.org/abs/nucl-th/0611087} {arXiv:nucl-th/0611087} \BibitemShut {NoStop}%
\bibitem [{\citenamefont {Borasoy}\ \emph {et~al.}(2008{\natexlab{a}})\citenamefont {Borasoy}, \citenamefont {Epelbaum}, \citenamefont {Krebs}, \citenamefont {Lee},\ and\ \citenamefont {Meissner}}]{Borasoy:2007vi}%
  \BibitemOpen
  \bibfield  {author} {\bibinfo {author} {\bibfnamefont {B.}~\bibnamefont {Borasoy}}, \bibinfo {author} {\bibfnamefont {E.}~\bibnamefont {Epelbaum}}, \bibinfo {author} {\bibfnamefont {H.}~\bibnamefont {Krebs}}, \bibinfo {author} {\bibfnamefont {D.}~\bibnamefont {Lee}},\ and\ \bibinfo {author} {\bibfnamefont {U.-G.}\ \bibnamefont {Meissner}},\ }\bibfield  {title} {\bibinfo {title} {{Chiral effective field theory on the lattice at next-to-leading order}},\ }\href {https://doi.org/10.1140/epja/i2008-10544-3} {\bibfield  {journal} {\bibinfo  {journal} {Eur. Phys. J. A}\ }\textbf {\bibinfo {volume} {35}},\ \bibinfo {pages} {343} (\bibinfo {year} {2008}{\natexlab{a}})},\ \Eprint {https://arxiv.org/abs/0712.2990} {arXiv:0712.2990 [nucl-th]} \BibitemShut {NoStop}%
\bibitem [{\citenamefont {Borasoy}\ \emph {et~al.}(2008{\natexlab{b}})\citenamefont {Borasoy}, \citenamefont {Epelbaum}, \citenamefont {Krebs}, \citenamefont {Lee},\ and\ \citenamefont {Meissner}}]{Borasoy:2007vk}%
  \BibitemOpen
  \bibfield  {author} {\bibinfo {author} {\bibfnamefont {B.}~\bibnamefont {Borasoy}}, \bibinfo {author} {\bibfnamefont {E.}~\bibnamefont {Epelbaum}}, \bibinfo {author} {\bibfnamefont {H.}~\bibnamefont {Krebs}}, \bibinfo {author} {\bibfnamefont {D.}~\bibnamefont {Lee}},\ and\ \bibinfo {author} {\bibfnamefont {U.-G.}\ \bibnamefont {Meissner}},\ }\bibfield  {title} {\bibinfo {title} {{Dilute neutron matter on the lattice at next-to-leading order in chiral effective field theory}},\ }\href {https://doi.org/10.1140/epja/i2008-10545-2} {\bibfield  {journal} {\bibinfo  {journal} {Eur. Phys. J. A}\ }\textbf {\bibinfo {volume} {35}},\ \bibinfo {pages} {357} (\bibinfo {year} {2008}{\natexlab{b}})},\ \Eprint {https://arxiv.org/abs/0712.2993} {arXiv:0712.2993 [nucl-th]} \BibitemShut {NoStop}%
\bibitem [{\citenamefont {Borasoy}\ \emph {et~al.}(2007{\natexlab{b}})\citenamefont {Borasoy}, \citenamefont {Epelbaum}, \citenamefont {Krebs}, \citenamefont {Lee},\ and\ \citenamefont {Meissner}}]{Borasoy:2007vy}%
  \BibitemOpen
  \bibfield  {author} {\bibinfo {author} {\bibfnamefont {B.}~\bibnamefont {Borasoy}}, \bibinfo {author} {\bibfnamefont {E.}~\bibnamefont {Epelbaum}}, \bibinfo {author} {\bibfnamefont {H.}~\bibnamefont {Krebs}}, \bibinfo {author} {\bibfnamefont {D.}~\bibnamefont {Lee}},\ and\ \bibinfo {author} {\bibfnamefont {U.-G.}\ \bibnamefont {Meissner}},\ }\bibfield  {title} {\bibinfo {title} {{Two-particle scattering on the lattice: Phase shifts, spin-orbit coupling, and mixing angles}},\ }\href {https://doi.org/10.1140/epja/i2007-10500-9} {\bibfield  {journal} {\bibinfo  {journal} {Eur. Phys. J. A}\ }\textbf {\bibinfo {volume} {34}},\ \bibinfo {pages} {185} (\bibinfo {year} {2007}{\natexlab{b}})},\ \Eprint {https://arxiv.org/abs/0708.1780} {arXiv:0708.1780 [nucl-th]} \BibitemShut {NoStop}%
\bibitem [{\citenamefont {Lee}\ and\ \citenamefont {Sch\"afer}(2006{\natexlab{a}})}]{Lee:2005is}%
  \BibitemOpen
  \bibfield  {author} {\bibinfo {author} {\bibfnamefont {D.}~\bibnamefont {Lee}}\ and\ \bibinfo {author} {\bibfnamefont {T.}~\bibnamefont {Sch\"afer}},\ }\bibfield  {title} {\bibinfo {title} {{Cold dilute neutron matter on the lattice. I. Lattice virial coefficients and large scattering lengths}},\ }\href {https://doi.org/10.1103/PhysRevC.73.015201} {\bibfield  {journal} {\bibinfo  {journal} {Phys. Rev. C}\ }\textbf {\bibinfo {volume} {73}},\ \bibinfo {pages} {015201} (\bibinfo {year} {2006}{\natexlab{a}})},\ \Eprint {https://arxiv.org/abs/nucl-th/0509017} {arXiv:nucl-th/0509017} \BibitemShut {NoStop}%
\bibitem [{\citenamefont {Lee}\ and\ \citenamefont {Sch\"afer}(2006{\natexlab{b}})}]{Lee:2005it}%
  \BibitemOpen
  \bibfield  {author} {\bibinfo {author} {\bibfnamefont {D.}~\bibnamefont {Lee}}\ and\ \bibinfo {author} {\bibfnamefont {T.}~\bibnamefont {Sch\"afer}},\ }\bibfield  {title} {\bibinfo {title} {{Cold dilute neutron matter on the lattice. II. Results in the unitary limit}},\ }\href {https://doi.org/10.1103/PhysRevC.73.015202} {\bibfield  {journal} {\bibinfo  {journal} {Phys. Rev. C}\ }\textbf {\bibinfo {volume} {73}},\ \bibinfo {pages} {015202} (\bibinfo {year} {2006}{\natexlab{b}})},\ \Eprint {https://arxiv.org/abs/nucl-th/0509018} {arXiv:nucl-th/0509018} \BibitemShut {NoStop}%
\bibitem [{\citenamefont {Lee}(2008)}]{Lee:2008xsa}%
  \BibitemOpen
  \bibfield  {author} {\bibinfo {author} {\bibfnamefont {D.}~\bibnamefont {Lee}},\ }\bibfield  {title} {\bibinfo {title} {{The Ground state energy at unitarity}},\ }\href {https://doi.org/10.1103/PhysRevC.78.024001} {\bibfield  {journal} {\bibinfo  {journal} {Phys.Rev.}\ }\textbf {\bibinfo {volume} {C78}},\ \bibinfo {pages} {024001} (\bibinfo {year} {2008})},\ \Eprint {https://arxiv.org/abs/0803.1280} {arXiv:0803.1280 [nucl-th]} \BibitemShut {NoStop}%
\bibitem [{\citenamefont {Lee}(2020)}]{Lee:2020meg}%
  \BibitemOpen
  \bibfield  {author} {\bibinfo {author} {\bibfnamefont {D.}~\bibnamefont {Lee}},\ }\bibfield  {title} {\bibinfo {title} {{Recent Progress in Nuclear Lattice Simulations}},\ }\href {https://doi.org/10.3389/fphy.2020.00174} {\bibfield  {journal} {\bibinfo  {journal} {Front. in Phys.}\ }\textbf {\bibinfo {volume} {8}},\ \bibinfo {pages} {174} (\bibinfo {year} {2020})}\BibitemShut {NoStop}%
\bibitem [{\citenamefont {Lee}(2021)}]{Lee:2021wey}%
  \BibitemOpen
  \bibfield  {author} {\bibinfo {author} {\bibfnamefont {D.}~\bibnamefont {Lee}},\ }\bibfield  {title} {\bibinfo {title} {{Chiral Effective Field Theory after Thirty Years: Nuclear Lattice Simulations}},\ }\href {https://doi.org/10.1007/s00601-021-01701-5} {\bibfield  {journal} {\bibinfo  {journal} {Few Body Syst.}\ }\textbf {\bibinfo {volume} {62}},\ \bibinfo {pages} {115} (\bibinfo {year} {2021})},\ \Eprint {https://arxiv.org/abs/2109.09582} {arXiv:2109.09582 [nucl-th]} \BibitemShut {NoStop}%
\bibitem [{\citenamefont {Epelbaum}\ \emph {et~al.}(2010{\natexlab{a}})\citenamefont {Epelbaum}, \citenamefont {Krebs}, \citenamefont {Lee},\ and\ \citenamefont {Meissner}}]{Epelbaum:2009pd}%
  \BibitemOpen
  \bibfield  {author} {\bibinfo {author} {\bibfnamefont {E.}~\bibnamefont {Epelbaum}}, \bibinfo {author} {\bibfnamefont {H.}~\bibnamefont {Krebs}}, \bibinfo {author} {\bibfnamefont {D.}~\bibnamefont {Lee}},\ and\ \bibinfo {author} {\bibfnamefont {U.-G.}\ \bibnamefont {Meissner}},\ }\bibfield  {title} {\bibinfo {title} {{Lattice effective field theory calculations for A = 3,4,6,12 nuclei}},\ }\href {https://doi.org/10.1103/PhysRevLett.104.142501} {\bibfield  {journal} {\bibinfo  {journal} {Phys. Rev. Lett.}\ }\textbf {\bibinfo {volume} {104}},\ \bibinfo {pages} {142501} (\bibinfo {year} {2010}{\natexlab{a}})},\ \Eprint {https://arxiv.org/abs/0912.4195} {arXiv:0912.4195 [nucl-th]} \BibitemShut {NoStop}%
\bibitem [{\citenamefont {Epelbaum}\ \emph {et~al.}(2010{\natexlab{b}})\citenamefont {Epelbaum}, \citenamefont {Krebs}, \citenamefont {Lee},\ and\ \citenamefont {Meissner}}]{Epelbaum:2010xt}%
  \BibitemOpen
  \bibfield  {author} {\bibinfo {author} {\bibfnamefont {E.}~\bibnamefont {Epelbaum}}, \bibinfo {author} {\bibfnamefont {H.}~\bibnamefont {Krebs}}, \bibinfo {author} {\bibfnamefont {D.}~\bibnamefont {Lee}},\ and\ \bibinfo {author} {\bibfnamefont {U.-G.}\ \bibnamefont {Meissner}},\ }\bibfield  {title} {\bibinfo {title} {{Lattice calculations for A=3,4,6,12 nuclei using chiral effective field theory}},\ }\href {https://doi.org/10.1140/epja/i2010-11009-x} {\bibfield  {journal} {\bibinfo  {journal} {Eur. Phys. J. A}\ }\textbf {\bibinfo {volume} {45}},\ \bibinfo {pages} {335} (\bibinfo {year} {2010}{\natexlab{b}})},\ \Eprint {https://arxiv.org/abs/1003.5697} {arXiv:1003.5697 [nucl-th]} \BibitemShut {NoStop}%
\bibitem [{\citenamefont {Epelbaum}\ \emph {et~al.}(2011)\citenamefont {Epelbaum}, \citenamefont {Krebs}, \citenamefont {Lee},\ and\ \citenamefont {Meissner}}]{Epelbaum:2011md}%
  \BibitemOpen
  \bibfield  {author} {\bibinfo {author} {\bibfnamefont {E.}~\bibnamefont {Epelbaum}}, \bibinfo {author} {\bibfnamefont {H.}~\bibnamefont {Krebs}}, \bibinfo {author} {\bibfnamefont {D.}~\bibnamefont {Lee}},\ and\ \bibinfo {author} {\bibfnamefont {U.-G.}\ \bibnamefont {Meissner}},\ }\bibfield  {title} {\bibinfo {title} {{Ab initio calculation of the Hoyle state}},\ }\href {https://doi.org/10.1103/PhysRevLett.106.192501} {\bibfield  {journal} {\bibinfo  {journal} {Phys. Rev. Lett.}\ }\textbf {\bibinfo {volume} {106}},\ \bibinfo {pages} {192501} (\bibinfo {year} {2011})},\ \Eprint {https://arxiv.org/abs/1101.2547} {arXiv:1101.2547 [nucl-th]} \BibitemShut {NoStop}%
\bibitem [{\citenamefont {Epelbaum}\ \emph {et~al.}(2009{\natexlab{b}})\citenamefont {Epelbaum}, \citenamefont {Krebs}, \citenamefont {Lee},\ and\ \citenamefont {Meissner}}]{Epelbaum:2009rkz}%
  \BibitemOpen
  \bibfield  {author} {\bibinfo {author} {\bibfnamefont {E.}~\bibnamefont {Epelbaum}}, \bibinfo {author} {\bibfnamefont {H.}~\bibnamefont {Krebs}}, \bibinfo {author} {\bibfnamefont {D.}~\bibnamefont {Lee}},\ and\ \bibinfo {author} {\bibfnamefont {U.-G.}\ \bibnamefont {Meissner}},\ }\bibfield  {title} {\bibinfo {title} {{Ground state energy of dilute neutron matter at next-to-leading order in lattice chiral effective field theory}},\ }\href {https://doi.org/10.1140/epja/i2009-10755-0} {\bibfield  {journal} {\bibinfo  {journal} {Eur. Phys. J. A}\ }\textbf {\bibinfo {volume} {40}},\ \bibinfo {pages} {199} (\bibinfo {year} {2009}{\natexlab{b}})},\ \Eprint {https://arxiv.org/abs/0812.3653} {arXiv:0812.3653 [nucl-th]} \BibitemShut {NoStop}%
\bibitem [{\citenamefont {L\"ahde}\ \emph {et~al.}(2015)\citenamefont {L\"ahde}, \citenamefont {Luu}, \citenamefont {Lee}, \citenamefont {Mei\ss{}ner}, \citenamefont {Epelbaum}, \citenamefont {Krebs},\ and\ \citenamefont {Rupak}}]{Lahde:2015ona}%
  \BibitemOpen
  \bibfield  {author} {\bibinfo {author} {\bibfnamefont {T.~A.}\ \bibnamefont {L\"ahde}}, \bibinfo {author} {\bibfnamefont {T.}~\bibnamefont {Luu}}, \bibinfo {author} {\bibfnamefont {D.}~\bibnamefont {Lee}}, \bibinfo {author} {\bibfnamefont {U.-G.}\ \bibnamefont {Mei\ss{}ner}}, \bibinfo {author} {\bibfnamefont {E.}~\bibnamefont {Epelbaum}}, \bibinfo {author} {\bibfnamefont {H.}~\bibnamefont {Krebs}},\ and\ \bibinfo {author} {\bibfnamefont {G.}~\bibnamefont {Rupak}},\ }\bibfield  {title} {\bibinfo {title} {{Nuclear Lattice Simulations using Symmetry-Sign Extrapolation}},\ }\href {https://doi.org/10.1140/epja/i2015-15092-1} {\bibfield  {journal} {\bibinfo  {journal} {Eur. Phys. J. A}\ }\textbf {\bibinfo {volume} {51}},\ \bibinfo {pages} {92} (\bibinfo {year} {2015})},\ \Eprint {https://arxiv.org/abs/1502.06787} {arXiv:1502.06787 [nucl-th]} \BibitemShut {NoStop}%
\bibitem [{\citenamefont {Klein}\ \emph {et~al.}(2015)\citenamefont {Klein}, \citenamefont {Lee}, \citenamefont {Liu},\ and\ \citenamefont {Mei\ss{}ner}}]{Klein:2015vna}%
  \BibitemOpen
  \bibfield  {author} {\bibinfo {author} {\bibfnamefont {N.}~\bibnamefont {Klein}}, \bibinfo {author} {\bibfnamefont {D.}~\bibnamefont {Lee}}, \bibinfo {author} {\bibfnamefont {W.}~\bibnamefont {Liu}},\ and\ \bibinfo {author} {\bibfnamefont {U.-G.}\ \bibnamefont {Mei\ss{}ner}},\ }\bibfield  {title} {\bibinfo {title} {{Regularization Methods for Nuclear Lattice Effective Field Theory}},\ }\href {https://doi.org/10.1016/j.physletb.2015.06.040} {\bibfield  {journal} {\bibinfo  {journal} {Phys. Lett. B}\ }\textbf {\bibinfo {volume} {747}},\ \bibinfo {pages} {511} (\bibinfo {year} {2015})},\ \Eprint {https://arxiv.org/abs/1505.07000} {arXiv:1505.07000 [nucl-th]} \BibitemShut {NoStop}%
\bibitem [{\citenamefont {Klein}\ \emph {et~al.}(2018{\natexlab{a}})\citenamefont {Klein}, \citenamefont {Lee},\ and\ \citenamefont {Mei\ss{}ner}}]{Klein:2018iqa}%
  \BibitemOpen
  \bibfield  {author} {\bibinfo {author} {\bibfnamefont {N.}~\bibnamefont {Klein}}, \bibinfo {author} {\bibfnamefont {D.}~\bibnamefont {Lee}},\ and\ \bibinfo {author} {\bibfnamefont {U.-G.}\ \bibnamefont {Mei\ss{}ner}},\ }\bibfield  {title} {\bibinfo {title} {{Lattice Improvement in Lattice Effective Field Theory}},\ }\href {https://doi.org/10.1140/epja/i2018-12676-1} {\bibfield  {journal} {\bibinfo  {journal} {Eur. Phys. J. A}\ }\textbf {\bibinfo {volume} {54}},\ \bibinfo {pages} {233} (\bibinfo {year} {2018}{\natexlab{a}})},\ \Eprint {https://arxiv.org/abs/1807.04234} {arXiv:1807.04234 [hep-lat]} \BibitemShut {NoStop}%
\bibitem [{\citenamefont {Klein}\ \emph {et~al.}(2018{\natexlab{b}})\citenamefont {Klein}, \citenamefont {Elhatisari}, \citenamefont {L\"ahde}, \citenamefont {Lee},\ and\ \citenamefont {Mei\ss{}ner}}]{Klein:2018lqz}%
  \BibitemOpen
  \bibfield  {author} {\bibinfo {author} {\bibfnamefont {N.}~\bibnamefont {Klein}}, \bibinfo {author} {\bibfnamefont {S.}~\bibnamefont {Elhatisari}}, \bibinfo {author} {\bibfnamefont {T.~A.}\ \bibnamefont {L\"ahde}}, \bibinfo {author} {\bibfnamefont {D.}~\bibnamefont {Lee}},\ and\ \bibinfo {author} {\bibfnamefont {U.-G.}\ \bibnamefont {Mei\ss{}ner}},\ }\bibfield  {title} {\bibinfo {title} {{The Tjon Band in Nuclear Lattice Effective Field Theory}},\ }\href {https://doi.org/10.1140/epja/i2018-12553-y} {\bibfield  {journal} {\bibinfo  {journal} {Eur. Phys. J. A}\ }\textbf {\bibinfo {volume} {54}},\ \bibinfo {pages} {121} (\bibinfo {year} {2018}{\natexlab{b}})},\ \Eprint {https://arxiv.org/abs/1803.04231} {arXiv:1803.04231 [nucl-th]} \BibitemShut {NoStop}%
\bibitem [{\citenamefont {Alarc\'on}\ \emph {et~al.}(2017)\citenamefont {Alarc\'on}, \citenamefont {Du}, \citenamefont {Klein}, \citenamefont {L\"ahde}, \citenamefont {Lee}, \citenamefont {Li}, \citenamefont {Lu}, \citenamefont {Luu},\ and\ \citenamefont {Mei\ss{}ner}}]{Alarcon:2017zcv}%
  \BibitemOpen
  \bibfield  {author} {\bibinfo {author} {\bibfnamefont {J.~M.}\ \bibnamefont {Alarc\'on}}, \bibinfo {author} {\bibfnamefont {D.}~\bibnamefont {Du}}, \bibinfo {author} {\bibfnamefont {N.}~\bibnamefont {Klein}}, \bibinfo {author} {\bibfnamefont {T.~A.}\ \bibnamefont {L\"ahde}}, \bibinfo {author} {\bibfnamefont {D.}~\bibnamefont {Lee}}, \bibinfo {author} {\bibfnamefont {N.}~\bibnamefont {Li}}, \bibinfo {author} {\bibfnamefont {B.-N.}\ \bibnamefont {Lu}}, \bibinfo {author} {\bibfnamefont {T.}~\bibnamefont {Luu}},\ and\ \bibinfo {author} {\bibfnamefont {U.-G.}\ \bibnamefont {Mei\ss{}ner}},\ }\bibfield  {title} {\bibinfo {title} {{Neutron-proton scattering at next-to-next-to-leading order in Nuclear Lattice Effective Field Theory}},\ }\href {https://doi.org/10.1140/epja/i2017-12273-x} {\bibfield  {journal} {\bibinfo  {journal} {Eur. Phys. J. A}\ }\textbf {\bibinfo {volume} {53}},\ \bibinfo {pages} {83} (\bibinfo {year} {2017})},\ \Eprint {https://arxiv.org/abs/1702.05319} {arXiv:1702.05319 [nucl-th]} \BibitemShut
  {NoStop}%
\bibitem [{\citenamefont {Li}\ \emph {et~al.}(2018)\citenamefont {Li}, \citenamefont {Elhatisari}, \citenamefont {Epelbaum}, \citenamefont {Lee}, \citenamefont {Lu},\ and\ \citenamefont {Mei\ss{}ner}}]{Li:2018ymw}%
  \BibitemOpen
  \bibfield  {author} {\bibinfo {author} {\bibfnamefont {N.}~\bibnamefont {Li}}, \bibinfo {author} {\bibfnamefont {S.}~\bibnamefont {Elhatisari}}, \bibinfo {author} {\bibfnamefont {E.}~\bibnamefont {Epelbaum}}, \bibinfo {author} {\bibfnamefont {D.}~\bibnamefont {Lee}}, \bibinfo {author} {\bibfnamefont {B.-N.}\ \bibnamefont {Lu}},\ and\ \bibinfo {author} {\bibfnamefont {U.-G.}\ \bibnamefont {Mei\ss{}ner}},\ }\bibfield  {title} {\bibinfo {title} {{Neutron-proton scattering with lattice chiral effective field theory at next-to-next-to-next-to-leading order}},\ }\href {https://doi.org/10.1103/PhysRevC.98.044002} {\bibfield  {journal} {\bibinfo  {journal} {Phys. Rev. C}\ }\textbf {\bibinfo {volume} {98}},\ \bibinfo {pages} {044002} (\bibinfo {year} {2018})},\ \Eprint {https://arxiv.org/abs/1806.07994} {arXiv:1806.07994 [nucl-th]} \BibitemShut {NoStop}%
\bibitem [{\citenamefont {Lu}\ \emph {et~al.}(2019)\citenamefont {Lu}, \citenamefont {Li}, \citenamefont {Elhatisari}, \citenamefont {Lee}, \citenamefont {Epelbaum},\ and\ \citenamefont {Mei\ss{}ner}}]{Lu:2018bat}%
  \BibitemOpen
  \bibfield  {author} {\bibinfo {author} {\bibfnamefont {B.-N.}\ \bibnamefont {Lu}}, \bibinfo {author} {\bibfnamefont {N.}~\bibnamefont {Li}}, \bibinfo {author} {\bibfnamefont {S.}~\bibnamefont {Elhatisari}}, \bibinfo {author} {\bibfnamefont {D.}~\bibnamefont {Lee}}, \bibinfo {author} {\bibfnamefont {E.}~\bibnamefont {Epelbaum}},\ and\ \bibinfo {author} {\bibfnamefont {U.-G.}\ \bibnamefont {Mei\ss{}ner}},\ }\bibfield  {title} {\bibinfo {title} {{Essential elements for nuclear binding}},\ }\href {https://doi.org/10.1016/j.physletb.2019.134863} {\bibfield  {journal} {\bibinfo  {journal} {Phys. Lett. B}\ }\textbf {\bibinfo {volume} {797}},\ \bibinfo {pages} {134863} (\bibinfo {year} {2019})},\ \Eprint {https://arxiv.org/abs/1812.10928} {arXiv:1812.10928 [nucl-th]} \BibitemShut {NoStop}%
\bibitem [{\citenamefont {Lu}\ \emph {et~al.}(2020{\natexlab{b}})\citenamefont {Lu}, \citenamefont {Li}, \citenamefont {Elhatisari}, \citenamefont {Lee}, \citenamefont {Drut}, \citenamefont {L\"ahde}, \citenamefont {Epelbaum},\ and\ \citenamefont {Mei\ss{}ner}}]{Lu:2019nbg}%
  \BibitemOpen
  \bibfield  {author} {\bibinfo {author} {\bibfnamefont {B.-N.}\ \bibnamefont {Lu}}, \bibinfo {author} {\bibfnamefont {N.}~\bibnamefont {Li}}, \bibinfo {author} {\bibfnamefont {S.}~\bibnamefont {Elhatisari}}, \bibinfo {author} {\bibfnamefont {D.}~\bibnamefont {Lee}}, \bibinfo {author} {\bibfnamefont {J.~E.}\ \bibnamefont {Drut}}, \bibinfo {author} {\bibfnamefont {T.~A.}\ \bibnamefont {L\"ahde}}, \bibinfo {author} {\bibfnamefont {E.}~\bibnamefont {Epelbaum}},\ and\ \bibinfo {author} {\bibfnamefont {U.-G.}\ \bibnamefont {Mei\ss{}ner}},\ }\bibfield  {title} {\bibinfo {title} {{$Ab Initio$ Nuclear Thermodynamics}},\ }\href {https://doi.org/10.1103/PhysRevLett.125.192502} {\bibfield  {journal} {\bibinfo  {journal} {Phys. Rev. Lett.}\ }\textbf {\bibinfo {volume} {125}},\ \bibinfo {pages} {192502} (\bibinfo {year} {2020}{\natexlab{b}})},\ \Eprint {https://arxiv.org/abs/1912.05105} {arXiv:1912.05105 [nucl-th]} \BibitemShut {NoStop}%
\bibitem [{\citenamefont {Lu}\ \emph {et~al.}(2022)\citenamefont {Lu}, \citenamefont {Li}, \citenamefont {Elhatisari}, \citenamefont {Ma}, \citenamefont {Lee},\ and\ \citenamefont {Mei\ss{}ner}}]{Lu:2021tab}%
  \BibitemOpen
  \bibfield  {author} {\bibinfo {author} {\bibfnamefont {B.-N.}\ \bibnamefont {Lu}}, \bibinfo {author} {\bibfnamefont {N.}~\bibnamefont {Li}}, \bibinfo {author} {\bibfnamefont {S.}~\bibnamefont {Elhatisari}}, \bibinfo {author} {\bibfnamefont {Y.-Z.}\ \bibnamefont {Ma}}, \bibinfo {author} {\bibfnamefont {D.}~\bibnamefont {Lee}},\ and\ \bibinfo {author} {\bibfnamefont {U.-G.}\ \bibnamefont {Mei\ss{}ner}},\ }\bibfield  {title} {\bibinfo {title} {{Perturbative Quantum Monte~Carlo Method for Nuclear Physics}},\ }\href {https://doi.org/10.1103/PhysRevLett.128.242501} {\bibfield  {journal} {\bibinfo  {journal} {Phys. Rev. Lett.}\ }\textbf {\bibinfo {volume} {128}},\ \bibinfo {pages} {242501} (\bibinfo {year} {2022})},\ \Eprint {https://arxiv.org/abs/2111.14191} {arXiv:2111.14191 [nucl-th]} \BibitemShut {NoStop}%
\bibitem [{\citenamefont {Elhatisari}\ \emph {et~al.}(2022)\citenamefont {Elhatisari} \emph {et~al.}}]{Elhatisari:2022qfr}%
  \BibitemOpen
  \bibfield  {author} {\bibinfo {author} {\bibfnamefont {S.}~\bibnamefont {Elhatisari}} \emph {et~al.},\ }\bibfield  {title} {\bibinfo {title} {{Wave function matching for the quantum many-body problem}},\ }\href@noop {} {\bibfield  {journal} {\bibinfo  {journal} {arXiv:2210.17488}\ } (\bibinfo {year} {2022})}\BibitemShut {NoStop}%
\bibitem [{\citenamefont {Bacca}\ \emph {et~al.}(2013)\citenamefont {Bacca}, \citenamefont {Barnea}, \citenamefont {Leidemann},\ and\ \citenamefont {Orlandini}}]{Bacca:2012xv}%
  \BibitemOpen
  \bibfield  {author} {\bibinfo {author} {\bibfnamefont {S.}~\bibnamefont {Bacca}}, \bibinfo {author} {\bibfnamefont {N.}~\bibnamefont {Barnea}}, \bibinfo {author} {\bibfnamefont {W.}~\bibnamefont {Leidemann}},\ and\ \bibinfo {author} {\bibfnamefont {G.}~\bibnamefont {Orlandini}},\ }\bibfield  {title} {\bibinfo {title} {{Isoscalar Monopole Resonance of the Alpha Particle: A Prism to Nuclear Hamiltonians}},\ }\href {https://doi.org/10.1103/PhysRevLett.110.042503} {\bibfield  {journal} {\bibinfo  {journal} {Phys. Rev. Lett.}\ }\textbf {\bibinfo {volume} {110}},\ \bibinfo {pages} {042503} (\bibinfo {year} {2013})},\ \Eprint {https://arxiv.org/abs/1210.7255} {arXiv:1210.7255 [nucl-th]} \BibitemShut {NoStop}%
\bibitem [{\citenamefont {Kegel}\ \emph {et~al.}(2023)\citenamefont {Kegel}, \citenamefont {Achenbach}, \citenamefont {Bacca}, \citenamefont {Barnea}, \citenamefont {Beri\ifmmode \check{c}\else \v{c}\fi{}i\ifmmode~\check{c}\else \v{c}\fi{}}, \citenamefont {Bosnar}, \citenamefont {Correa}, \citenamefont {Distler}, \citenamefont {Esser}, \citenamefont {Fonvieille}, \citenamefont {Fri\ifmmode \check{s}\else \v{s}\fi{}\ifmmode \check{c}\else \v{c}\fi{}i\ifmmode~\acute{c}\else \'{c}\fi{}}, \citenamefont {Heilig}, \citenamefont {Herrmann}, \citenamefont {Hoek}, \citenamefont {Klag}, \citenamefont {Kolar}, \citenamefont {Leidemann}, \citenamefont {Merkel}, \citenamefont {Mihovilovi\ifmmode~\check{c}\else \v{c}\fi{}}, \citenamefont {M\"uller}, \citenamefont {M\"uller}, \citenamefont {Orlandini}, \citenamefont {Pochodzalla}, \citenamefont {Schlimme}, \citenamefont {Schoth}, \citenamefont {Schulz}, \citenamefont {Sfienti}, \citenamefont {\ifmmode~\check{S}\else \v{S}\fi{}irca}, \citenamefont {Spreckels}, \citenamefont
  {St\"ottinger}, \citenamefont {Thiel}, \citenamefont {Tyukin}, \citenamefont {Walcher},\ and\ \citenamefont {Weber}}]{PhysRevLett.130.152502}%
  \BibitemOpen
  \bibfield  {author} {\bibinfo {author} {\bibfnamefont {S.}~\bibnamefont {Kegel}}, \bibinfo {author} {\bibfnamefont {P.}~\bibnamefont {Achenbach}}, \bibinfo {author} {\bibfnamefont {S.}~\bibnamefont {Bacca}}, \bibinfo {author} {\bibfnamefont {N.}~\bibnamefont {Barnea}}, \bibinfo {author} {\bibfnamefont {J.}~\bibnamefont {Beri\ifmmode \check{c}\else \v{c}\fi{}i\ifmmode~\check{c}\else \v{c}\fi{}}}, \bibinfo {author} {\bibfnamefont {D.}~\bibnamefont {Bosnar}}, \bibinfo {author} {\bibfnamefont {L.}~\bibnamefont {Correa}}, \bibinfo {author} {\bibfnamefont {M.~O.}\ \bibnamefont {Distler}}, \bibinfo {author} {\bibfnamefont {A.}~\bibnamefont {Esser}}, \bibinfo {author} {\bibfnamefont {H.}~\bibnamefont {Fonvieille}}, \bibinfo {author} {\bibfnamefont {I.}~\bibnamefont {Fri\ifmmode \check{s}\else \v{s}\fi{}\ifmmode \check{c}\else \v{c}\fi{}i\ifmmode~\acute{c}\else \'{c}\fi{}}}, \bibinfo {author} {\bibfnamefont {M.}~\bibnamefont {Heilig}}, \bibinfo {author} {\bibfnamefont {P.}~\bibnamefont {Herrmann}}, \bibinfo {author}
  {\bibfnamefont {M.}~\bibnamefont {Hoek}}, \bibinfo {author} {\bibfnamefont {P.}~\bibnamefont {Klag}}, \bibinfo {author} {\bibfnamefont {T.}~\bibnamefont {Kolar}}, \bibinfo {author} {\bibfnamefont {W.}~\bibnamefont {Leidemann}}, \bibinfo {author} {\bibfnamefont {H.}~\bibnamefont {Merkel}}, \bibinfo {author} {\bibfnamefont {M.}~\bibnamefont {Mihovilovi\ifmmode~\check{c}\else \v{c}\fi{}}}, \bibinfo {author} {\bibfnamefont {J.}~\bibnamefont {M\"uller}}, \bibinfo {author} {\bibfnamefont {U.}~\bibnamefont {M\"uller}}, \bibinfo {author} {\bibfnamefont {G.}~\bibnamefont {Orlandini}}, \bibinfo {author} {\bibfnamefont {J.}~\bibnamefont {Pochodzalla}}, \bibinfo {author} {\bibfnamefont {B.~S.}\ \bibnamefont {Schlimme}}, \bibinfo {author} {\bibfnamefont {M.}~\bibnamefont {Schoth}}, \bibinfo {author} {\bibfnamefont {F.}~\bibnamefont {Schulz}}, \bibinfo {author} {\bibfnamefont {C.}~\bibnamefont {Sfienti}}, \bibinfo {author} {\bibfnamefont {S.}~\bibnamefont {\ifmmode~\check{S}\else \v{S}\fi{}irca}}, \bibinfo {author}
  {\bibfnamefont {R.}~\bibnamefont {Spreckels}}, \bibinfo {author} {\bibfnamefont {Y.}~\bibnamefont {St\"ottinger}}, \bibinfo {author} {\bibfnamefont {M.}~\bibnamefont {Thiel}}, \bibinfo {author} {\bibfnamefont {A.}~\bibnamefont {Tyukin}}, \bibinfo {author} {\bibfnamefont {T.}~\bibnamefont {Walcher}},\ and\ \bibinfo {author} {\bibfnamefont {A.}~\bibnamefont {Weber}},\ }\bibfield  {title} {\bibinfo {title} {Measurement of the $\ensuremath{\alpha}$-particle monopole transition form factor challenges theory: A low-energy puzzle for nuclear forces?},\ }\href {https://doi.org/10.1103/PhysRevLett.130.152502} {\bibfield  {journal} {\bibinfo  {journal} {Phys. Rev. Lett.}\ }\textbf {\bibinfo {volume} {130}},\ \bibinfo {pages} {152502} (\bibinfo {year} {2023})}\BibitemShut {NoStop}%
\bibitem [{\citenamefont {Mei\ss{}ner}\ \emph {et~al.}(2023)\citenamefont {Mei\ss{}ner}, \citenamefont {Shen}, \citenamefont {Elhatisari},\ and\ \citenamefont {Lee}}]{Meissner:2023cvo}%
  \BibitemOpen
  \bibfield  {author} {\bibinfo {author} {\bibfnamefont {U.-G.}\ \bibnamefont {Mei\ss{}ner}}, \bibinfo {author} {\bibfnamefont {S.}~\bibnamefont {Shen}}, \bibinfo {author} {\bibfnamefont {S.}~\bibnamefont {Elhatisari}},\ and\ \bibinfo {author} {\bibfnamefont {D.}~\bibnamefont {Lee}},\ }\bibfield  {title} {\bibinfo {title} {{Ab initio calculation of the alpha-particle monopole transition form factor: No puzzle for nuclear forces}},\ }\href@noop {} {\bibfield  {journal} {\bibinfo  {journal} {arXiv:2309.01558}\ } (\bibinfo {year} {2023})}\BibitemShut {NoStop}%
\bibitem [{\citenamefont {Eliyahu}\ \emph {et~al.}(2020)\citenamefont {Eliyahu}, \citenamefont {Bazak},\ and\ \citenamefont {Barnea}}]{Eliyahu:2019nkz}%
  \BibitemOpen
  \bibfield  {author} {\bibinfo {author} {\bibfnamefont {M.}~\bibnamefont {Eliyahu}}, \bibinfo {author} {\bibfnamefont {B.}~\bibnamefont {Bazak}},\ and\ \bibinfo {author} {\bibfnamefont {N.}~\bibnamefont {Barnea}},\ }\bibfield  {title} {\bibinfo {title} {{Extrapolating Lattice QCD Results using Effective Field Theory}},\ }\href {https://doi.org/10.1103/PhysRevC.102.044003} {\bibfield  {journal} {\bibinfo  {journal} {Phys. Rev. C}\ }\textbf {\bibinfo {volume} {102}},\ \bibinfo {pages} {044003} (\bibinfo {year} {2020})},\ \Eprint {https://arxiv.org/abs/1912.07017} {arXiv:1912.07017 [nucl-th]} \BibitemShut {NoStop}%
\bibitem [{\citenamefont {Beane}\ \emph {et~al.}(2013)\citenamefont {Beane}, \citenamefont {Chang}, \citenamefont {Cohen}, \citenamefont {Detmold}, \citenamefont {Lin}, \citenamefont {Luu}, \citenamefont {Orginos}, \citenamefont {Parreno}, \citenamefont {Savage},\ and\ \citenamefont {Walker-Loud}}]{NPLQCD:2012mex}%
  \BibitemOpen
  \bibfield  {author} {\bibinfo {author} {\bibfnamefont {S.~R.}\ \bibnamefont {Beane}}, \bibinfo {author} {\bibfnamefont {E.}~\bibnamefont {Chang}}, \bibinfo {author} {\bibfnamefont {S.~D.}\ \bibnamefont {Cohen}}, \bibinfo {author} {\bibfnamefont {W.}~\bibnamefont {Detmold}}, \bibinfo {author} {\bibfnamefont {H.~W.}\ \bibnamefont {Lin}}, \bibinfo {author} {\bibfnamefont {T.~C.}\ \bibnamefont {Luu}}, \bibinfo {author} {\bibfnamefont {K.}~\bibnamefont {Orginos}}, \bibinfo {author} {\bibfnamefont {A.}~\bibnamefont {Parreno}}, \bibinfo {author} {\bibfnamefont {M.~J.}\ \bibnamefont {Savage}},\ and\ \bibinfo {author} {\bibfnamefont {A.}~\bibnamefont {Walker-Loud}} (\bibinfo {collaboration} {NPLQCD}),\ }\bibfield  {title} {\bibinfo {title} {{Light Nuclei and Hypernuclei from Quantum Chromodynamics in the Limit of SU(3) Flavor Symmetry}},\ }\href {https://doi.org/10.1103/PhysRevD.87.034506} {\bibfield  {journal} {\bibinfo  {journal} {Phys. Rev. D}\ }\textbf {\bibinfo {volume} {87}},\ \bibinfo {pages} {034506}
  (\bibinfo {year} {2013})},\ \Eprint {https://arxiv.org/abs/1206.5219} {arXiv:1206.5219 [hep-lat]} \BibitemShut {NoStop}%
\bibitem [{\citenamefont {Barnea}\ \emph {et~al.}(2015)\citenamefont {Barnea}, \citenamefont {Contessi}, \citenamefont {Gazit}, \citenamefont {Pederiva},\ and\ \citenamefont {van Kolck}}]{PhysRevLett.114.052501}%
  \BibitemOpen
  \bibfield  {author} {\bibinfo {author} {\bibfnamefont {N.}~\bibnamefont {Barnea}}, \bibinfo {author} {\bibfnamefont {L.}~\bibnamefont {Contessi}}, \bibinfo {author} {\bibfnamefont {D.}~\bibnamefont {Gazit}}, \bibinfo {author} {\bibfnamefont {F.}~\bibnamefont {Pederiva}},\ and\ \bibinfo {author} {\bibfnamefont {U.}~\bibnamefont {van Kolck}},\ }\bibfield  {title} {\bibinfo {title} {Effective field theory for lattice nuclei},\ }\href {https://doi.org/10.1103/PhysRevLett.114.052501} {\bibfield  {journal} {\bibinfo  {journal} {Phys. Rev. Lett.}\ }\textbf {\bibinfo {volume} {114}},\ \bibinfo {pages} {052501} (\bibinfo {year} {2015})}\BibitemShut {NoStop}%
\bibitem [{\citenamefont {Detmold}\ and\ \citenamefont {Shanahan}(2021)}]{Detmold:2021oro}%
  \BibitemOpen
  \bibfield  {author} {\bibinfo {author} {\bibfnamefont {W.}~\bibnamefont {Detmold}}\ and\ \bibinfo {author} {\bibfnamefont {P.~E.}\ \bibnamefont {Shanahan}},\ }\bibfield  {title} {\bibinfo {title} {{Few-nucleon matrix elements in pionless effective field theory in a finite volume}},\ }\href {https://doi.org/10.1103/PhysRevD.103.074503} {\bibfield  {journal} {\bibinfo  {journal} {Phys. Rev. D}\ }\textbf {\bibinfo {volume} {103}},\ \bibinfo {pages} {074503} (\bibinfo {year} {2021})},\ \Eprint {https://arxiv.org/abs/2102.04329} {arXiv:2102.04329 [nucl-th]} \BibitemShut {NoStop}%
\bibitem [{\citenamefont {Sun}\ \emph {et~al.}(2022)\citenamefont {Sun}, \citenamefont {Detmold}, \citenamefont {Luo},\ and\ \citenamefont {Shanahan}}]{Sun:2022frr}%
  \BibitemOpen
  \bibfield  {author} {\bibinfo {author} {\bibfnamefont {X.}~\bibnamefont {Sun}}, \bibinfo {author} {\bibfnamefont {W.}~\bibnamefont {Detmold}}, \bibinfo {author} {\bibfnamefont {D.}~\bibnamefont {Luo}},\ and\ \bibinfo {author} {\bibfnamefont {P.~E.}\ \bibnamefont {Shanahan}},\ }\bibfield  {title} {\bibinfo {title} {{Finite-volume pionless effective field theory for few-nucleon systems with differentiable programming}},\ }\href {https://doi.org/10.1103/PhysRevD.105.074508} {\bibfield  {journal} {\bibinfo  {journal} {Phys. Rev. D}\ }\textbf {\bibinfo {volume} {105}},\ \bibinfo {pages} {074508} (\bibinfo {year} {2022})},\ \Eprint {https://arxiv.org/abs/2202.03530} {arXiv:2202.03530 [nucl-th]} \BibitemShut {NoStop}%
\bibitem [{\citenamefont {Detmold}\ \emph {et~al.}(2023)\citenamefont {Detmold}, \citenamefont {Romero-L\'opez},\ and\ \citenamefont {Shanahan}}]{Detmold:2023lwn}%
  \BibitemOpen
  \bibfield  {author} {\bibinfo {author} {\bibfnamefont {W.}~\bibnamefont {Detmold}}, \bibinfo {author} {\bibfnamefont {F.}~\bibnamefont {Romero-L\'opez}},\ and\ \bibinfo {author} {\bibfnamefont {P.~E.}\ \bibnamefont {Shanahan}},\ }\bibfield  {title} {\bibinfo {title} {{Constraint of pionless EFT using two-nucleon spectra from lattice QCD}},\ }\href {https://doi.org/10.1103/PhysRevD.108.034509} {\bibfield  {journal} {\bibinfo  {journal} {Phys. Rev. D}\ }\textbf {\bibinfo {volume} {108}},\ \bibinfo {pages} {034509} (\bibinfo {year} {2023})},\ \Eprint {https://arxiv.org/abs/2305.06313} {arXiv:2305.06313 [nucl-th]} \BibitemShut {NoStop}%
\bibitem [{\citenamefont {Bazak}\ \emph {et~al.}(2022)\citenamefont {Bazak}, \citenamefont {Sch\"afer}, \citenamefont {Yaron},\ and\ \citenamefont {Barnea}}]{Bazak:2022mjh}%
  \BibitemOpen
  \bibfield  {author} {\bibinfo {author} {\bibfnamefont {B.}~\bibnamefont {Bazak}}, \bibinfo {author} {\bibfnamefont {M.}~\bibnamefont {Sch\"afer}}, \bibinfo {author} {\bibfnamefont {R.}~\bibnamefont {Yaron}},\ and\ \bibinfo {author} {\bibfnamefont {N.}~\bibnamefont {Barnea}},\ }\bibfield  {title} {\bibinfo {title} {{Spectrum of few-body systems in a finite volume}},\ }\href {https://doi.org/10.1051/epjconf/202227101011} {\bibfield  {journal} {\bibinfo  {journal} {EPJ Web Conf.}\ }\textbf {\bibinfo {volume} {271}},\ \bibinfo {pages} {01011} (\bibinfo {year} {2022})},\ \Eprint {https://arxiv.org/abs/2206.04497} {arXiv:2206.04497 [nucl-th]} \BibitemShut {NoStop}%
\bibitem [{\citenamefont {Lepage}(1997)}]{Lepage:1997cs}%
  \BibitemOpen
  \bibfield  {author} {\bibinfo {author} {\bibfnamefont {G.~P.}\ \bibnamefont {Lepage}},\ }\bibfield  {title} {\bibinfo {title} {{How to renormalize the Schrodinger equation}},\ }in\ \href@noop {} {\emph {\bibinfo {booktitle} {{8th Jorge Andre Swieca Summer School on Nuclear Physics}}}}\ (\bibinfo {year} {1997})\ pp.\ \bibinfo {pages} {135--180},\ \Eprint {https://arxiv.org/abs/nucl-th/9706029} {arXiv:nucl-th/9706029} \BibitemShut {NoStop}%
\bibitem [{\citenamefont {Joyce}\ and\ \citenamefont {Zucker}(2005)}]{joyce2005evaluation}%
  \BibitemOpen
  \bibfield  {author} {\bibinfo {author} {\bibfnamefont {G.}~\bibnamefont {Joyce}}\ and\ \bibinfo {author} {\bibfnamefont {I.}~\bibnamefont {Zucker}},\ }\bibfield  {title} {\bibinfo {title} {On the evaluation of generalized watson integrals},\ }\href@noop {} {\bibfield  {journal} {\bibinfo  {journal} {Proceedings of the American Mathematical Society}\ }\textbf {\bibinfo {volume} {133}},\ \bibinfo {pages} {71} (\bibinfo {year} {2005})}\BibitemShut {NoStop}%
\bibitem [{\citenamefont {Braaten}\ and\ \citenamefont {Hammer}(2006)}]{Braaten:2004rn}%
  \BibitemOpen
  \bibfield  {author} {\bibinfo {author} {\bibfnamefont {E.}~\bibnamefont {Braaten}}\ and\ \bibinfo {author} {\bibfnamefont {H.~W.}\ \bibnamefont {Hammer}},\ }\bibfield  {title} {\bibinfo {title} {{Universality in few-body systems with large scattering length}},\ }\href {https://doi.org/10.1016/j.physrep.2006.03.001} {\bibfield  {journal} {\bibinfo  {journal} {Phys. Rept.}\ }\textbf {\bibinfo {volume} {428}},\ \bibinfo {pages} {259} (\bibinfo {year} {2006})},\ \Eprint {https://arxiv.org/abs/cond-mat/0410417} {arXiv:cond-mat/0410417} \BibitemShut {NoStop}%
\bibitem [{\citenamefont {Blankleider}\ and\ \citenamefont {Gegelia}(2000)}]{Blankleider:2000vi}%
  \BibitemOpen
  \bibfield  {author} {\bibinfo {author} {\bibfnamefont {B.}~\bibnamefont {Blankleider}}\ and\ \bibinfo {author} {\bibfnamefont {J.}~\bibnamefont {Gegelia}},\ }\bibfield  {title} {\bibinfo {title} {{Three body system in leading order effective field theory without three body forces}},\ }\href@noop {} {\bibfield  {journal} {\bibinfo  {journal} {arXiv:nucl-th/0009007}\ } (\bibinfo {year} {2000})},\ \Eprint {https://arxiv.org/abs/nucl-th/0009007} {arXiv:nucl-th/0009007} \BibitemShut {NoStop}%
\bibitem [{\citenamefont {Platter}\ \emph {et~al.}(2004)\citenamefont {Platter}, \citenamefont {Hammer},\ and\ \citenamefont {Meissner}}]{Platter:2004he}%
  \BibitemOpen
  \bibfield  {author} {\bibinfo {author} {\bibfnamefont {L.}~\bibnamefont {Platter}}, \bibinfo {author} {\bibfnamefont {H.~W.}\ \bibnamefont {Hammer}},\ and\ \bibinfo {author} {\bibfnamefont {U.-G.}\ \bibnamefont {Meissner}},\ }\bibfield  {title} {\bibinfo {title} {{The Four boson system with short range interactions}},\ }\href {https://doi.org/10.1103/PhysRevA.70.052101} {\bibfield  {journal} {\bibinfo  {journal} {Phys. Rev. A}\ }\textbf {\bibinfo {volume} {70}},\ \bibinfo {pages} {052101} (\bibinfo {year} {2004})},\ \Eprint {https://arxiv.org/abs/cond-mat/0404313} {arXiv:cond-mat/0404313} \BibitemShut {NoStop}%
\bibitem [{\citenamefont {Bazak}\ \emph {et~al.}(2019)\citenamefont {Bazak}, \citenamefont {Kirscher}, \citenamefont {K\"onig}, \citenamefont {Pav\'on~Valderrama}, \citenamefont {Barnea},\ and\ \citenamefont {van Kolck}}]{Bazak:2018qnu}%
  \BibitemOpen
  \bibfield  {author} {\bibinfo {author} {\bibfnamefont {B.}~\bibnamefont {Bazak}}, \bibinfo {author} {\bibfnamefont {J.}~\bibnamefont {Kirscher}}, \bibinfo {author} {\bibfnamefont {S.}~\bibnamefont {K\"onig}}, \bibinfo {author} {\bibfnamefont {M.}~\bibnamefont {Pav\'on~Valderrama}}, \bibinfo {author} {\bibfnamefont {N.}~\bibnamefont {Barnea}},\ and\ \bibinfo {author} {\bibfnamefont {U.}~\bibnamefont {van Kolck}},\ }\bibfield  {title} {\bibinfo {title} {{Four-Body Scale in Universal Few-Boson Systems}},\ }\href {https://doi.org/10.1103/PhysRevLett.122.143001} {\bibfield  {journal} {\bibinfo  {journal} {Phys. Rev. Lett.}\ }\textbf {\bibinfo {volume} {122}},\ \bibinfo {pages} {143001} (\bibinfo {year} {2019})},\ \Eprint {https://arxiv.org/abs/1812.00387} {arXiv:1812.00387 [cond-mat.quant-gas]} \BibitemShut {NoStop}%
\bibitem [{\citenamefont {Lin}(2023)}]{Lin:2023zqw}%
  \BibitemOpen
  \bibfield  {author} {\bibinfo {author} {\bibfnamefont {X.}~\bibnamefont {Lin}},\ }\bibfield  {title} {\bibinfo {title} {{Four-Body Systems at Large Cutoffs in Effective Field Theory}},\ }\href@noop {} {\bibfield  {journal} {\bibinfo  {journal} {arXiv:2304.06172}\ } (\bibinfo {year} {2023})}\BibitemShut {NoStop}%
\bibitem [{\citenamefont {Michael}(1985)}]{Michael:1985ne}%
  \BibitemOpen
  \bibfield  {author} {\bibinfo {author} {\bibfnamefont {C.}~\bibnamefont {Michael}},\ }\bibfield  {title} {\bibinfo {title} {{Adjoint Sources in Lattice Gauge Theory}},\ }\href {https://doi.org/10.1016/0550-3213(85)90297-4} {\bibfield  {journal} {\bibinfo  {journal} {Nucl. Phys. B}\ }\textbf {\bibinfo {volume} {259}},\ \bibinfo {pages} {58} (\bibinfo {year} {1985})}\BibitemShut {NoStop}%
\bibitem [{\citenamefont {Luscher}\ and\ \citenamefont {Wolff}(1990)}]{Luscher:1990ck}%
  \BibitemOpen
  \bibfield  {author} {\bibinfo {author} {\bibfnamefont {M.}~\bibnamefont {Luscher}}\ and\ \bibinfo {author} {\bibfnamefont {U.}~\bibnamefont {Wolff}},\ }\bibfield  {title} {\bibinfo {title} {{How to Calculate the Elastic Scattering Matrix in Two-dimensional Quantum Field Theories by Numerical Simulation}},\ }\href {https://doi.org/10.1016/0550-3213(90)90540-T} {\bibfield  {journal} {\bibinfo  {journal} {Nucl. Phys. B}\ }\textbf {\bibinfo {volume} {339}},\ \bibinfo {pages} {222} (\bibinfo {year} {1990})}\BibitemShut {NoStop}%
\bibitem [{\citenamefont {Basak}\ \emph {et~al.}(2005)\citenamefont {Basak}, \citenamefont {Edwards}, \citenamefont {Fleming}, \citenamefont {Heller}, \citenamefont {Morningstar}, \citenamefont {Richards}, \citenamefont {Sato},\ and\ \citenamefont {Wallace}}]{Basak:2005aq}%
  \BibitemOpen
  \bibfield  {author} {\bibinfo {author} {\bibfnamefont {S.}~\bibnamefont {Basak}}, \bibinfo {author} {\bibfnamefont {R.~G.}\ \bibnamefont {Edwards}}, \bibinfo {author} {\bibfnamefont {G.~T.}\ \bibnamefont {Fleming}}, \bibinfo {author} {\bibfnamefont {U.~M.}\ \bibnamefont {Heller}}, \bibinfo {author} {\bibfnamefont {C.}~\bibnamefont {Morningstar}}, \bibinfo {author} {\bibfnamefont {D.}~\bibnamefont {Richards}}, \bibinfo {author} {\bibfnamefont {I.}~\bibnamefont {Sato}},\ and\ \bibinfo {author} {\bibfnamefont {S.}~\bibnamefont {Wallace}},\ }\bibfield  {title} {\bibinfo {title} {{Group-theoretical construction of extended baryon operators in lattice QCD}},\ }\href {https://doi.org/10.1103/PhysRevD.72.094506} {\bibfield  {journal} {\bibinfo  {journal} {Phys. Rev. D}\ }\textbf {\bibinfo {volume} {72}},\ \bibinfo {pages} {094506} (\bibinfo {year} {2005})},\ \Eprint {https://arxiv.org/abs/hep-lat/0506029} {arXiv:hep-lat/0506029} \BibitemShut {NoStop}%
\bibitem [{\citenamefont {He}\ \emph {et~al.}(2022)\citenamefont {He}, \citenamefont {Brantley}, \citenamefont {Chang}, \citenamefont {Chernyshev}, \citenamefont {Berkowitz}, \citenamefont {Howarth}, \citenamefont {K\"orber}, \citenamefont {Meyer}, \citenamefont {Monge-Camacho}, \citenamefont {Rinaldi}, \citenamefont {Bouchard}, \citenamefont {Clark}, \citenamefont {Gambhir}, \citenamefont {Monahan}, \citenamefont {Nicholson}, \citenamefont {Vranas},\ and\ \citenamefont {Walker-Loud}}]{PhysRevC.105.065203}%
  \BibitemOpen
  \bibfield  {author} {\bibinfo {author} {\bibfnamefont {J.}~\bibnamefont {He}}, \bibinfo {author} {\bibfnamefont {D.~A.}\ \bibnamefont {Brantley}}, \bibinfo {author} {\bibfnamefont {C.~C.}\ \bibnamefont {Chang}}, \bibinfo {author} {\bibfnamefont {I.}~\bibnamefont {Chernyshev}}, \bibinfo {author} {\bibfnamefont {E.}~\bibnamefont {Berkowitz}}, \bibinfo {author} {\bibfnamefont {D.}~\bibnamefont {Howarth}}, \bibinfo {author} {\bibfnamefont {C.}~\bibnamefont {K\"orber}}, \bibinfo {author} {\bibfnamefont {A.~S.}\ \bibnamefont {Meyer}}, \bibinfo {author} {\bibfnamefont {H.}~\bibnamefont {Monge-Camacho}}, \bibinfo {author} {\bibfnamefont {E.}~\bibnamefont {Rinaldi}}, \bibinfo {author} {\bibfnamefont {C.}~\bibnamefont {Bouchard}}, \bibinfo {author} {\bibfnamefont {M.~A.}\ \bibnamefont {Clark}}, \bibinfo {author} {\bibfnamefont {A.~S.}\ \bibnamefont {Gambhir}}, \bibinfo {author} {\bibfnamefont {C.~J.}\ \bibnamefont {Monahan}}, \bibinfo {author} {\bibfnamefont {A.}~\bibnamefont {Nicholson}}, \bibinfo {author}
  {\bibfnamefont {P.}~\bibnamefont {Vranas}},\ and\ \bibinfo {author} {\bibfnamefont {A.}~\bibnamefont {Walker-Loud}},\ }\bibfield  {title} {\bibinfo {title} {Detailed analysis of excited-state systematics in a lattice qcd calculation of ${g}_{A}$},\ }\href {https://doi.org/10.1103/PhysRevC.105.065203} {\bibfield  {journal} {\bibinfo  {journal} {Phys. Rev. C}\ }\textbf {\bibinfo {volume} {105}},\ \bibinfo {pages} {065203} (\bibinfo {year} {2022})}\BibitemShut {NoStop}%
\bibitem [{\citenamefont {Naidon}\ and\ \citenamefont {Endo}(2017)}]{Naidon:2016dpf}%
  \BibitemOpen
  \bibfield  {author} {\bibinfo {author} {\bibfnamefont {P.}~\bibnamefont {Naidon}}\ and\ \bibinfo {author} {\bibfnamefont {S.}~\bibnamefont {Endo}},\ }\bibfield  {title} {\bibinfo {title} {{Efimov Physics: a review}},\ }\href {https://doi.org/10.1088/1361-6633/aa50e8} {\bibfield  {journal} {\bibinfo  {journal} {Rept. Prog. Phys.}\ }\textbf {\bibinfo {volume} {80}},\ \bibinfo {pages} {056001} (\bibinfo {year} {2017})},\ \Eprint {https://arxiv.org/abs/1610.09805} {arXiv:1610.09805 [quant-ph]} \BibitemShut {NoStop}%
\bibitem [{\citenamefont {Kievsky}\ \emph {et~al.}(2021)\citenamefont {Kievsky}, \citenamefont {Girlanda}, \citenamefont {Gattobigio},\ and\ \citenamefont {Viviani}}]{Kievsky:2021ghz}%
  \BibitemOpen
  \bibfield  {author} {\bibinfo {author} {\bibfnamefont {A.}~\bibnamefont {Kievsky}}, \bibinfo {author} {\bibfnamefont {L.}~\bibnamefont {Girlanda}}, \bibinfo {author} {\bibfnamefont {M.}~\bibnamefont {Gattobigio}},\ and\ \bibinfo {author} {\bibfnamefont {M.}~\bibnamefont {Viviani}},\ }\bibfield  {title} {\bibinfo {title} {{Efimov Physics and Connections to Nuclear Physics}},\ }\href {https://doi.org/10.1146/annurev-nucl-102419-032845} {\bibfield  {journal} {\bibinfo  {journal} {Ann. Rev. Nucl. Part. Sci.}\ }\textbf {\bibinfo {volume} {71}},\ \bibinfo {pages} {465} (\bibinfo {year} {2021})},\ \Eprint {https://arxiv.org/abs/2102.13504} {arXiv:2102.13504 [nucl-th]} \BibitemShut {NoStop}%
\bibitem [{\citenamefont {de~Forcrand}\ \emph {et~al.}(2005)\citenamefont {de~Forcrand}, \citenamefont {Lucini},\ and\ \citenamefont {Vettorazzo}}]{deForcrand:2004jt}%
  \BibitemOpen
  \bibfield  {author} {\bibinfo {author} {\bibfnamefont {P.}~\bibnamefont {de~Forcrand}}, \bibinfo {author} {\bibfnamefont {B.}~\bibnamefont {Lucini}},\ and\ \bibinfo {author} {\bibfnamefont {M.}~\bibnamefont {Vettorazzo}},\ }\bibfield  {title} {\bibinfo {title} {{Measuring interface tensions in 4d SU(N) lattice gauge theories}},\ }\href {https://doi.org/10.1016/j.nuclphysbps.2004.11.260} {\bibfield  {journal} {\bibinfo  {journal} {Nucl. Phys. B Proc. Suppl.}\ }\textbf {\bibinfo {volume} {140}},\ \bibinfo {pages} {647} (\bibinfo {year} {2005})},\ \Eprint {https://arxiv.org/abs/hep-lat/0409148} {arXiv:hep-lat/0409148} \BibitemShut {NoStop}%
\end{thebibliography}%

\end{document}